\documentclass[aps,prl,reprint,balancelastpage,nofootinbib,preprintnumbers,amsmath,amssymb,superscriptaddress]{revtex4-1}

\usepackage{tensor}     % manipulate tensors
\usepackage{graphicx}   % include figures
\usepackage[
colorlinks=true,        % color link
citecolor=blue,         % cite color
linkcolor=blue,         % link color
urlcolor=blue           % url color
]{hyperref}             % create hyperlinks
\usepackage{bm}         % \bm{<text>} Bold math symbols
\usepackage{xcolor}     % textcolor
\usepackage{tabulary}
\usepackage{color}      % \textcolor{declared-color}{text}
\usepackage[utf8]{inputenc} % accented characters for .bib file using XeLaTex
\usepackage[section]{placeins} % figures processing
\usepackage{graphics,appendix,afterpage,makecell} 
\usepackage{orcidlink}
\newcommand{\nc}{\newcommand*}

\nc{\al}{\alpha}
\nc{\s}{\sigma}
\nc{\dt}{\delta}
\nc{\Dt}{\Delta}
\nc{\Ld}{\Lambda}
\nc{\p}{\partial}
\nc{\om}{\omega}
\nc{\Om}{\Omega}
\nc{\rd}{\mathrm{d}}
\nc{\Od}[1]{\mathcal{O}(#1)} % order operator
\nc{\kp}{\kappa}

\def\({\left(}
\def\){\right)}
\def\[{\left[}
\def\]{\right]}
\def\e{\begin{equation}}
\def\q{\end{equation}}
\def\m{\begin{eqnarray}}
\def\n{\end{eqnarray}}

\nc{\Eq}[1]{Eq.~\eqref{#1}}     % equation
\nc{\Fig}[1]{Fig.~\ref{#1}}     % figure
\nc{\Table}[1]{Table~\ref{#1}}  % table
\nc{\Sec}[1]{Sec.~\ref{#1}}     % section

\nc{\Msun}{M_\odot}             % solar mass
\nc{\fpbh}{f_{\mathrm{pbh}}}    % f_pbh
\nc{\fpbhn}{f_{\mathrm{pbh0}}}    % f_pbh
\nc{\mR}{\mathcal{R}} % merger rate density
\nc{\seq}{\sigma_{\mathrm{eq}}}
\nc{\ogw}{\Omega_{\mathrm{GW}}}
\nc{\gpcyr}{\mathrm{Gpc}^{-3}\,\mathrm{yr}^{-1}}
\nc{\lvc}{LIGO/Virgo} % LIGO-VIRGO collaboration
\nc{\SNR}{\mathrm{SNR}} % signal to noise ratio
\nc{\mmin}{{m_{\mathrm{min}}}}
\nc{\mmax}{{m_{\mathrm{max}}}}
\nc{\Mmin}{{M_{\mathrm{min}}}}
\nc{\fmin}{{f_{\mathrm{min}}}}
\nc{\VT}{\mathrm{VT}}
\nc{\rhoGW}{\rho_{\mathrm{GW}}}
\nc{\vth}{\vec{\theta}}
\nc{\vd}{\vec{d}}
\nc{\vla}{\vec{\lambda}}
\nc{\Nobs}{N_{\mathrm{obs}}}
\nc{\av}[1]{\langle #1 \rangle} % average bracket
\nc{\km}{\mathrm{km}}
\nc{\Mpc}{\mathrm{Mpc}}
\nc{\Tobs}{T_{\mathrm{obs}}}
\nc{\Ntemp}{N_{\mathrm{temp}}}

\nc{\addref}{[\textcolor{red}{add ref}] } % placeholder of references
\nc{\eg}{\textit{e.g.~}}
\nc{\app}{\approx}
\nc{\hf}{\frac{1}{2}}
\nc{\discuss}{\textcolor{red}{Add discussion here!}}

\nc{\mpbh}{m_{\rm{pbh}}}
\nc{\cR}{\mathcal{R}}
\nc{\mU}{{\mathcal{U}}}
\nc{\Mc}{{M_\mathrm{c}}}
\nc{\Mf}{{M_\mathrm{f}}}
\nc{\red}[1]{\textcolor{red}{#1}}
\nc{\yellow}[1]{\textcolor{yellow}{#1}}
\nc{\green}[1]{\textcolor{green}{#1}}
\nc{\blue}[1]{\textcolor{blue}{#1}}
\nc{\fnl}{F_{\mathrm{NL}}}
\nc{\gnl}{G_{\mathrm{NL}}}
\nc{\MG}{\mathcal{M}_{\mathrm{G}}}
\nc{\MNG}{\mathcal{M}_{\mathrm{NG}}}
\begin{document}
	
\title{Probing the speed of scalar-induced gravitational waves with pulsar timing arrays} 

%%%%%%%%%%%%%%%%%%%%%%%%%%%%%%%%%%%% author %%%%%%%%%%%%%%%%%%%%%%%%%%%%%%%%%%%%
\author{Zu-Cheng Chen\orcidlink{0000-0001-7016-9934}}
\email{zuchengchen@gmail.com}
\affiliation{Department of Physics and Synergetic Innovation Center for Quantum Effects and Applications, Hunan Normal University, Changsha, Hunan 410081, China}
\affiliation{Institute of Interdisciplinary Studies, Hunan Normal University, Changsha, Hunan 410081, China}

%%%%%%%%%%%%%%%%%%%%%%%%%%%%%%%%%%%% author %%%%%%%%%%%%%%%%%%%%%%%%%%%%%%%%%%%%
\author{Jun~Li\orcidlink{0000-0001-5173-271X}}
\email{lijun@qust.edu.cn}
\affiliation{School of Mathematics and Physics, Qingdao University of Science and Technology, Qingdao 266061, China}
\affiliation{CAS Key Laboratory of Theoretical Physics, Institute of Theoretical Physics, Chinese Academy of Sciences, Beijing 100190, China}

%%%%%%%%%%%%%%%%%%%%%%%%%%%%%%%%%%%%%%%%%%%%%%%%%%%%%%%%%%%%%%%%%
% \author[0000-0002-0297-9633]{Lang~Liu}
\author{Lang~Liu\orcidlink{0000-0002-0297-9633}}
\email{Corresponding author: liulang@bnu.edu.cn}	
\affiliation{Department of Astronomy, Beijing Normal University, Beijing 100875, China}
\affiliation{Advanced Institute of Natural Sciences, Beijing Normal University, Zhuhai 519087, China}

%%%%%%%%%%%%%%%%%%%%%%%%%%%%%%%%%%%% author %%%%%%%%%%%%%%%%%%%%%%%%%%%%%%%%%%%%
\author{Zhu~Yi \orcidlink{0000-0001-7770-9542}}
\email{yz@bnu.edu.cn}
\affiliation{Advanced Institute of Natural Sciences, Beijing Normal University, Zhuhai 519087, China}
	
\begin{abstract}
Recently, several regional pulsar timing array collaborations, including CPTA, EPTA, PPTA, and NANOGrav, have individually reported compelling evidence for a stochastic signal at nanohertz frequencies. This signal originates potentially from scalar-induced gravitational waves associated with significant primordial curvature perturbations on small scales. In this letter, we employ data from the EPTA DR2, PPTA DR3, and NANOGrav 15-year data set, to explore the speed of scalar-induced gravitational waves using a comprehensive Bayesian analysis. Our results suggest that, to be consistent with pulsar timing array observations, the speed of scalar-induced gravitational waves should be $c_g \gtrsim 0.61$ at a $95\%$ credible interval for a lognormal power spectrum of curvature perturbations. Additionally, this constraint aligns with the prediction of general relativity that $c_g=1$ within a $90\%$ credible interval. Our findings underscore the capacity of pulsar timing arrays as a powerful tool for probing the speed of scalar-induced gravitational waves.
\end{abstract}
	
\maketitle
%%%%%%%%%%%%%%%%%%%%%%%%%%%%%%%%%%%%%%%%%%%%%%%%%%%%%%%%%%%%%%%%%
\textbf{Introduction.} 	
The detection of gravitational waves (GWs) from compact binary coalescences by ground-based detectors~\cite{LIGOScientific:2018mvr,LIGOScientific:2020ibl,LIGOScientific:2021djp}, has transformed our understanding of the Universe and significantly advanced our ability to scrutinize theories of gravity in the strong gravitational field regime~\cite{LIGOScientific:2019fpa,LIGOScientific:2020tif,LIGOScientific:2021sio}. These detections not only confirmed the existence of GWs but also provided a wealth of information about the properties and astrophysical characteristics of the sources involved~\cite{LIGOScientific:2018jsj,Chen:2018rzo,Chen:2019irf,LIGOScientific:2020kqk,Chen:2021nxo,KAGRA:2021duu,Chen:2022fda,Liu:2022iuf,Zheng:2022wqo,You:2023ouk}.

The detection of individual GW events is a remarkable achievement; nevertheless, the observation of another GW source, the stochastic gravitational wave background (SGWB), remains an ongoing quest. In contrast to the well-localized and characterized signals emanating from compact binary mergers, the SGWB represents a continuous and diffuse background of GWs permeating the Universe. The detection and characterization of the SGWB are of paramount importance, offering potentially crucial insights into various cosmological processes in the early Universe and astrophysical phenomena.

To detect the SGWB in the nanohertz frequency range, the pulsar timing array (PTA) has emerged as an indispensable tool. By regularly monitoring the correlated fluctuations caused by GWs on the time of arrivals (TOAs) of radio pulses emitted by an array of pulsars~\cite{1978SvA....22...36S,Detweiler:1979wn,1990ApJ...361..300F}, a PTA offers a unique and powerful approach to detecting SGWBs. The nanohertz frequency range targeted by PTAs aligns with the characteristic frequencies of GWs from various cosmological sources, making PTAs an excellent tool for exploring the SGWB that originated from the early Universe or new physics~\cite{Li:2019vlb,Vagnozzi:2020gtf,Chen:2021wdo,Wu:2021kmd,Chen:2021ncc,Sakharov:2021dim,Benetti:2021uea,Chen:2022azo,Ashoorioon:2022raz,PPTA:2022eul,Wu:2023pbt,IPTA:2023ero,Wu:2023dnp,Dandoy:2023jot,Madge:2023cak,Yi:2023npi,Wu:2023rib,Bi:2023ewq,Chen:2023uiz}.

Recently, the PTA community has achieved significant progress, with multiple collaborations presenting compelling evidence supporting the existence of a stochastic signal in the frequency range of approximately $10^{-9} \sim 10^{-7}~\mathrm{Hz}$. Notable contributions have been made by several PTA collaborations, including the Chinese PTA (CPTA)~\cite{Xu:2023wog}, the European PTA (EPTA) along with the Indian PTA (InPTA)~\cite{EPTA:2023sfo,Antoniadis:2023ott}, the Parkes PTA (PPTA)~\cite{Zic:2023gta,Reardon:2023gzh}, and the North American Nanoherz Observatory for GWs (NANOGrav)~\cite{NANOGrav:2023hde,NANOGrav:2023gor}. This collective effort has marked a remarkable milestone in the detection of SGWBs in the nanohertz frequency range, generating substantial interest due to its profound implications (see \eg~\cite{NANOGrav:2023hvm,Antoniadis:2023xlr,Bi:2023tib, Zhao:2023joc,Wang:2023ost,Liu:2023pau,Vagnozzi:2023lwo,Fu:2023aab,Han:2023olf,Kitajima:2023cek,Franciolini:2023pbf,Cai:2023dls,Inomata:2023zup,Li:2023bxy,Liu:2023ymk,Abe:2023yrw,Ghosh:2023aum,Figueroa:2023zhu,Yi:2023mbm,Wu:2023hsa,You:2023rmn,Antusch:2023zjk,HosseiniMansoori:2023mqh,Jin:2023wri,Zhang:2023nrs,Choudhury:2023kam,Gorji:2023sil,Das:2023nmm,Yi:2023tdk,Ellis:2023oxs,He:2023ado,Balaji:2023ehk,Cannizzaro:2023mgc,Maji:2023fhv,Bhaumik:2023wmw,Zhu:2023lbf,Basilakos:2023xof,Huang:2023chx,Jiang:2023gfe,InternationalPulsarTimingArray:2023mzf,Harigaya:2023pmw,Lozanov:2023rcd,Choudhury:2023fwk,Cang:2023ysz,Mu:2023wdt,Chen:2023bms,Liu:2023hpw,Chao:2023lox,Fei:2023iel,Maiti:2024nhv}).

Although the inferred amplitude and spectrum of the PTA signal are consistent with astrophysical predictions for a signal originating from the population of supermassive black hole binaries (SMBHBs), the search for new physics within this observational window remains an exciting possibility. The nanohertz frequency band encompasses a broad range of cosmological phenomena that could serve as sources of the SGWB. One such source is the enhanced scalar perturbations at small scales during inflation, which may give rise to the formation of primordial black holes (PBHs). PBHs have attracted considerable attention as viable candidates for dark matter in recent years~\cite{Belotsky:2014kca,Wang:2016ana,Carr:2016drx,Garcia-Bellido:2017mdw,Carr:2017jsz,Germani:2017bcs,Liu:2018ess,Chen:2018czv,Liu:2019rnx,Fu:2019ttf,Wang:2019kaf,Liu:2019lul,Cai:2019bmk,Liu:2020cds,Fu:2020lob,Wu:2020drm,DeLuca:2020sae,Vaskonen:2020lbd,DeLuca:2020agl,Domenech:2020ssp,Domenech:2020ers,Hutsi:2020sol,Kawai:2021edk,Braglia:2021wwa,Cai:2021wzd,Liu:2021jnw,Braglia:2022icu,Chen:2022qvg,Inomata:2022yte,Guo:2023hyp,Cai:2023uhc,Meng:2022low,Gu:2023mmd} (see also reviews~\cite{Sasaki:2018dmp,Carr:2020gox,Carr:2020xqk}). On the other hand, these scalar perturbations can generate scalar-induced gravitational waves (SIGWs) \cite{Ananda:2006af,Baumann:2007zm,Garcia-Bellido:2016dkw,Inomata:2016rbd,Garcia-Bellido:2017aan,Kohri:2018awv,Cai:2018dig,Lu:2019sti,Yuan:2019wwo,Chen:2019xse,Xu:2019bdp,Cai:2019cdl,Yuan:2019fwv,Yi:2020kmq,Yi:2020cut,Liu:2020oqe,Gao:2020tsa,Yi:2021lxc,Yi:2022anu,Yi:2022ymw,Yuan:2019udt,Inui:2023qsd} (for a recent review, see \eg \cite{Domenech:2021ztg}).
In addition, Bayesian analysis of the NANOGrav data indicates that the SIGW model with a lognormal power spectrum achieves a Bayes factor of $57 \pm 3$ in comparison to the SMBHB model~\cite{NANOGrav:2023hvm}. This statistical evidence suggests that SIGWs present a plausible explanation for the signal observed, although this interpretation is subject to the underlying assumptions about the cosmic SMBHB population~\cite{NANOGrav:2023hvm}.

In the framework of general relativity, GWs are predicted to propagate at the speed of light, a prediction consistent with the observed speeds of GWs detected by LIGO and Virgo~\cite{LIGOScientific:2017vwq,LIGOScientific:2017zic}. 
However, alternative theories of gravity propose modifications to general relativity, including variations in the speed of GWs. For instance, certain modified gravity theories, such as massive gravity~\cite{deRham:2010kj,Hassan:2011hr} and bigravity~\cite{Hassan:2011zd,Schmidt-May:2015vnx}, hypothesize that the gravitons possess a nonzero mass. This assumption leads to a frequency-dependent dispersion relation, resulting in variable GW propagation speeds at different frequencies. Similarly, theories with extra dimensions, such as the Dvali-Gabadadze-Porrati model~\cite{Dvali:2000hr}, also anticipate frequency-dependent changes in GW velocities.
In this sense, it is crucial to recognize that the velocity constraints observed at high frequencies may not necessarily apply to the lower frequency range. Therefore, a thorough examination of velocity constraints within the lower frequency band, accessible through PTAs becomes imperative.
In this letter, we assume that the stochastic signal observed by PTAs originates from SIGWs. Under this assumption, we conduct a comprehensive analysis by jointly utilizing PTA data from the EPTA DR2, PPTA DR3, and NANOGrav 15-year data set to investigate the speed of SIGWs for a lognormal power spectrum of curvature perturbations. 
% Throughout this letter, we set $G=c=1$ for convenience.  
% The rest of this letter is organized as follows. In Section~\ref{SIGW}, we provide an overview of the energy density of SIGWs with nontrivial speeds. 
% In Section~\ref{Data}, we describe in detail the methodology employed for the data analysis and present the results obtained using the recent PTA data sets. Finally, in Section~\ref{Con}, we summarize our findings.

%%%%%%%%%%%%%%%%%%%%%%%%%%%%%%%%%%%%%%%%%%%%%%%%%%%%%%%%%%%%%%%%%
\textbf{Scalar-induced gravitational waves.}
In this section, we will introduce the formalism of SIGWs with nontrivial speeds. To begin, let us consider the metric in the conformal Newtonian gauge, which describes the perturbations around a Friedmann-Robertson-Walker (FRW) background:
\e
\label{metric}
\mathrm{d}s^2=a^2(\eta)\left\{-(1+2\Phi)\mathrm{d}\eta^2+\left[(1-2\Phi)\delta_{ij}+\frac{h_{ij}}{2}\right]\mathrm{d}x^i\mathrm{d}x^j \right\},      
\q
where $\eta$ is the conformal time, $a(\eta)$ is the scale factor, and $\Phi$ and $h_{ij}$ represent the Bardeen potential and the tensor perturbations, respectively. In our analysis, we choose to disregard the influences of first-order GWs, vector perturbations, and anisotropic stress. This decision is based on the findings of previous investigations (\cite{Baumann:2007zm, Weinberg:2003ur, Watanabe:2006qe}), which have shown that the impact of these factors is negligible. In Fourier space, the tensor perturbation $h_{ij}$ can be expressed as
\e
h_{ij}(\eta,\mathbf{x})=\int\frac{\mathrm{d}^3k}{(2\pi)^{3/2}}\Bigg(e_{ij}^{+}(\mathbf{k})h_{\mathbf{k}}^+(\eta)+e_{ij}^{\times}(\mathbf{k})h_{\mathbf{k}}^{\times}(\eta)\Bigg)e^{i\mathbf{k}\cdot\mathbf{x}},
\q
where the plus and cross polarization tensors are defined as
\begin{gather}
e_{ij}^{+}(\mathbf{k})\equiv\frac{1}{\sqrt{2}}\Bigg(e_i(\mathbf{k})e_j(\mathbf{k})-\bar{e}_i(\mathbf{k})\bar{e}_j(\mathbf{k})\Bigg),\\
e_{ij}^{\times}(\mathbf{k})\equiv\frac{1}{\sqrt{2}}\Bigg(e_i(\mathbf{k})\bar{e}_j(\mathbf{k})+\bar{e}_i(\mathbf{k})e_j(\mathbf{k})\Bigg).
\end{gather}
Here, the normalized vectors $e_i(\mathbf{k})$ and $\bar{e}_i(\mathbf{k})$ are orthogonal to each other and to $\mathbf{k}$. The factor of $1/\sqrt{2}$ ensures the normalization of the polarization tensors. Working in Fourier space provides a convenient way to analyze and study tensor perturbations, particularly in the context of GWs and cosmology. When accounting for the source originating from the second order of linear scalar perturbations, the tensor perturbations with either polarization in Fourier space satisfy the following equation \cite{Ananda:2006af,Baumann:2007zm,Li:2023uhu}
\e
h_{\mathbf{k}}^{\prime\prime}(\eta)+2\mathcal{H}h_{\mathbf{k}}^\prime(\eta)+c_g^2k^2h_{\mathbf{k}}(\eta)=4S_{\mathbf{k}}(\eta). \label{gw}
\q
Here, the prime represents the derivative with respect to conformal time, $\mathcal{H} \equiv a'/a$  is the comoving Hubble parameter, and $c_g$ is the speed of SIGWs. The equation \eqref{gw} describes the evolution of the tensor perturbations as they propagate through the expanding universe. It relates the second derivative of the perturbation $h_{\mathbf{k}}$ to its first derivative and the perturbation itself. The term $c_g^2k^2h_{\mathbf{k}}$ represents the kinetic term of the perturbation, while the source term $4S_{\mathbf{k}}(\eta)$ accounts for the influence of the second-order scalar perturbations. In addition, the source term $S_{\mathbf{k}}$ is given by the integral expression \cite{Ananda:2006af,Baumann:2007zm}
\begin{equation}
\begin{aligned}
  S_{\mathbf{k}}=&\int\frac{\mathrm{d}^3q}{(2\pi)^{3/2}} \sum_{P=+,\times} e^{P}_{ij}(\mathbf{k})q_iq_j\Bigg(2\Phi_{\mathbf{q}}\Phi_{\mathbf{k-q}}\\ &+\left(\mathcal{H}^{-1}\Phi^{\prime}_{\mathbf{q}}+\Phi_{\mathbf{q}}\right)\left(\mathcal{H}^{-1}\Phi^{\prime}_{\mathbf{k-q}}+\Phi_{\mathbf{k-q}}\right)\Bigg).  
\end{aligned}
\end{equation}

%%%%%%%%%%%%%%%%%%%%%%%%%%%%%%%%%%%%%%%%%%%%%%%%%%%%%%%%%%%%%%%%%
\begin{figure}[tbp!]
	\centering
	\includegraphics[width=0.5\textwidth]{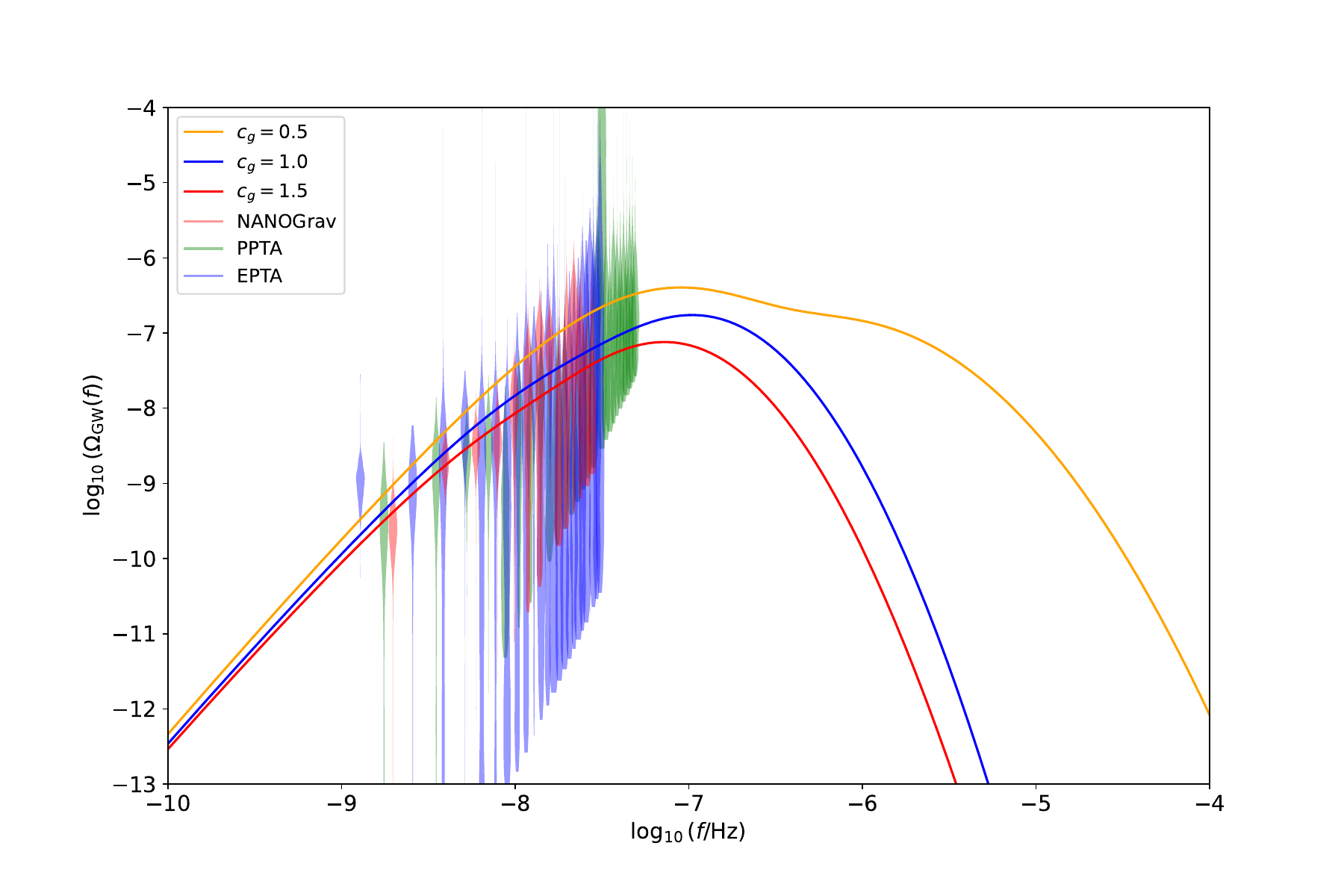}
	\caption{\label{ogw_log}Energy density spectra of SIGW with different speeds $c_g$ as a function of GW frequency. As an illustration, we have fixed other parameters as $f_* = 10^{-7}\,\mathrm{Hz}$, $A = 0.2$, and $\Delta=1$.}
\end{figure}

In Fourier space, the Bardeen potential $\Phi_{\mathbf{k}}$ is related to the primordial curvature perturbations $\zeta_{\mathbf{k}}$ as $\Phi_{\mathbf{k}} = 2\zeta_{\mathbf{k}}/3$. This relationship connects the scalar perturbations to the gravitational potential. The solution for tensor perturbations can be expressed as
\e
h_{\mathbf{k}}(\eta)=\frac{4}{a(\eta)}\int^{\eta}{\mathrm{d}}\bar{\eta}G_{\mathbf{k}}(\eta,\bar{\eta})a(\bar{\eta})S_{\mathbf{k}}(\bar{\eta}),
\q
where the Green function $G_{\mathbf{k}}(\eta,\bar{\eta})$ is given by \cite{Li:2023uhu}
\e
G_{\mathbf{k}}(\eta,\bar{\eta})=\frac{1}{k}\sin[c_gk(\eta-\bar{\eta})].
\q
The power spectrum of tensor perturbations is described by the following expression
\e
\langle h_{\mathbf{k}}(\eta) h_{\mathbf{k}^{\prime}}(\eta)\rangle=\frac{2\pi^2}{k^3}\delta^{(3)}(\mathbf{k}+\mathbf{k}^{\prime})\mathcal{P}_h(\eta, k),
\q
where the delta function $\delta^{(3)}(\mathbf{k}+\mathbf{k}^{\prime})$ enforces momentum conservation, ensuring that the total momentum of the perturbations is conserved and the factor of $2\pi^2/k^3$ accounts for the normalization of the power spectrum. Therefore, the power spectrum of tensor perturbations takes the form \cite{Ananda:2006af,Baumann:2007zm,Kohri:2018awv,Espinosa:2018eve}
\begin{equation}
    \begin{aligned}
\mathcal{P}_h(k,\eta)&=4\int_0^\infty\mathrm{d}v\int_{\vert1-v\vert}^{1+v}\mathrm{d}u \bigg[ \left(\frac{4v^2-(1+v^2-u^2)^2}{4vu}\right)^2 \\
& \times  I_{\rm RD}^2(u, v, c_g)\mathcal{P}_{\zeta}(kv)\mathcal{P}_{\zeta}(ku)\bigg], \label{ph}        
\end{aligned}
\end{equation}
where $u=|\mathbf{k}-\tilde{\mathbf{k}}|/k$, $v=\tilde k/k$ and $I_{\rm RD}$ is the integral kernel which can be found in Ref. \cite{Li:2023uhu}. The characterization of SGWBs often involves quantifying the energy density per logarithmic frequency interval relative to the critical density $\rho_c(\eta)$. This quantity, denoted as $\Omega_{\mathrm{GW}}(k, \eta)$, can be expressed as~\cite{Thrane:2013oya}
\begin{equation}
\Omega_{\mathrm{GW}}(k, \eta) = \frac{1}{\rho_c(\eta)} \frac{\mathrm{d} \rho_{\mathrm{GW}}(k, \eta)}{\mathrm{d} \ln k} = \frac{k^3}{48 \pi^2}\left(\frac{k}{\mathcal{H}}\right)^2 \overline{\left\langle\left|h_{\mathbf{k}}(\eta)\right|^2\right\rangle}.
\end{equation}

Notice that the term $\overline{\left\langle\left|h_{\mathbf{k}}(\eta)\right|^2\right\rangle}$ corresponds to an average taken over a few wavelengths. In the radiation-dominated era, GWs are produced by curvature perturbations. At the epoch of matter-radiation equality, the density parameter of GW is denoted as $\Omega_{\mathrm{GW}}(k)=\Omega_{\mathrm{GW}}(k,\eta\rightarrow\infty)$. Therefore, the expression for the energy density of GWs $\Omega_{\mathrm{GW}}(k)$ can be written as~\cite{Espinosa:2018eve,Li:2023uhu}
\begin{equation}
\Omega_{\mathrm{GW}}(k) = \int_0^{\infty} \mathrm{d} v \int_{|1-v|}^{1+v} \mathrm{d} u \mathcal{T}(u, v, c_g) {P}_{\zeta}(v k) {P}_{\zeta}(u k),
\end{equation}
where the transfer function $\mathcal{T}$ is given by
\begin{equation}
\begin{aligned}
\mathcal{T}(u,v,& c_g)  =  \frac{3 \left[4 v^2-\left(v^2-u^2+1\right)^2\right]^2\left(v^2+u^2-3c_g^2\right)^2 }{1024 v^8 u^8} \\
\times &\Bigg\{ \left[\left(v^2+u^2-3c_g^2\right) \ln \left(\left|\frac{3c_g^2-(v+u)^2}{3c_g^2-(v-u)^2}\right|\right)-4 v u\right]^2  \\
& ~~~ + \pi^2\left(v^2+u^2-3c_g^2\right)^2 \Theta(v+u-\sqrt{3}c_g)\Bigg\}.
\end{aligned}
\end{equation}
Here, $\Theta$ denotes the Heaviside theta function. In the case $c_g=1$, the SIGWs propagate at the speed of light and the result is reduced to the results in Ref.~\cite{Kohri:2018awv} as predicted by general relativity. Using the relationship between frequency $f$ and the wavenumber $k$, $f = k/2\pi$, one can derive the spectrum of SIGWs at the present time to be
\begin{equation}
\Omega_{\mathrm{GW}, 0}(f)=\Omega_{\mathrm{r}, 0}\left[\frac{g_{*, r}(T)}{g_{*, r}\left(T_{\mathrm{eq}}\right)}\right]\left[\frac{g_{*, s}\left(T_{\mathrm{eq}}\right)}{g_{*, s}(T)}\right]^{\frac{4}{3}} \Omega_{\mathrm{GW}}(k),
\end{equation}
where $g_{*,s}$ and $g_{*,r}$ correspond to the effective degrees of freedom for entropy density and radiation, respectively. Additionally, $\Omega_{r,0}$ is the present energy density fraction of radiation. To demonstrate the method, we consider a common power spectrum for the curvature perturbation ${P}_{\zeta}$,  which is characterized by a lognormal shape
\begin{equation}
{P}_{\zeta}(k) = \frac{A}{\sqrt{2\pi} \Delta} \exp\left(-\frac{\ln^2(k/k_*)}{2\Delta^2}\right),
\end{equation}
 where $A$ represents the amplitude of the power spectrum, $k_*$ is the characteristic scale that determines the peak location, and $\Delta$ corresponds to the width of the spectrum.

%%%%%%%%%%%%%%%%%%%%%%%%%%%%%%%%%%%%%%%%%%%%%%%%%%%%%%%%%%%%%%%%%
\textbf{Methodology and result.}
In this study, we jointly employ PTA data from the EPTA DR2~\cite{EPTA:2023sfo}, PPTA DR3~\cite{Zic:2023gta}, and NANOGrav 15-year data set~\cite{NANOGrav:2023hde}, to rigorously constrain the speed of SIGWs using the Bayesian inference method. In particular, we make use of the amplitude values from the free spectrum obtained by each PTA without considering spatial correlations, namely from the free spectrum of a common uncorrelated red noise signal including only the auto-correlations. 
It's worth noting that deviations in the speed of GWs result in variations in the Hellings-Downs~\cite{Hellings:1983fr} correlations. While a comprehensive analysis should consider this effect, the current sensitivity of PTAs poses challenges in distinguishing the deviation from Hellings-Downs correlations due to varying speeds of gravitational waves, as discussed in~\cite{Bi:2023ewq,Bernardo:2023zna,Bernardo:2023mxc}. Consequently, we concentrate on dominant information from auto-correlations, neglecting cross-correlations, and emphasize the energy density spectrum's alteration due to the speed deviation of SIGWs from the speed of light.

The dedicated efforts of PTA collaborations have spanned over a decade. Specifically, the EPTA DR2 includes data from $25$ pulsars over a timespan of $24.7$ years~\cite{EPTA:2023sfo}. The PPTA DR3 comprises observations of $32$ pulsars spanning up to $18$ years~\cite{Zic:2023gta}, while the NANOGrav 15-year data set encompasses observations from $68$ pulsars over a timespan of $16.03$ years~\cite{NANOGrav:2023hde}. 
These PTA data sets collectively reveal a stochastic signal consistent with the anticipated Hellings-Downs spatial correlations for an SGWB. Assuming this signal originates from GWs, it is expected to share similar characteristics across these PTAs. Therefore, we merge observations from NANOGrav, PPTA, and EPTA to enhance precision in estimating model parameters, opting for a collaborative approach rather than analyzing each PTA individually.
The sensitivity of a PTA initiates at a frequency of $1/T_{\mathrm{obs}}$, where $T_{\mathrm{obs}}$ denotes the observational timespan. The EPTA, PPTA, and NANOGrav collaborations adopt $24$~\cite{Antoniadis:2023rey}, $28$~\cite{Reardon:2023gzh}, and $14$~\cite{NANOGrav:2023gor} frequencies in their search for the SGWB signal, respectively.
Through the combination of data from these PTAs, we analyze a total of $66$ frequency components within the free spectrum, covering the range from $1.28$\,nHz to $49.1$\,nHz. 
In \Fig{ogw_log}, we present a visual representation of the data utilized in our analyses, illustrating the energy density of SIGWs for various values of $c_g$, while fixing $f_* = 10^{-7}\,\mathrm{Hz}$, $A = 0.2$, and $\Delta=1$.

We initiate our analysis by utilizing the posterior data of time delay $d(f)$, provided by each PTA. The power spectrum $S(f)$ is linked to the time delay through
\begin{equation}
S(f) = d(f)^2\, T_{\mathrm{obs}}.
\end{equation}
With the provided time delay data, the energy density of the free spectrum can be computed by
\begin{equation}
\hat{\Omega}_{\mathrm{GW}}(f)=\frac{2 \pi^2}{3 H_0^2} f^2 h_c^2(f) = \frac{8\pi^4}{H_0^2} T_{\mathrm{obs}} f^5 d^2(f),
\end{equation}
where $H_0$ represents the Hubble constant with value taken from Planck~\cite{Planck:2018vyg}.
The characteristic strain, $h_c(f)$, is given by
\begin{equation}
h_c^2(f)=12 \pi^2 f^3 S(f).
\end{equation}
For every observed frequency $f_i$, we utilize the obtained posteriors of $\hat{\Omega}_{\mathrm{GW}}(f_i)$ as described above, to estimate the corresponding kernel density, $\mathcal{L}_i$.
Hence, the overall log-likelihood is the sum of the individual log-likelihoods, and it is expressed as~\cite{Moore:2021ibq,Lamb:2023jls,Liu:2023ymk,Wu:2023hsa,Jin:2023wri,Liu:2023pau}
\begin{equation}
\ln \mathcal{L}(\Lambda) = \sum_{i=1}^{66} \ln \mathcal{L}_i(\Omega_{\mathrm{GW}}(f_i, \Lambda)).
\end{equation}
Here, the set of four model parameters is denoted as $\Lambda\equiv \{f_*, A, \Delta, c_g\}$.
To explore the parameter space, we utilize the \texttt{dynesty}~\cite{Speagle:2019ivv} sampler, which is accessible in the \texttt{Bilby}~\cite{Ashton:2018jfp,Romero-Shaw:2020owr} package.

%%%%%%%%%%%%%%%%%%%%%%%%%%%%%%%%%%%%%%%%%%%%%%%%%%%%%%%%%%%%%%%%%
\begin{table}
    \centering
 \begin{tabular}{c|ccccc}
		%\hline
  \hline
		Parameter & $\log_{10} (f_*/\mathrm{Hz})$ & $\log_{10} A$ & $\Delta$ &  $c_g$  \\[1pt]
		\hline
		 Prior& $\mU(-8, -2)$ & $\mU(-2, 2)$ & $\mU(0.05, 5)$  & $\mU(0.1, 2)$ \\[1pt]
        % \hline
		Result  & $\gtrsim -6.4$ & $0.61^{+0.74}_{-1.09}$ & $3.57^{+0.84}_{-1.58}$ & $\gtrsim 0.61$ \\[1pt]
  \hline
	\end{tabular}
	\caption{\label{tab:priors}Priors and results for the model parameters $\Lambda\equiv \{f_*, A, \Delta, c_g\}$. We use $\mU$ to denote the uniform distribution and quote the result of each parameter with the median and a $90\%$ equal-tail credible interval.}
\end{table}

%%%%%%%%%%%%%%%%%%%%%%%%%%%%%%%%%%%%%%%%%%%%%%%%%%%%%%%%%%%%%%%%%
\begin{figure}[tbp!]
	\centering
 \includegraphics[width=0.5\textwidth]{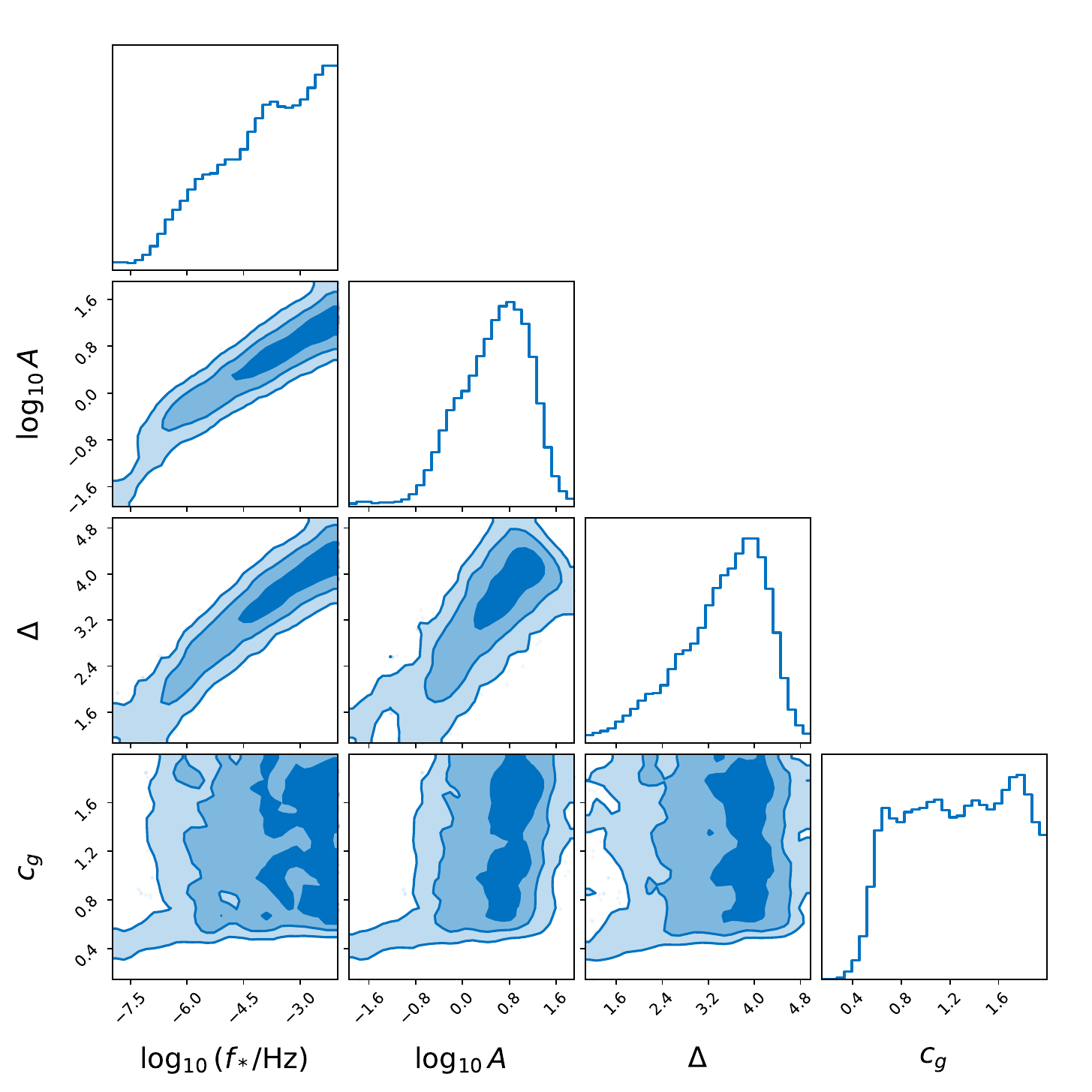}
	\caption{\label{posts_log}Posterior distributions for the model parameters $\Lambda\equiv \{f_*, A, \Delta, c_g\}$ obtained from the combined PTA data of EPTA DR2, PPTA DR3, and NANOGrav 15-yr data set. We show the $1 \sigma$, $2 \sigma$, and $3 \sigma$
credible regions as contours in the two-dimensional plot.}
\end{figure}

In our analysis, we employ uniform priors for each parameter: $\log_{10} (f_*/\mathrm{Hz})$ in the range $[-8, -2]$, $\log_{10} A$ in the range $[-2, 2]$, $\Delta$ in the range $[0.05, 5]$, and $c_g$ in the range $[0.1, 2]$. The posterior distributions for these model parameters are illustrated in~\Fig{posts_log}. 
Specifically, we find $\log_{10} A = 0.61^{+0.74}_{-1.09}$ and $\Delta = 3.57^{+0.84}_{-1.58}$. Unless specified, we quote results using the median value and $90\%$ equal-tail credible interval. Additionally, we find $\log_{10} (f_*/\mathrm{Hz}) \gtrsim -6.4$ and $c_g \gtrsim 0.61$ at a $95\%$ credible interval. A concise summary of priors and results for the model parameters is provided in~\Table{tab:priors}. It is important to note that the obtained constraint is consistent with the prediction of general relativity ($c_g=1$) at a $90\%$ credible interval.

%%%%%%%%%%%%%%%%%%%%%%%%%%%%%%%%%%%%%%%%%%%%%%%%%%%%%%%%%%%%%%%%%
\textbf{Conclusion and discussion.}
The speed of GWs has been a topic of great interest in the field of astrophysics and cosmology due to its profound implications for our understanding of the fundamental laws of the Universe. It represents a key parameter that directly influences the behavior and propagation of GW. 
In the theory of general relativity, GWs are predicted to propagate at the speed of light. Notably, the binary merger event GW170817 observed by LIGO and Virgo has constrained the propagation speed of GWs as $|1-v| \lesssim 10^{-15}$ at the high frequency of $f \sim 100 \mathrm{~Hz}$. However, the velocity constraint observed at high frequencies may not necessarily apply to the lower frequency range. 
Thus, an independent examination of velocity constraints within the lower frequency band, accessible through PTAs, becomes crucial.

In this letter, we conduct a comprehensive analysis by jointly utilizing PTA data from the EPTA DR2~\cite{EPTA:2023sfo}, PPTA DR3~\cite{Zic:2023gta}, and NANOGrav 15-year data set~\cite{NANOGrav:2023hde}  to investigate the speed of the SIGWs, assuming that the stochastic signal observed by recent PTA collaborations originates from the SIGWs. 
The analysis suggests that, to be consistent with PTA observations, the speed of SIGWs should be $c_g \gtrsim 0.61$ at a $95\%$ credible interval. Moreover, the obtained constraint is consistent with the prediction of general relativity ($c_g=1$) at a $90\%$ credible interval.

While LIGO's constraints on the speed of GW propagation are notably more stringent than those derived from PTAs by several orders of magnitude, it is critical to note that these measurements pertain to distinct frequency domains. LIGO's observations are confined to the high-frequency regime at the kilohertz band, whereas PTAs explore the nanohertz frequency band. The high-frequency constraints from LIGO affirm that the speed of GW propagation closely mirrors the speed of light, yet this does not preclude the potential for discrepancies at significantly lower frequencies. Although current PTA-derived constraints may not be as restrictive as those from LIGO, they furnish invaluable independent constraints within a separate frequency range, thereby offering profound insights into the nature of gravity and the fundamental principles of physics.
For instance, within the context of models inspired by an effective field theory~\cite{deRham:2018red,LISACosmologyWorkingGroup:2022wjo}, the frequency-dependent GW propagation speed can be described by
\begin{equation}
\label{cgf}
c_g(f)=\left[1+\frac{f_{\star}^2}{f^2}-\frac{f_{\star}^2}{f^2} \sqrt{1+2\left(1-c_0^2\right) \frac{f^2}{f_{\star}^2}}\right]^{1 / 2},
\end{equation}
implying a rapid transition from a velocity of $c_0$ at low frequencies to $c_g = 1$ at higher frequencies. Our analysis suggests $c_0 \gtrsim 0.61$ at the $95\%$ credible interval. This finding is significant, as it provides constraints on the speed of GWs in a low-frequency range, contributing to our broader understanding of gravitational theories and the potential for new physics beyond the standard model.

The results derived from the current PTA data exhibit large uncertainties, underscoring the necessity for further investigation to refine our understanding of the fundamental nature of gravity. With the continuous advancement of GW detection technology and precision, PTAs will continue to play a crucial role in unraveling the mysteries of the cosmos and exploring the boundaries of our current theories. 
The ongoing development of next-generation radio telescopes, such as the Square Kilometre Array~\cite{Lazio:2013mea}, will significantly improve PTA sensitivity.
Our findings have underscored the significant capacity of PTAs in probing the speed of SIGWs.

% The ongoing development of next-generation PTA projects, such as the Square Kilometre Array~\cite{Lazio:2013mea}, holds immense promise. 

%%%%%%%%%%%%%%%%%%%%%%%%%%%%%%%%%%%%%%%%%%%%%%%%%%%%%%%%%%%%%%%%%
\emph{Acknowledgments.}
ZCC is supported by the National Natural Science Foundation of China (Grant No.~12247176 and No.~12247112) and the innovative research group of Hunan Province under Grant No. 2024JJ1006.
JL is supported by the Natural Science Foundation of Shandong Province (grant No.
ZR2021QA073) and Research Start-up Fund of QUST (grant No. 1203043003587).
LL is supported by the National Natural Science Foundation of China (Grant No.~12247112 and No.~12247176) and the China Postdoctoral Science Foundation Fellowship No. 2023M730300.
ZY is supported by the National Natural Science Foundation of China under Grant No. 12205015.

\bibliography{ref}

\begin{thebibliography}{179}%
\makeatletter
\providecommand \@ifxundefined [1]{%
 \@ifx{#1\undefined}
}%
\providecommand \@ifnum [1]{%
 \ifnum #1\expandafter \@firstoftwo
 \else \expandafter \@secondoftwo
 \fi
}%
\providecommand \@ifx [1]{%
 \ifx #1\expandafter \@firstoftwo
 \else \expandafter \@secondoftwo
 \fi
}%
\providecommand \natexlab [1]{#1}%
\providecommand \enquote  [1]{``#1''}%
\providecommand \bibnamefont  [1]{#1}%
\providecommand \bibfnamefont [1]{#1}%
\providecommand \citenamefont [1]{#1}%
\providecommand \href@noop [0]{\@secondoftwo}%
\providecommand \href [0]{\begingroup \@sanitize@url \@href}%
\providecommand \@href[1]{\@@startlink{#1}\@@href}%
\providecommand \@@href[1]{\endgroup#1\@@endlink}%
\providecommand \@sanitize@url [0]{\catcode `\\12\catcode `\$12\catcode `\&12\catcode `\#12\catcode `\^12\catcode `\_12\catcode `\%12\relax}%
\providecommand \@@startlink[1]{}%
\providecommand \@@endlink[0]{}%
\providecommand \url  [0]{\begingroup\@sanitize@url \@url }%
\providecommand \@url [1]{\endgroup\@href {#1}{\urlprefix }}%
\providecommand \urlprefix  [0]{URL }%
\providecommand \Eprint [0]{\href }%
\providecommand \doibase [0]{http://dx.doi.org/}%
\providecommand \selectlanguage [0]{\@gobble}%
\providecommand \bibinfo  [0]{\@secondoftwo}%
\providecommand \bibfield  [0]{\@secondoftwo}%
\providecommand \translation [1]{[#1]}%
\providecommand \BibitemOpen [0]{}%
\providecommand \bibitemStop [0]{}%
\providecommand \bibitemNoStop [0]{.\EOS\space}%
\providecommand \EOS [0]{\spacefactor3000\relax}%
\providecommand \BibitemShut  [1]{\csname bibitem#1\endcsname}%
\let\auto@bib@innerbib\@empty
%</preamble>
\bibitem [{\citenamefont {Abbott}\ \emph {et~al.}(2019{\natexlab{a}})\citenamefont {Abbott} \emph {et~al.}}]{LIGOScientific:2018mvr}%
  \BibitemOpen
  \bibfield  {author} {\bibinfo {author} {\bibfnamefont {B.~P.}\ \bibnamefont {Abbott}} \emph {et~al.} (\bibinfo {collaboration} {LIGO Scientific, Virgo}),\ }\href {\doibase 10.1103/PhysRevX.9.031040} {\bibfield  {journal} {\bibinfo  {journal} {Phys. Rev. X}\ }\textbf {\bibinfo {volume} {9}},\ \bibinfo {pages} {031040} (\bibinfo {year} {2019}{\natexlab{a}})},\ \Eprint {http://arxiv.org/abs/1811.12907} {arXiv:1811.12907 [astro-ph.HE]} \BibitemShut {NoStop}%
\bibitem [{\citenamefont {Abbott}\ \emph {et~al.}(2021{\natexlab{a}})\citenamefont {Abbott} \emph {et~al.}}]{LIGOScientific:2020ibl}%
  \BibitemOpen
  \bibfield  {author} {\bibinfo {author} {\bibfnamefont {R.}~\bibnamefont {Abbott}} \emph {et~al.} (\bibinfo {collaboration} {LIGO Scientific, Virgo}),\ }\href {\doibase 10.1103/PhysRevX.11.021053} {\bibfield  {journal} {\bibinfo  {journal} {Phys. Rev. X}\ }\textbf {\bibinfo {volume} {11}},\ \bibinfo {pages} {021053} (\bibinfo {year} {2021}{\natexlab{a}})},\ \Eprint {http://arxiv.org/abs/2010.14527} {arXiv:2010.14527 [gr-qc]} \BibitemShut {NoStop}%
\bibitem [{\citenamefont {Abbott}\ \emph {et~al.}(2023{\natexlab{a}})\citenamefont {Abbott} \emph {et~al.}}]{LIGOScientific:2021djp}%
  \BibitemOpen
  \bibfield  {author} {\bibinfo {author} {\bibfnamefont {R.}~\bibnamefont {Abbott}} \emph {et~al.} (\bibinfo {collaboration} {KAGRA, VIRGO, LIGO Scientific}),\ }\href {\doibase 10.1103/PhysRevX.13.041039} {\bibfield  {journal} {\bibinfo  {journal} {Phys. Rev. X}\ }\textbf {\bibinfo {volume} {13}},\ \bibinfo {pages} {041039} (\bibinfo {year} {2023}{\natexlab{a}})},\ \Eprint {http://arxiv.org/abs/2111.03606} {arXiv:2111.03606 [gr-qc]} \BibitemShut {NoStop}%
\bibitem [{\citenamefont {Abbott}\ \emph {et~al.}(2019{\natexlab{b}})\citenamefont {Abbott} \emph {et~al.}}]{LIGOScientific:2019fpa}%
  \BibitemOpen
  \bibfield  {author} {\bibinfo {author} {\bibfnamefont {B.~P.}\ \bibnamefont {Abbott}} \emph {et~al.} (\bibinfo {collaboration} {LIGO Scientific, Virgo}),\ }\href {\doibase 10.1103/PhysRevD.100.104036} {\bibfield  {journal} {\bibinfo  {journal} {Phys. Rev. D}\ }\textbf {\bibinfo {volume} {100}},\ \bibinfo {pages} {104036} (\bibinfo {year} {2019}{\natexlab{b}})},\ \Eprint {http://arxiv.org/abs/1903.04467} {arXiv:1903.04467 [gr-qc]} \BibitemShut {NoStop}%
\bibitem [{\citenamefont {Abbott}\ \emph {et~al.}(2021{\natexlab{b}})\citenamefont {Abbott} \emph {et~al.}}]{LIGOScientific:2020tif}%
  \BibitemOpen
  \bibfield  {author} {\bibinfo {author} {\bibfnamefont {R.}~\bibnamefont {Abbott}} \emph {et~al.} (\bibinfo {collaboration} {LIGO Scientific, Virgo}),\ }\href {\doibase 10.1103/PhysRevD.103.122002} {\bibfield  {journal} {\bibinfo  {journal} {Phys. Rev. D}\ }\textbf {\bibinfo {volume} {103}},\ \bibinfo {pages} {122002} (\bibinfo {year} {2021}{\natexlab{b}})},\ \Eprint {http://arxiv.org/abs/2010.14529} {arXiv:2010.14529 [gr-qc]} \BibitemShut {NoStop}%
\bibitem [{\citenamefont {Abbott}\ \emph {et~al.}(2021{\natexlab{c}})\citenamefont {Abbott} \emph {et~al.}}]{LIGOScientific:2021sio}%
  \BibitemOpen
  \bibfield  {author} {\bibinfo {author} {\bibfnamefont {R.}~\bibnamefont {Abbott}} \emph {et~al.} (\bibinfo {collaboration} {LIGO Scientific, VIRGO, KAGRA}),\ }\href@noop {} {\  (\bibinfo {year} {2021}{\natexlab{c}})},\ \Eprint {http://arxiv.org/abs/2112.06861} {arXiv:2112.06861 [gr-qc]} \BibitemShut {NoStop}%
\bibitem [{\citenamefont {Abbott}\ \emph {et~al.}(2019{\natexlab{c}})\citenamefont {Abbott} \emph {et~al.}}]{LIGOScientific:2018jsj}%
  \BibitemOpen
  \bibfield  {author} {\bibinfo {author} {\bibfnamefont {B.~P.}\ \bibnamefont {Abbott}} \emph {et~al.} (\bibinfo {collaboration} {LIGO Scientific, Virgo}),\ }\href {\doibase 10.3847/2041-8213/ab3800} {\bibfield  {journal} {\bibinfo  {journal} {Astrophys. J. Lett.}\ }\textbf {\bibinfo {volume} {882}},\ \bibinfo {pages} {L24} (\bibinfo {year} {2019}{\natexlab{c}})},\ \Eprint {http://arxiv.org/abs/1811.12940} {arXiv:1811.12940 [astro-ph.HE]} \BibitemShut {NoStop}%
\bibitem [{\citenamefont {Chen}\ \emph {et~al.}(2019)\citenamefont {Chen}, \citenamefont {Huang},\ and\ \citenamefont {Huang}}]{Chen:2018rzo}%
  \BibitemOpen
  \bibfield  {author} {\bibinfo {author} {\bibfnamefont {Z.-C.}\ \bibnamefont {Chen}}, \bibinfo {author} {\bibfnamefont {F.}~\bibnamefont {Huang}}, \ and\ \bibinfo {author} {\bibfnamefont {Q.-G.}\ \bibnamefont {Huang}},\ }\href {\doibase 10.3847/1538-4357/aaf581} {\bibfield  {journal} {\bibinfo  {journal} {Astrophys. J.}\ }\textbf {\bibinfo {volume} {871}},\ \bibinfo {pages} {97} (\bibinfo {year} {2019})},\ \Eprint {http://arxiv.org/abs/1809.10360} {arXiv:1809.10360 [gr-qc]} \BibitemShut {NoStop}%
\bibitem [{\citenamefont {Chen}\ and\ \citenamefont {Huang}(2020)}]{Chen:2019irf}%
  \BibitemOpen
  \bibfield  {author} {\bibinfo {author} {\bibfnamefont {Z.-C.}\ \bibnamefont {Chen}}\ and\ \bibinfo {author} {\bibfnamefont {Q.-G.}\ \bibnamefont {Huang}},\ }\href {\doibase 10.1088/1475-7516/2020/08/039} {\bibfield  {journal} {\bibinfo  {journal} {JCAP}\ }\textbf {\bibinfo {volume} {08}},\ \bibinfo {pages} {039} (\bibinfo {year} {2020})},\ \Eprint {http://arxiv.org/abs/1904.02396} {arXiv:1904.02396 [astro-ph.CO]} \BibitemShut {NoStop}%
\bibitem [{\citenamefont {Abbott}\ \emph {et~al.}(2021{\natexlab{d}})\citenamefont {Abbott} \emph {et~al.}}]{LIGOScientific:2020kqk}%
  \BibitemOpen
  \bibfield  {author} {\bibinfo {author} {\bibfnamefont {R.}~\bibnamefont {Abbott}} \emph {et~al.} (\bibinfo {collaboration} {LIGO Scientific, Virgo}),\ }\href {\doibase 10.3847/2041-8213/abe949} {\bibfield  {journal} {\bibinfo  {journal} {Astrophys. J. Lett.}\ }\textbf {\bibinfo {volume} {913}},\ \bibinfo {pages} {L7} (\bibinfo {year} {2021}{\natexlab{d}})},\ \Eprint {http://arxiv.org/abs/2010.14533} {arXiv:2010.14533 [astro-ph.HE]} \BibitemShut {NoStop}%
\bibitem [{\citenamefont {Chen}\ \emph {et~al.}(2022{\natexlab{a}})\citenamefont {Chen}, \citenamefont {Yuan},\ and\ \citenamefont {Huang}}]{Chen:2021nxo}%
  \BibitemOpen
  \bibfield  {author} {\bibinfo {author} {\bibfnamefont {Z.-C.}\ \bibnamefont {Chen}}, \bibinfo {author} {\bibfnamefont {C.}~\bibnamefont {Yuan}}, \ and\ \bibinfo {author} {\bibfnamefont {Q.-G.}\ \bibnamefont {Huang}},\ }\href {\doibase 10.1016/j.physletb.2022.137040} {\bibfield  {journal} {\bibinfo  {journal} {Phys. Lett. B}\ }\textbf {\bibinfo {volume} {829}},\ \bibinfo {pages} {137040} (\bibinfo {year} {2022}{\natexlab{a}})},\ \Eprint {http://arxiv.org/abs/2108.11740} {arXiv:2108.11740 [astro-ph.CO]} \BibitemShut {NoStop}%
\bibitem [{\citenamefont {Abbott}\ \emph {et~al.}(2023{\natexlab{b}})\citenamefont {Abbott} \emph {et~al.}}]{KAGRA:2021duu}%
  \BibitemOpen
  \bibfield  {author} {\bibinfo {author} {\bibfnamefont {R.}~\bibnamefont {Abbott}} \emph {et~al.} (\bibinfo {collaboration} {KAGRA, VIRGO, LIGO Scientific}),\ }\href {\doibase 10.1103/PhysRevX.13.011048} {\bibfield  {journal} {\bibinfo  {journal} {Phys. Rev. X}\ }\textbf {\bibinfo {volume} {13}},\ \bibinfo {pages} {011048} (\bibinfo {year} {2023}{\natexlab{b}})},\ \Eprint {http://arxiv.org/abs/2111.03634} {arXiv:2111.03634 [astro-ph.HE]} \BibitemShut {NoStop}%
\bibitem [{\citenamefont {Chen}\ \emph {et~al.}(2023{\natexlab{a}})\citenamefont {Chen}, \citenamefont {Du}, \citenamefont {Huang},\ and\ \citenamefont {You}}]{Chen:2022fda}%
  \BibitemOpen
  \bibfield  {author} {\bibinfo {author} {\bibfnamefont {Z.-C.}\ \bibnamefont {Chen}}, \bibinfo {author} {\bibfnamefont {S.-S.}\ \bibnamefont {Du}}, \bibinfo {author} {\bibfnamefont {Q.-G.}\ \bibnamefont {Huang}}, \ and\ \bibinfo {author} {\bibfnamefont {Z.-Q.}\ \bibnamefont {You}},\ }\href {\doibase 10.1088/1475-7516/2023/03/024} {\bibfield  {journal} {\bibinfo  {journal} {JCAP}\ }\textbf {\bibinfo {volume} {03}},\ \bibinfo {pages} {024} (\bibinfo {year} {2023}{\natexlab{a}})},\ \Eprint {http://arxiv.org/abs/2205.11278} {arXiv:2205.11278 [astro-ph.CO]} \BibitemShut {NoStop}%
\bibitem [{\citenamefont {Liu}\ \emph {et~al.}(2023{\natexlab{a}})\citenamefont {Liu}, \citenamefont {You}, \citenamefont {Wu},\ and\ \citenamefont {Chen}}]{Liu:2022iuf}%
  \BibitemOpen
  \bibfield  {author} {\bibinfo {author} {\bibfnamefont {L.}~\bibnamefont {Liu}}, \bibinfo {author} {\bibfnamefont {Z.-Q.}\ \bibnamefont {You}}, \bibinfo {author} {\bibfnamefont {Y.}~\bibnamefont {Wu}}, \ and\ \bibinfo {author} {\bibfnamefont {Z.-C.}\ \bibnamefont {Chen}},\ }\href {\doibase 10.1103/PhysRevD.107.063035} {\bibfield  {journal} {\bibinfo  {journal} {Phys. Rev. D}\ }\textbf {\bibinfo {volume} {107}},\ \bibinfo {pages} {063035} (\bibinfo {year} {2023}{\natexlab{a}})},\ \Eprint {http://arxiv.org/abs/2210.16094} {arXiv:2210.16094 [astro-ph.CO]} \BibitemShut {NoStop}%
\bibitem [{\citenamefont {Zheng}\ \emph {et~al.}(2023)\citenamefont {Zheng}, \citenamefont {Li}, \citenamefont {Chen}, \citenamefont {Zhou},\ and\ \citenamefont {Zhu}}]{Zheng:2022wqo}%
  \BibitemOpen
  \bibfield  {author} {\bibinfo {author} {\bibfnamefont {L.-M.}\ \bibnamefont {Zheng}}, \bibinfo {author} {\bibfnamefont {Z.}~\bibnamefont {Li}}, \bibinfo {author} {\bibfnamefont {Z.-C.}\ \bibnamefont {Chen}}, \bibinfo {author} {\bibfnamefont {H.}~\bibnamefont {Zhou}}, \ and\ \bibinfo {author} {\bibfnamefont {Z.-H.}\ \bibnamefont {Zhu}},\ }\href {\doibase 10.1016/j.physletb.2023.137720} {\bibfield  {journal} {\bibinfo  {journal} {Phys. Lett. B}\ }\textbf {\bibinfo {volume} {838}},\ \bibinfo {pages} {137720} (\bibinfo {year} {2023})},\ \Eprint {http://arxiv.org/abs/2212.05516} {arXiv:2212.05516 [astro-ph.CO]} \BibitemShut {NoStop}%
\bibitem [{\citenamefont {You}\ \emph {et~al.}(2023{\natexlab{a}})\citenamefont {You}, \citenamefont {Chen}, \citenamefont {Liu}, \citenamefont {Yi}, \citenamefont {Liu}, \citenamefont {Wu},\ and\ \citenamefont {Gong}}]{You:2023ouk}%
  \BibitemOpen
  \bibfield  {author} {\bibinfo {author} {\bibfnamefont {Z.-Q.}\ \bibnamefont {You}}, \bibinfo {author} {\bibfnamefont {Z.-C.}\ \bibnamefont {Chen}}, \bibinfo {author} {\bibfnamefont {L.}~\bibnamefont {Liu}}, \bibinfo {author} {\bibfnamefont {Z.}~\bibnamefont {Yi}}, \bibinfo {author} {\bibfnamefont {X.-J.}\ \bibnamefont {Liu}}, \bibinfo {author} {\bibfnamefont {Y.}~\bibnamefont {Wu}}, \ and\ \bibinfo {author} {\bibfnamefont {Y.}~\bibnamefont {Gong}},\ }\href@noop {} {\  (\bibinfo {year} {2023}{\natexlab{a}})},\ \Eprint {http://arxiv.org/abs/2306.12950} {arXiv:2306.12950 [astro-ph.CO]} \BibitemShut {NoStop}%
\bibitem [{\citenamefont {{Sazhin}}(1978)}]{1978SvA....22...36S}%
  \BibitemOpen
  \bibfield  {author} {\bibinfo {author} {\bibfnamefont {M.~V.}\ \bibnamefont {{Sazhin}}},\ }\href@noop {} {\bibfield  {journal} {\bibinfo  {journal} {Soviet Astronomy}\ }\textbf {\bibinfo {volume} {22}},\ \bibinfo {pages} {36} (\bibinfo {year} {1978})}\BibitemShut {NoStop}%
\bibitem [{\citenamefont {Detweiler}(1979)}]{Detweiler:1979wn}%
  \BibitemOpen
  \bibfield  {author} {\bibinfo {author} {\bibfnamefont {S.~L.}\ \bibnamefont {Detweiler}},\ }\href {\doibase 10.1086/157593} {\bibfield  {journal} {\bibinfo  {journal} {Astrophys. J.}\ }\textbf {\bibinfo {volume} {234}},\ \bibinfo {pages} {1100} (\bibinfo {year} {1979})}\BibitemShut {NoStop}%
\bibitem [{\citenamefont {{Foster}}\ and\ \citenamefont {{Backer}}(1990)}]{1990ApJ...361..300F}%
  \BibitemOpen
  \bibfield  {author} {\bibinfo {author} {\bibfnamefont {R.~S.}\ \bibnamefont {{Foster}}}\ and\ \bibinfo {author} {\bibfnamefont {D.~C.}\ \bibnamefont {{Backer}}},\ }\href {\doibase 10.1086/169195} {\bibfield  {journal} {\bibinfo  {journal} {Astrophys. J.}\ }\textbf {\bibinfo {volume} {361}},\ \bibinfo {pages} {300} (\bibinfo {year} {1990})}\BibitemShut {NoStop}%
\bibitem [{\citenamefont {Li}\ \emph {et~al.}(2019)\citenamefont {Li}, \citenamefont {Chen},\ and\ \citenamefont {Huang}}]{Li:2019vlb}%
  \BibitemOpen
  \bibfield  {author} {\bibinfo {author} {\bibfnamefont {J.}~\bibnamefont {Li}}, \bibinfo {author} {\bibfnamefont {Z.-C.}\ \bibnamefont {Chen}}, \ and\ \bibinfo {author} {\bibfnamefont {Q.-G.}\ \bibnamefont {Huang}},\ }\href {\doibase 10.1007/s11433-019-9605-5} {\bibfield  {journal} {\bibinfo  {journal} {Sci. China Phys. Mech. Astron.}\ }\textbf {\bibinfo {volume} {62}},\ \bibinfo {pages} {110421} (\bibinfo {year} {2019})},\ \bibinfo {note} {[Erratum: Sci.China Phys.Mech.Astron. 64, 250451 (2021)]},\ \Eprint {http://arxiv.org/abs/1907.09794} {arXiv:1907.09794 [astro-ph.CO]} \BibitemShut {NoStop}%
\bibitem [{\citenamefont {Vagnozzi}(2021)}]{Vagnozzi:2020gtf}%
  \BibitemOpen
  \bibfield  {author} {\bibinfo {author} {\bibfnamefont {S.}~\bibnamefont {Vagnozzi}},\ }\href {\doibase 10.1093/mnrasl/slaa203} {\bibfield  {journal} {\bibinfo  {journal} {Mon. Not. Roy. Astron. Soc.}\ }\textbf {\bibinfo {volume} {502}},\ \bibinfo {pages} {L11} (\bibinfo {year} {2021})},\ \Eprint {http://arxiv.org/abs/2009.13432} {arXiv:2009.13432 [astro-ph.CO]} \BibitemShut {NoStop}%
\bibitem [{\citenamefont {Chen}\ \emph {et~al.}(2021)\citenamefont {Chen}, \citenamefont {Yuan},\ and\ \citenamefont {Huang}}]{Chen:2021wdo}%
  \BibitemOpen
  \bibfield  {author} {\bibinfo {author} {\bibfnamefont {Z.-C.}\ \bibnamefont {Chen}}, \bibinfo {author} {\bibfnamefont {C.}~\bibnamefont {Yuan}}, \ and\ \bibinfo {author} {\bibfnamefont {Q.-G.}\ \bibnamefont {Huang}},\ }\href {\doibase 10.1007/s11433-021-1797-y} {\bibfield  {journal} {\bibinfo  {journal} {Sci. China Phys. Mech. Astron.}\ }\textbf {\bibinfo {volume} {64}},\ \bibinfo {pages} {120412} (\bibinfo {year} {2021})},\ \Eprint {http://arxiv.org/abs/2101.06869} {arXiv:2101.06869 [astro-ph.CO]} \BibitemShut {NoStop}%
\bibitem [{\citenamefont {Wu}\ \emph {et~al.}(2022{\natexlab{a}})\citenamefont {Wu}, \citenamefont {Chen},\ and\ \citenamefont {Huang}}]{Wu:2021kmd}%
  \BibitemOpen
  \bibfield  {author} {\bibinfo {author} {\bibfnamefont {Y.-M.}\ \bibnamefont {Wu}}, \bibinfo {author} {\bibfnamefont {Z.-C.}\ \bibnamefont {Chen}}, \ and\ \bibinfo {author} {\bibfnamefont {Q.-G.}\ \bibnamefont {Huang}},\ }\href {\doibase 10.3847/1538-4357/ac35cc} {\bibfield  {journal} {\bibinfo  {journal} {Astrophys. J.}\ }\textbf {\bibinfo {volume} {925}},\ \bibinfo {pages} {37} (\bibinfo {year} {2022}{\natexlab{a}})},\ \Eprint {http://arxiv.org/abs/2108.10518} {arXiv:2108.10518 [astro-ph.CO]} \BibitemShut {NoStop}%
\bibitem [{\citenamefont {Chen}\ \emph {et~al.}(2022{\natexlab{b}})\citenamefont {Chen}, \citenamefont {Wu},\ and\ \citenamefont {Huang}}]{Chen:2021ncc}%
  \BibitemOpen
  \bibfield  {author} {\bibinfo {author} {\bibfnamefont {Z.-C.}\ \bibnamefont {Chen}}, \bibinfo {author} {\bibfnamefont {Y.-M.}\ \bibnamefont {Wu}}, \ and\ \bibinfo {author} {\bibfnamefont {Q.-G.}\ \bibnamefont {Huang}},\ }\href {\doibase 10.1088/1572-9494/ac7cdf} {\bibfield  {journal} {\bibinfo  {journal} {Commun. Theor. Phys.}\ }\textbf {\bibinfo {volume} {74}},\ \bibinfo {pages} {105402} (\bibinfo {year} {2022}{\natexlab{b}})},\ \Eprint {http://arxiv.org/abs/2109.00296} {arXiv:2109.00296 [astro-ph.CO]} \BibitemShut {NoStop}%
\bibitem [{\citenamefont {Sakharov}\ \emph {et~al.}(2021)\citenamefont {Sakharov}, \citenamefont {Eroshenko},\ and\ \citenamefont {Rubin}}]{Sakharov:2021dim}%
  \BibitemOpen
  \bibfield  {author} {\bibinfo {author} {\bibfnamefont {A.~S.}\ \bibnamefont {Sakharov}}, \bibinfo {author} {\bibfnamefont {Y.~N.}\ \bibnamefont {Eroshenko}}, \ and\ \bibinfo {author} {\bibfnamefont {S.~G.}\ \bibnamefont {Rubin}},\ }\href {\doibase 10.1103/PhysRevD.104.043005} {\bibfield  {journal} {\bibinfo  {journal} {Phys. Rev. D}\ }\textbf {\bibinfo {volume} {104}},\ \bibinfo {pages} {043005} (\bibinfo {year} {2021})},\ \Eprint {http://arxiv.org/abs/2104.08750} {arXiv:2104.08750 [hep-ph]} \BibitemShut {NoStop}%
\bibitem [{\citenamefont {Benetti}\ \emph {et~al.}(2022)\citenamefont {Benetti}, \citenamefont {Graef},\ and\ \citenamefont {Vagnozzi}}]{Benetti:2021uea}%
  \BibitemOpen
  \bibfield  {author} {\bibinfo {author} {\bibfnamefont {M.}~\bibnamefont {Benetti}}, \bibinfo {author} {\bibfnamefont {L.~L.}\ \bibnamefont {Graef}}, \ and\ \bibinfo {author} {\bibfnamefont {S.}~\bibnamefont {Vagnozzi}},\ }\href {\doibase 10.1103/PhysRevD.105.043520} {\bibfield  {journal} {\bibinfo  {journal} {Phys. Rev. D}\ }\textbf {\bibinfo {volume} {105}},\ \bibinfo {pages} {043520} (\bibinfo {year} {2022})},\ \Eprint {http://arxiv.org/abs/2111.04758} {arXiv:2111.04758 [astro-ph.CO]} \BibitemShut {NoStop}%
\bibitem [{\citenamefont {Chen}\ \emph {et~al.}(2022{\natexlab{c}})\citenamefont {Chen}, \citenamefont {Wu},\ and\ \citenamefont {Huang}}]{Chen:2022azo}%
  \BibitemOpen
  \bibfield  {author} {\bibinfo {author} {\bibfnamefont {Z.-C.}\ \bibnamefont {Chen}}, \bibinfo {author} {\bibfnamefont {Y.-M.}\ \bibnamefont {Wu}}, \ and\ \bibinfo {author} {\bibfnamefont {Q.-G.}\ \bibnamefont {Huang}},\ }\href {\doibase 10.3847/1538-4357/ac86cb} {\bibfield  {journal} {\bibinfo  {journal} {Astrophys. J.}\ }\textbf {\bibinfo {volume} {936}},\ \bibinfo {pages} {20} (\bibinfo {year} {2022}{\natexlab{c}})},\ \Eprint {http://arxiv.org/abs/2205.07194} {arXiv:2205.07194 [astro-ph.CO]} \BibitemShut {NoStop}%
\bibitem [{\citenamefont {Ashoorioon}\ \emph {et~al.}(2022)\citenamefont {Ashoorioon}, \citenamefont {Rezazadeh},\ and\ \citenamefont {Rostami}}]{Ashoorioon:2022raz}%
  \BibitemOpen
  \bibfield  {author} {\bibinfo {author} {\bibfnamefont {A.}~\bibnamefont {Ashoorioon}}, \bibinfo {author} {\bibfnamefont {K.}~\bibnamefont {Rezazadeh}}, \ and\ \bibinfo {author} {\bibfnamefont {A.}~\bibnamefont {Rostami}},\ }\href {\doibase 10.1016/j.physletb.2022.137542} {\bibfield  {journal} {\bibinfo  {journal} {Phys. Lett. B}\ }\textbf {\bibinfo {volume} {835}},\ \bibinfo {pages} {137542} (\bibinfo {year} {2022})},\ \Eprint {http://arxiv.org/abs/2202.01131} {arXiv:2202.01131 [astro-ph.CO]} \BibitemShut {NoStop}%
\bibitem [{\citenamefont {Wu}\ \emph {et~al.}(2022{\natexlab{b}})\citenamefont {Wu}, \citenamefont {Chen}, \citenamefont {Huang}, \citenamefont {Zhu}, \citenamefont {Bhat}, \citenamefont {Feng}, \citenamefont {Hobbs}, \citenamefont {Manchester}, \citenamefont {Russell},\ and\ \citenamefont {Shannon}}]{PPTA:2022eul}%
  \BibitemOpen
  \bibfield  {author} {\bibinfo {author} {\bibfnamefont {Y.-M.}\ \bibnamefont {Wu}}, \bibinfo {author} {\bibfnamefont {Z.-C.}\ \bibnamefont {Chen}}, \bibinfo {author} {\bibfnamefont {Q.-G.}\ \bibnamefont {Huang}}, \bibinfo {author} {\bibfnamefont {X.}~\bibnamefont {Zhu}}, \bibinfo {author} {\bibfnamefont {N.~D.~R.}\ \bibnamefont {Bhat}}, \bibinfo {author} {\bibfnamefont {Y.}~\bibnamefont {Feng}}, \bibinfo {author} {\bibfnamefont {G.}~\bibnamefont {Hobbs}}, \bibinfo {author} {\bibfnamefont {R.~N.}\ \bibnamefont {Manchester}}, \bibinfo {author} {\bibfnamefont {C.~J.}\ \bibnamefont {Russell}}, \ and\ \bibinfo {author} {\bibfnamefont {R.~M.}\ \bibnamefont {Shannon}} (\bibinfo {collaboration} {PPTA}),\ }\href {\doibase 10.1103/PhysRevD.106.L081101} {\bibfield  {journal} {\bibinfo  {journal} {Phys. Rev. D}\ }\textbf {\bibinfo {volume} {106}},\ \bibinfo {pages} {L081101} (\bibinfo {year} {2022}{\natexlab{b}})},\ \Eprint {http://arxiv.org/abs/2210.03880} {arXiv:2210.03880 [astro-ph.CO]} \BibitemShut {NoStop}%
\bibitem [{\citenamefont {Wu}\ \emph {et~al.}(2023{\natexlab{a}})\citenamefont {Wu}, \citenamefont {Chen},\ and\ \citenamefont {Huang}}]{Wu:2023pbt}%
  \BibitemOpen
  \bibfield  {author} {\bibinfo {author} {\bibfnamefont {Y.-M.}\ \bibnamefont {Wu}}, \bibinfo {author} {\bibfnamefont {Z.-C.}\ \bibnamefont {Chen}}, \ and\ \bibinfo {author} {\bibfnamefont {Q.-G.}\ \bibnamefont {Huang}},\ }\href {\doibase 10.1103/PhysRevD.107.042003} {\bibfield  {journal} {\bibinfo  {journal} {Phys. Rev. D}\ }\textbf {\bibinfo {volume} {107}},\ \bibinfo {pages} {042003} (\bibinfo {year} {2023}{\natexlab{a}})},\ \Eprint {http://arxiv.org/abs/2302.00229} {arXiv:2302.00229 [gr-qc]} \BibitemShut {NoStop}%
\bibitem [{\citenamefont {Falxa}\ \emph {et~al.}(2023)\citenamefont {Falxa} \emph {et~al.}}]{IPTA:2023ero}%
  \BibitemOpen
  \bibfield  {author} {\bibinfo {author} {\bibfnamefont {M.}~\bibnamefont {Falxa}} \emph {et~al.} (\bibinfo {collaboration} {IPTA}),\ }\href {\doibase 10.1093/mnras/stad812} {\bibfield  {journal} {\bibinfo  {journal} {Mon. Not. Roy. Astron. Soc.}\ }\textbf {\bibinfo {volume} {521}},\ \bibinfo {pages} {5077} (\bibinfo {year} {2023})},\ \Eprint {http://arxiv.org/abs/2303.10767} {arXiv:2303.10767 [gr-qc]} \BibitemShut {NoStop}%
\bibitem [{\citenamefont {Wu}\ \emph {et~al.}(2023{\natexlab{b}})\citenamefont {Wu}, \citenamefont {Chen},\ and\ \citenamefont {Huang}}]{Wu:2023dnp}%
  \BibitemOpen
  \bibfield  {author} {\bibinfo {author} {\bibfnamefont {Y.-M.}\ \bibnamefont {Wu}}, \bibinfo {author} {\bibfnamefont {Z.-C.}\ \bibnamefont {Chen}}, \ and\ \bibinfo {author} {\bibfnamefont {Q.-G.}\ \bibnamefont {Huang}},\ }\href {\doibase 10.1088/1475-7516/2023/09/021} {\bibfield  {journal} {\bibinfo  {journal} {JCAP}\ }\textbf {\bibinfo {volume} {09}},\ \bibinfo {pages} {021} (\bibinfo {year} {2023}{\natexlab{b}})},\ \Eprint {http://arxiv.org/abs/2305.08091} {arXiv:2305.08091 [hep-ph]} \BibitemShut {NoStop}%
\bibitem [{\citenamefont {Dandoy}\ \emph {et~al.}(2023)\citenamefont {Dandoy}, \citenamefont {Domcke},\ and\ \citenamefont {Rompineve}}]{Dandoy:2023jot}%
  \BibitemOpen
  \bibfield  {author} {\bibinfo {author} {\bibfnamefont {V.}~\bibnamefont {Dandoy}}, \bibinfo {author} {\bibfnamefont {V.}~\bibnamefont {Domcke}}, \ and\ \bibinfo {author} {\bibfnamefont {F.}~\bibnamefont {Rompineve}},\ }\href {\doibase 10.21468/SciPostPhysCore.6.3.060} {\bibfield  {journal} {\bibinfo  {journal} {SciPost Phys. Core}\ }\textbf {\bibinfo {volume} {6}},\ \bibinfo {pages} {060} (\bibinfo {year} {2023})},\ \Eprint {http://arxiv.org/abs/2302.07901} {arXiv:2302.07901 [astro-ph.CO]} \BibitemShut {NoStop}%
\bibitem [{\citenamefont {Madge}\ \emph {et~al.}(2023)\citenamefont {Madge}, \citenamefont {Morgante}, \citenamefont {Puchades-Ib\'a\~nez}, \citenamefont {Ramberg}, \citenamefont {Ratzinger}, \citenamefont {Schenk},\ and\ \citenamefont {Schwaller}}]{Madge:2023cak}%
  \BibitemOpen
  \bibfield  {author} {\bibinfo {author} {\bibfnamefont {E.}~\bibnamefont {Madge}}, \bibinfo {author} {\bibfnamefont {E.}~\bibnamefont {Morgante}}, \bibinfo {author} {\bibfnamefont {C.}~\bibnamefont {Puchades-Ib\'a\~nez}}, \bibinfo {author} {\bibfnamefont {N.}~\bibnamefont {Ramberg}}, \bibinfo {author} {\bibfnamefont {W.}~\bibnamefont {Ratzinger}}, \bibinfo {author} {\bibfnamefont {S.}~\bibnamefont {Schenk}}, \ and\ \bibinfo {author} {\bibfnamefont {P.}~\bibnamefont {Schwaller}},\ }\href {\doibase 10.1007/JHEP10(2023)171} {\bibfield  {journal} {\bibinfo  {journal} {JHEP}\ }\textbf {\bibinfo {volume} {10}},\ \bibinfo {pages} {171} (\bibinfo {year} {2023})},\ \Eprint {http://arxiv.org/abs/2306.14856} {arXiv:2306.14856 [hep-ph]} \BibitemShut {NoStop}%
\bibitem [{\citenamefont {Yi}\ \emph {et~al.}(2023{\natexlab{a}})\citenamefont {Yi}, \citenamefont {You}, \citenamefont {Wu}, \citenamefont {Chen},\ and\ \citenamefont {Liu}}]{Yi:2023npi}%
  \BibitemOpen
  \bibfield  {author} {\bibinfo {author} {\bibfnamefont {Z.}~\bibnamefont {Yi}}, \bibinfo {author} {\bibfnamefont {Z.-Q.}\ \bibnamefont {You}}, \bibinfo {author} {\bibfnamefont {Y.}~\bibnamefont {Wu}}, \bibinfo {author} {\bibfnamefont {Z.-C.}\ \bibnamefont {Chen}}, \ and\ \bibinfo {author} {\bibfnamefont {L.}~\bibnamefont {Liu}},\ }\href@noop {} {\  (\bibinfo {year} {2023}{\natexlab{a}})},\ \Eprint {http://arxiv.org/abs/2308.14688} {arXiv:2308.14688 [astro-ph.CO]} \BibitemShut {NoStop}%
\bibitem [{\citenamefont {Wu}\ \emph {et~al.}(2024{\natexlab{a}})\citenamefont {Wu}, \citenamefont {Chen}, \citenamefont {Bi},\ and\ \citenamefont {Huang}}]{Wu:2023rib}%
  \BibitemOpen
  \bibfield  {author} {\bibinfo {author} {\bibfnamefont {Y.-M.}\ \bibnamefont {Wu}}, \bibinfo {author} {\bibfnamefont {Z.-C.}\ \bibnamefont {Chen}}, \bibinfo {author} {\bibfnamefont {Y.-C.}\ \bibnamefont {Bi}}, \ and\ \bibinfo {author} {\bibfnamefont {Q.-G.}\ \bibnamefont {Huang}},\ }\href {\doibase 10.1088/1361-6382/ad2a9b} {\bibfield  {journal} {\bibinfo  {journal} {Class. Quant. Grav.}\ }\textbf {\bibinfo {volume} {41}},\ \bibinfo {pages} {075002} (\bibinfo {year} {2024}{\natexlab{a}})},\ \Eprint {http://arxiv.org/abs/2310.07469} {arXiv:2310.07469 [astro-ph.CO]} \BibitemShut {NoStop}%
\bibitem [{\citenamefont {Bi}\ \emph {et~al.}(2024)\citenamefont {Bi}, \citenamefont {Wu}, \citenamefont {Chen},\ and\ \citenamefont {Huang}}]{Bi:2023ewq}%
  \BibitemOpen
  \bibfield  {author} {\bibinfo {author} {\bibfnamefont {Y.-C.}\ \bibnamefont {Bi}}, \bibinfo {author} {\bibfnamefont {Y.-M.}\ \bibnamefont {Wu}}, \bibinfo {author} {\bibfnamefont {Z.-C.}\ \bibnamefont {Chen}}, \ and\ \bibinfo {author} {\bibfnamefont {Q.-G.}\ \bibnamefont {Huang}},\ }\href {\doibase 10.1103/PhysRevD.109.L061101} {\bibfield  {journal} {\bibinfo  {journal} {Phys. Rev. D}\ }\textbf {\bibinfo {volume} {109}},\ \bibinfo {pages} {L061101} (\bibinfo {year} {2024})},\ \Eprint {http://arxiv.org/abs/2310.08366} {arXiv:2310.08366 [astro-ph.CO]} \BibitemShut {NoStop}%
\bibitem [{\citenamefont {Chen}\ \emph {et~al.}(2023{\natexlab{b}})\citenamefont {Chen}, \citenamefont {Wu}, \citenamefont {Bi},\ and\ \citenamefont {Huang}}]{Chen:2023uiz}%
  \BibitemOpen
  \bibfield  {author} {\bibinfo {author} {\bibfnamefont {Z.-C.}\ \bibnamefont {Chen}}, \bibinfo {author} {\bibfnamefont {Y.-M.}\ \bibnamefont {Wu}}, \bibinfo {author} {\bibfnamefont {Y.-C.}\ \bibnamefont {Bi}}, \ and\ \bibinfo {author} {\bibfnamefont {Q.-G.}\ \bibnamefont {Huang}},\ }\href@noop {} {\  (\bibinfo {year} {2023}{\natexlab{b}})},\ \Eprint {http://arxiv.org/abs/2310.11238} {arXiv:2310.11238 [astro-ph.CO]} \BibitemShut {NoStop}%
\bibitem [{\citenamefont {Xu}\ \emph {et~al.}(2023)\citenamefont {Xu} \emph {et~al.}}]{Xu:2023wog}%
  \BibitemOpen
  \bibfield  {author} {\bibinfo {author} {\bibfnamefont {H.}~\bibnamefont {Xu}} \emph {et~al.},\ }\href {\doibase 10.1088/1674-4527/acdfa5} {\bibfield  {journal} {\bibinfo  {journal} {Res. Astron. Astrophys.}\ }\textbf {\bibinfo {volume} {23}},\ \bibinfo {pages} {075024} (\bibinfo {year} {2023})},\ \Eprint {http://arxiv.org/abs/2306.16216} {arXiv:2306.16216 [astro-ph.HE]} \BibitemShut {NoStop}%
\bibitem [{\citenamefont {Antoniadis}\ \emph {et~al.}(2023{\natexlab{a}})\citenamefont {Antoniadis} \emph {et~al.}}]{EPTA:2023sfo}%
  \BibitemOpen
  \bibfield  {author} {\bibinfo {author} {\bibfnamefont {J.}~\bibnamefont {Antoniadis}} \emph {et~al.} (\bibinfo {collaboration} {EPTA}),\ }\href {\doibase 10.1051/0004-6361/202346841} {\bibfield  {journal} {\bibinfo  {journal} {Astron. Astrophys.}\ }\textbf {\bibinfo {volume} {678}},\ \bibinfo {pages} {A48} (\bibinfo {year} {2023}{\natexlab{a}})},\ \Eprint {http://arxiv.org/abs/2306.16224} {arXiv:2306.16224 [astro-ph.HE]} \BibitemShut {NoStop}%
\bibitem [{\citenamefont {Antoniadis}\ \emph {et~al.}(2023{\natexlab{b}})\citenamefont {Antoniadis} \emph {et~al.}}]{Antoniadis:2023ott}%
  \BibitemOpen
  \bibfield  {author} {\bibinfo {author} {\bibfnamefont {J.}~\bibnamefont {Antoniadis}} \emph {et~al.} (\bibinfo {collaboration} {EPTA, InPTA:}),\ }\href {\doibase 10.1051/0004-6361/202346844} {\bibfield  {journal} {\bibinfo  {journal} {Astron. Astrophys.}\ }\textbf {\bibinfo {volume} {678}},\ \bibinfo {pages} {A50} (\bibinfo {year} {2023}{\natexlab{b}})},\ \Eprint {http://arxiv.org/abs/2306.16214} {arXiv:2306.16214 [astro-ph.HE]} \BibitemShut {NoStop}%
\bibitem [{\citenamefont {Zic}\ \emph {et~al.}(2023)\citenamefont {Zic} \emph {et~al.}}]{Zic:2023gta}%
  \BibitemOpen
  \bibfield  {author} {\bibinfo {author} {\bibfnamefont {A.}~\bibnamefont {Zic}} \emph {et~al.},\ }\href {\doibase 10.1017/pasa.2023.36} {\bibfield  {journal} {\bibinfo  {journal} {Publ. Astron. Soc. Austral.}\ }\textbf {\bibinfo {volume} {40}},\ \bibinfo {pages} {e049} (\bibinfo {year} {2023})},\ \Eprint {http://arxiv.org/abs/2306.16230} {arXiv:2306.16230 [astro-ph.HE]} \BibitemShut {NoStop}%
\bibitem [{\citenamefont {Reardon}\ \emph {et~al.}(2023)\citenamefont {Reardon} \emph {et~al.}}]{Reardon:2023gzh}%
  \BibitemOpen
  \bibfield  {author} {\bibinfo {author} {\bibfnamefont {D.~J.}\ \bibnamefont {Reardon}} \emph {et~al.},\ }\href {\doibase 10.3847/2041-8213/acdd02} {\bibfield  {journal} {\bibinfo  {journal} {Astrophys. J. Lett.}\ }\textbf {\bibinfo {volume} {951}},\ \bibinfo {pages} {L6} (\bibinfo {year} {2023})},\ \Eprint {http://arxiv.org/abs/2306.16215} {arXiv:2306.16215 [astro-ph.HE]} \BibitemShut {NoStop}%
\bibitem [{\citenamefont {Agazie}\ \emph {et~al.}(2023{\natexlab{a}})\citenamefont {Agazie} \emph {et~al.}}]{NANOGrav:2023hde}%
  \BibitemOpen
  \bibfield  {author} {\bibinfo {author} {\bibfnamefont {G.}~\bibnamefont {Agazie}} \emph {et~al.} (\bibinfo {collaboration} {NANOGrav}),\ }\href {\doibase 10.3847/2041-8213/acda9a} {\bibfield  {journal} {\bibinfo  {journal} {Astrophys. J. Lett.}\ }\textbf {\bibinfo {volume} {951}},\ \bibinfo {pages} {L9} (\bibinfo {year} {2023}{\natexlab{a}})},\ \Eprint {http://arxiv.org/abs/2306.16217} {arXiv:2306.16217 [astro-ph.HE]} \BibitemShut {NoStop}%
\bibitem [{\citenamefont {Agazie}\ \emph {et~al.}(2023{\natexlab{b}})\citenamefont {Agazie} \emph {et~al.}}]{NANOGrav:2023gor}%
  \BibitemOpen
  \bibfield  {author} {\bibinfo {author} {\bibfnamefont {G.}~\bibnamefont {Agazie}} \emph {et~al.} (\bibinfo {collaboration} {NANOGrav}),\ }\href {\doibase 10.3847/2041-8213/acdac6} {\bibfield  {journal} {\bibinfo  {journal} {Astrophys. J. Lett.}\ }\textbf {\bibinfo {volume} {951}},\ \bibinfo {pages} {L8} (\bibinfo {year} {2023}{\natexlab{b}})},\ \Eprint {http://arxiv.org/abs/2306.16213} {arXiv:2306.16213 [astro-ph.HE]} \BibitemShut {NoStop}%
\bibitem [{\citenamefont {Afzal}\ \emph {et~al.}(2023)\citenamefont {Afzal} \emph {et~al.}}]{NANOGrav:2023hvm}%
  \BibitemOpen
  \bibfield  {author} {\bibinfo {author} {\bibfnamefont {A.}~\bibnamefont {Afzal}} \emph {et~al.} (\bibinfo {collaboration} {NANOGrav}),\ }\href {\doibase 10.3847/2041-8213/acdc91} {\bibfield  {journal} {\bibinfo  {journal} {Astrophys. J. Lett.}\ }\textbf {\bibinfo {volume} {951}},\ \bibinfo {pages} {L11} (\bibinfo {year} {2023})},\ \Eprint {http://arxiv.org/abs/2306.16219} {arXiv:2306.16219 [astro-ph.HE]} \BibitemShut {NoStop}%
\bibitem [{\citenamefont {Antoniadis}\ \emph {et~al.}(2023{\natexlab{c}})\citenamefont {Antoniadis} \emph {et~al.}}]{Antoniadis:2023xlr}%
  \BibitemOpen
  \bibfield  {author} {\bibinfo {author} {\bibfnamefont {J.}~\bibnamefont {Antoniadis}} \emph {et~al.} (\bibinfo {collaboration} {EPTA}),\ }\href@noop {} {\  (\bibinfo {year} {2023}{\natexlab{c}})},\ \Eprint {http://arxiv.org/abs/2306.16227} {arXiv:2306.16227 [astro-ph.CO]} \BibitemShut {NoStop}%
\bibitem [{\citenamefont {Bi}\ \emph {et~al.}(2023)\citenamefont {Bi}, \citenamefont {Wu}, \citenamefont {Chen},\ and\ \citenamefont {Huang}}]{Bi:2023tib}%
  \BibitemOpen
  \bibfield  {author} {\bibinfo {author} {\bibfnamefont {Y.-C.}\ \bibnamefont {Bi}}, \bibinfo {author} {\bibfnamefont {Y.-M.}\ \bibnamefont {Wu}}, \bibinfo {author} {\bibfnamefont {Z.-C.}\ \bibnamefont {Chen}}, \ and\ \bibinfo {author} {\bibfnamefont {Q.-G.}\ \bibnamefont {Huang}},\ }\href {\doibase 10.1007/s11433-023-2252-4} {\bibfield  {journal} {\bibinfo  {journal} {Sci. China Phys. Mech. Astron.}\ }\textbf {\bibinfo {volume} {66}},\ \bibinfo {pages} {120402} (\bibinfo {year} {2023})},\ \Eprint {http://arxiv.org/abs/2307.00722} {arXiv:2307.00722 [astro-ph.CO]} \BibitemShut {NoStop}%
\bibitem [{\citenamefont {Zhao}\ \emph {et~al.}(2023)\citenamefont {Zhao}, \citenamefont {Zhu}, \citenamefont {Wang},\ and\ \citenamefont {Zhang}}]{Zhao:2023joc}%
  \BibitemOpen
  \bibfield  {author} {\bibinfo {author} {\bibfnamefont {Z.-C.}\ \bibnamefont {Zhao}}, \bibinfo {author} {\bibfnamefont {Q.-H.}\ \bibnamefont {Zhu}}, \bibinfo {author} {\bibfnamefont {S.}~\bibnamefont {Wang}}, \ and\ \bibinfo {author} {\bibfnamefont {X.}~\bibnamefont {Zhang}},\ }\href@noop {} {\  (\bibinfo {year} {2023})},\ \Eprint {http://arxiv.org/abs/2307.13574} {arXiv:2307.13574 [astro-ph.CO]} \BibitemShut {NoStop}%
\bibitem [{\citenamefont {Wang}\ \emph {et~al.}(2024)\citenamefont {Wang}, \citenamefont {Zhao}, \citenamefont {Li},\ and\ \citenamefont {Zhu}}]{Wang:2023ost}%
  \BibitemOpen
  \bibfield  {author} {\bibinfo {author} {\bibfnamefont {S.}~\bibnamefont {Wang}}, \bibinfo {author} {\bibfnamefont {Z.-C.}\ \bibnamefont {Zhao}}, \bibinfo {author} {\bibfnamefont {J.-P.}\ \bibnamefont {Li}}, \ and\ \bibinfo {author} {\bibfnamefont {Q.-H.}\ \bibnamefont {Zhu}},\ }\href {\doibase 10.1103/PhysRevResearch.6.L012060} {\bibfield  {journal} {\bibinfo  {journal} {Phys. Rev. Res.}\ }\textbf {\bibinfo {volume} {6}},\ \bibinfo {pages} {L012060} (\bibinfo {year} {2024})},\ \Eprint {http://arxiv.org/abs/2307.00572} {arXiv:2307.00572 [astro-ph.CO]} \BibitemShut {NoStop}%
\bibitem [{\citenamefont {Liu}\ \emph {et~al.}(2023{\natexlab{b}})\citenamefont {Liu}, \citenamefont {Chen},\ and\ \citenamefont {Huang}}]{Liu:2023pau}%
  \BibitemOpen
  \bibfield  {author} {\bibinfo {author} {\bibfnamefont {L.}~\bibnamefont {Liu}}, \bibinfo {author} {\bibfnamefont {Z.-C.}\ \bibnamefont {Chen}}, \ and\ \bibinfo {author} {\bibfnamefont {Q.-G.}\ \bibnamefont {Huang}},\ }\href {\doibase 10.1088/1475-7516/2023/11/071} {\bibfield  {journal} {\bibinfo  {journal} {JCAP}\ }\textbf {\bibinfo {volume} {11}},\ \bibinfo {pages} {071} (\bibinfo {year} {2023}{\natexlab{b}})},\ \Eprint {http://arxiv.org/abs/2307.14911} {arXiv:2307.14911 [astro-ph.CO]} \BibitemShut {NoStop}%
\bibitem [{\citenamefont {Vagnozzi}(2023)}]{Vagnozzi:2023lwo}%
  \BibitemOpen
  \bibfield  {author} {\bibinfo {author} {\bibfnamefont {S.}~\bibnamefont {Vagnozzi}},\ }\href {\doibase 10.1016/j.jheap.2023.07.001} {\bibfield  {journal} {\bibinfo  {journal} {JHEAp}\ }\textbf {\bibinfo {volume} {39}},\ \bibinfo {pages} {81} (\bibinfo {year} {2023})},\ \Eprint {http://arxiv.org/abs/2306.16912} {arXiv:2306.16912 [astro-ph.CO]} \BibitemShut {NoStop}%
\bibitem [{\citenamefont {Fu}\ \emph {et~al.}(2024)\citenamefont {Fu}, \citenamefont {Liu}, \citenamefont {Yang}, \citenamefont {Yu},\ and\ \citenamefont {Zhang}}]{Fu:2023aab}%
  \BibitemOpen
  \bibfield  {author} {\bibinfo {author} {\bibfnamefont {C.}~\bibnamefont {Fu}}, \bibinfo {author} {\bibfnamefont {J.}~\bibnamefont {Liu}}, \bibinfo {author} {\bibfnamefont {X.-Y.}\ \bibnamefont {Yang}}, \bibinfo {author} {\bibfnamefont {W.-W.}\ \bibnamefont {Yu}}, \ and\ \bibinfo {author} {\bibfnamefont {Y.}~\bibnamefont {Zhang}},\ }\href {\doibase 10.1103/PhysRevD.109.063526} {\bibfield  {journal} {\bibinfo  {journal} {Phys. Rev. D}\ }\textbf {\bibinfo {volume} {109}},\ \bibinfo {pages} {063526} (\bibinfo {year} {2024})},\ \Eprint {http://arxiv.org/abs/2308.15329} {arXiv:2308.15329 [astro-ph.CO]} \BibitemShut {NoStop}%
\bibitem [{\citenamefont {Han}\ \emph {et~al.}(2023)\citenamefont {Han}, \citenamefont {Xie}, \citenamefont {Yang},\ and\ \citenamefont {Zhang}}]{Han:2023olf}%
  \BibitemOpen
  \bibfield  {author} {\bibinfo {author} {\bibfnamefont {C.}~\bibnamefont {Han}}, \bibinfo {author} {\bibfnamefont {K.-P.}\ \bibnamefont {Xie}}, \bibinfo {author} {\bibfnamefont {J.~M.}\ \bibnamefont {Yang}}, \ and\ \bibinfo {author} {\bibfnamefont {M.}~\bibnamefont {Zhang}},\ }\href@noop {} {\  (\bibinfo {year} {2023})},\ \Eprint {http://arxiv.org/abs/2306.16966} {arXiv:2306.16966 [hep-ph]} \BibitemShut {NoStop}%
\bibitem [{\citenamefont {Kitajima}\ \emph {et~al.}(2024)\citenamefont {Kitajima}, \citenamefont {Lee}, \citenamefont {Murai}, \citenamefont {Takahashi},\ and\ \citenamefont {Yin}}]{Kitajima:2023cek}%
  \BibitemOpen
  \bibfield  {author} {\bibinfo {author} {\bibfnamefont {N.}~\bibnamefont {Kitajima}}, \bibinfo {author} {\bibfnamefont {J.}~\bibnamefont {Lee}}, \bibinfo {author} {\bibfnamefont {K.}~\bibnamefont {Murai}}, \bibinfo {author} {\bibfnamefont {F.}~\bibnamefont {Takahashi}}, \ and\ \bibinfo {author} {\bibfnamefont {W.}~\bibnamefont {Yin}},\ }\href {\doibase 10.1016/j.physletb.2024.138586} {\bibfield  {journal} {\bibinfo  {journal} {Phys. Lett. B}\ }\textbf {\bibinfo {volume} {851}},\ \bibinfo {pages} {138586} (\bibinfo {year} {2024})},\ \Eprint {http://arxiv.org/abs/2306.17146} {arXiv:2306.17146 [hep-ph]} \BibitemShut {NoStop}%
\bibitem [{\citenamefont {Franciolini}\ \emph {et~al.}(2023)\citenamefont {Franciolini}, \citenamefont {Iovino}, \citenamefont {Vaskonen},\ and\ \citenamefont {Veermae}}]{Franciolini:2023pbf}%
  \BibitemOpen
  \bibfield  {author} {\bibinfo {author} {\bibfnamefont {G.}~\bibnamefont {Franciolini}}, \bibinfo {author} {\bibfnamefont {A.}~\bibnamefont {Iovino}, \bibfnamefont {Junior.}}, \bibinfo {author} {\bibfnamefont {V.}~\bibnamefont {Vaskonen}}, \ and\ \bibinfo {author} {\bibfnamefont {H.}~\bibnamefont {Veermae}},\ }\href {\doibase 10.1103/PhysRevLett.131.201401} {\bibfield  {journal} {\bibinfo  {journal} {Phys. Rev. Lett.}\ }\textbf {\bibinfo {volume} {131}},\ \bibinfo {pages} {201401} (\bibinfo {year} {2023})},\ \Eprint {http://arxiv.org/abs/2306.17149} {arXiv:2306.17149 [astro-ph.CO]} \BibitemShut {NoStop}%
\bibitem [{\citenamefont {Cai}\ \emph {et~al.}(2023{\natexlab{a}})\citenamefont {Cai}, \citenamefont {He}, \citenamefont {Ma}, \citenamefont {Yan},\ and\ \citenamefont {Yuan}}]{Cai:2023dls}%
  \BibitemOpen
  \bibfield  {author} {\bibinfo {author} {\bibfnamefont {Y.-F.}\ \bibnamefont {Cai}}, \bibinfo {author} {\bibfnamefont {X.-C.}\ \bibnamefont {He}}, \bibinfo {author} {\bibfnamefont {X.-H.}\ \bibnamefont {Ma}}, \bibinfo {author} {\bibfnamefont {S.-F.}\ \bibnamefont {Yan}}, \ and\ \bibinfo {author} {\bibfnamefont {G.-W.}\ \bibnamefont {Yuan}},\ }\href {\doibase 10.1016/j.scib.2023.10.027} {\bibfield  {journal} {\bibinfo  {journal} {Sci. Bull.}\ }\textbf {\bibinfo {volume} {68}},\ \bibinfo {pages} {2929} (\bibinfo {year} {2023}{\natexlab{a}})},\ \Eprint {http://arxiv.org/abs/2306.17822} {arXiv:2306.17822 [gr-qc]} \BibitemShut {NoStop}%
\bibitem [{\citenamefont {Inomata}\ \emph {et~al.}(2024)\citenamefont {Inomata}, \citenamefont {Kohri},\ and\ \citenamefont {Terada}}]{Inomata:2023zup}%
  \BibitemOpen
  \bibfield  {author} {\bibinfo {author} {\bibfnamefont {K.}~\bibnamefont {Inomata}}, \bibinfo {author} {\bibfnamefont {K.}~\bibnamefont {Kohri}}, \ and\ \bibinfo {author} {\bibfnamefont {T.}~\bibnamefont {Terada}},\ }\href {\doibase 10.1103/PhysRevD.109.063506} {\bibfield  {journal} {\bibinfo  {journal} {Phys. Rev. D}\ }\textbf {\bibinfo {volume} {109}},\ \bibinfo {pages} {063506} (\bibinfo {year} {2024})},\ \Eprint {http://arxiv.org/abs/2306.17834} {arXiv:2306.17834 [astro-ph.CO]} \BibitemShut {NoStop}%
\bibitem [{\citenamefont {Li}\ and\ \citenamefont {Xie}(2023)}]{Li:2023bxy}%
  \BibitemOpen
  \bibfield  {author} {\bibinfo {author} {\bibfnamefont {S.-P.}\ \bibnamefont {Li}}\ and\ \bibinfo {author} {\bibfnamefont {K.-P.}\ \bibnamefont {Xie}},\ }\href {\doibase 10.1103/PhysRevD.108.055018} {\bibfield  {journal} {\bibinfo  {journal} {Phys. Rev. D}\ }\textbf {\bibinfo {volume} {108}},\ \bibinfo {pages} {055018} (\bibinfo {year} {2023})},\ \Eprint {http://arxiv.org/abs/2307.01086} {arXiv:2307.01086 [hep-ph]} \BibitemShut {NoStop}%
\bibitem [{\citenamefont {Liu}\ \emph {et~al.}(2024{\natexlab{a}})\citenamefont {Liu}, \citenamefont {Chen},\ and\ \citenamefont {Huang}}]{Liu:2023ymk}%
  \BibitemOpen
  \bibfield  {author} {\bibinfo {author} {\bibfnamefont {L.}~\bibnamefont {Liu}}, \bibinfo {author} {\bibfnamefont {Z.-C.}\ \bibnamefont {Chen}}, \ and\ \bibinfo {author} {\bibfnamefont {Q.-G.}\ \bibnamefont {Huang}},\ }\href {\doibase 10.1103/PhysRevD.109.L061301} {\bibfield  {journal} {\bibinfo  {journal} {Phys. Rev. D}\ }\textbf {\bibinfo {volume} {109}},\ \bibinfo {pages} {L061301} (\bibinfo {year} {2024}{\natexlab{a}})},\ \Eprint {http://arxiv.org/abs/2307.01102} {arXiv:2307.01102 [astro-ph.CO]} \BibitemShut {NoStop}%
\bibitem [{\citenamefont {Abe}\ and\ \citenamefont {Tada}(2023)}]{Abe:2023yrw}%
  \BibitemOpen
  \bibfield  {author} {\bibinfo {author} {\bibfnamefont {K.~T.}\ \bibnamefont {Abe}}\ and\ \bibinfo {author} {\bibfnamefont {Y.}~\bibnamefont {Tada}},\ }\href {\doibase 10.1103/PhysRevD.108.L101304} {\bibfield  {journal} {\bibinfo  {journal} {Phys. Rev. D}\ }\textbf {\bibinfo {volume} {108}},\ \bibinfo {pages} {L101304} (\bibinfo {year} {2023})},\ \Eprint {http://arxiv.org/abs/2307.01653} {arXiv:2307.01653 [astro-ph.CO]} \BibitemShut {NoStop}%
\bibitem [{\citenamefont {Ghosh}\ \emph {et~al.}(2023)\citenamefont {Ghosh}, \citenamefont {Ghoshal}, \citenamefont {Guo}, \citenamefont {Hajkarim}, \citenamefont {King}, \citenamefont {Sinha}, \citenamefont {Wang},\ and\ \citenamefont {White}}]{Ghosh:2023aum}%
  \BibitemOpen
  \bibfield  {author} {\bibinfo {author} {\bibfnamefont {T.}~\bibnamefont {Ghosh}}, \bibinfo {author} {\bibfnamefont {A.}~\bibnamefont {Ghoshal}}, \bibinfo {author} {\bibfnamefont {H.-K.}\ \bibnamefont {Guo}}, \bibinfo {author} {\bibfnamefont {F.}~\bibnamefont {Hajkarim}}, \bibinfo {author} {\bibfnamefont {S.~F.}\ \bibnamefont {King}}, \bibinfo {author} {\bibfnamefont {K.}~\bibnamefont {Sinha}}, \bibinfo {author} {\bibfnamefont {X.}~\bibnamefont {Wang}}, \ and\ \bibinfo {author} {\bibfnamefont {G.}~\bibnamefont {White}},\ }\href@noop {} {\  (\bibinfo {year} {2023})},\ \Eprint {http://arxiv.org/abs/2307.02259} {arXiv:2307.02259 [astro-ph.HE]} \BibitemShut {NoStop}%
\bibitem [{\citenamefont {Figueroa}\ \emph {et~al.}(2023)\citenamefont {Figueroa}, \citenamefont {Pieroni}, \citenamefont {Ricciardone},\ and\ \citenamefont {Simakachorn}}]{Figueroa:2023zhu}%
  \BibitemOpen
  \bibfield  {author} {\bibinfo {author} {\bibfnamefont {D.~G.}\ \bibnamefont {Figueroa}}, \bibinfo {author} {\bibfnamefont {M.}~\bibnamefont {Pieroni}}, \bibinfo {author} {\bibfnamefont {A.}~\bibnamefont {Ricciardone}}, \ and\ \bibinfo {author} {\bibfnamefont {P.}~\bibnamefont {Simakachorn}},\ }\href@noop {} {\  (\bibinfo {year} {2023})},\ \Eprint {http://arxiv.org/abs/2307.02399} {arXiv:2307.02399 [astro-ph.CO]} \BibitemShut {NoStop}%
\bibitem [{\citenamefont {Yi}\ \emph {et~al.}(2023{\natexlab{b}})\citenamefont {Yi}, \citenamefont {Gao}, \citenamefont {Gong}, \citenamefont {Wang},\ and\ \citenamefont {Zhang}}]{Yi:2023mbm}%
  \BibitemOpen
  \bibfield  {author} {\bibinfo {author} {\bibfnamefont {Z.}~\bibnamefont {Yi}}, \bibinfo {author} {\bibfnamefont {Q.}~\bibnamefont {Gao}}, \bibinfo {author} {\bibfnamefont {Y.}~\bibnamefont {Gong}}, \bibinfo {author} {\bibfnamefont {Y.}~\bibnamefont {Wang}}, \ and\ \bibinfo {author} {\bibfnamefont {F.}~\bibnamefont {Zhang}},\ }\href {\doibase 10.1007/s11433-023-2266-1} {\bibfield  {journal} {\bibinfo  {journal} {Sci. China Phys. Mech. Astron.}\ }\textbf {\bibinfo {volume} {66}},\ \bibinfo {pages} {120404} (\bibinfo {year} {2023}{\natexlab{b}})},\ \Eprint {http://arxiv.org/abs/2307.02467} {arXiv:2307.02467 [gr-qc]} \BibitemShut {NoStop}%
\bibitem [{\citenamefont {Wu}\ \emph {et~al.}(2024{\natexlab{b}})\citenamefont {Wu}, \citenamefont {Chen},\ and\ \citenamefont {Huang}}]{Wu:2023hsa}%
  \BibitemOpen
  \bibfield  {author} {\bibinfo {author} {\bibfnamefont {Y.-M.}\ \bibnamefont {Wu}}, \bibinfo {author} {\bibfnamefont {Z.-C.}\ \bibnamefont {Chen}}, \ and\ \bibinfo {author} {\bibfnamefont {Q.-G.}\ \bibnamefont {Huang}},\ }\href {\doibase 10.1007/s11433-023-2298-7} {\bibfield  {journal} {\bibinfo  {journal} {Sci. China Phys. Mech. Astron.}\ }\textbf {\bibinfo {volume} {67}},\ \bibinfo {pages} {240412} (\bibinfo {year} {2024}{\natexlab{b}})},\ \Eprint {http://arxiv.org/abs/2307.03141} {arXiv:2307.03141 [astro-ph.CO]} \BibitemShut {NoStop}%
\bibitem [{\citenamefont {You}\ \emph {et~al.}(2023{\natexlab{b}})\citenamefont {You}, \citenamefont {Yi},\ and\ \citenamefont {Wu}}]{You:2023rmn}%
  \BibitemOpen
  \bibfield  {author} {\bibinfo {author} {\bibfnamefont {Z.-Q.}\ \bibnamefont {You}}, \bibinfo {author} {\bibfnamefont {Z.}~\bibnamefont {Yi}}, \ and\ \bibinfo {author} {\bibfnamefont {Y.}~\bibnamefont {Wu}},\ }\href {\doibase 10.1088/1475-7516/2023/11/065} {\bibfield  {journal} {\bibinfo  {journal} {JCAP}\ }\textbf {\bibinfo {volume} {11}},\ \bibinfo {pages} {065} (\bibinfo {year} {2023}{\natexlab{b}})},\ \Eprint {http://arxiv.org/abs/2307.04419} {arXiv:2307.04419 [gr-qc]} \BibitemShut {NoStop}%
\bibitem [{\citenamefont {Antusch}\ \emph {et~al.}(2023)\citenamefont {Antusch}, \citenamefont {Hinze}, \citenamefont {Saad},\ and\ \citenamefont {Steiner}}]{Antusch:2023zjk}%
  \BibitemOpen
  \bibfield  {author} {\bibinfo {author} {\bibfnamefont {S.}~\bibnamefont {Antusch}}, \bibinfo {author} {\bibfnamefont {K.}~\bibnamefont {Hinze}}, \bibinfo {author} {\bibfnamefont {S.}~\bibnamefont {Saad}}, \ and\ \bibinfo {author} {\bibfnamefont {J.}~\bibnamefont {Steiner}},\ }\href {\doibase 10.1103/PhysRevD.108.095053} {\bibfield  {journal} {\bibinfo  {journal} {Phys. Rev. D}\ }\textbf {\bibinfo {volume} {108}},\ \bibinfo {pages} {095053} (\bibinfo {year} {2023})},\ \Eprint {http://arxiv.org/abs/2307.04595} {arXiv:2307.04595 [hep-ph]} \BibitemShut {NoStop}%
\bibitem [{\citenamefont {Hosseini~Mansoori}\ \emph {et~al.}(2023)\citenamefont {Hosseini~Mansoori}, \citenamefont {Felegray}, \citenamefont {Talebian},\ and\ \citenamefont {Sami}}]{HosseiniMansoori:2023mqh}%
  \BibitemOpen
  \bibfield  {author} {\bibinfo {author} {\bibfnamefont {S.~A.}\ \bibnamefont {Hosseini~Mansoori}}, \bibinfo {author} {\bibfnamefont {F.}~\bibnamefont {Felegray}}, \bibinfo {author} {\bibfnamefont {A.}~\bibnamefont {Talebian}}, \ and\ \bibinfo {author} {\bibfnamefont {M.}~\bibnamefont {Sami}},\ }\href {\doibase 10.1088/1475-7516/2023/08/067} {\bibfield  {journal} {\bibinfo  {journal} {JCAP}\ }\textbf {\bibinfo {volume} {08}},\ \bibinfo {pages} {067} (\bibinfo {year} {2023})},\ \Eprint {http://arxiv.org/abs/2307.06757} {arXiv:2307.06757 [astro-ph.CO]} \BibitemShut {NoStop}%
\bibitem [{\citenamefont {Jin}\ \emph {et~al.}(2023)\citenamefont {Jin}, \citenamefont {Chen}, \citenamefont {Yi}, \citenamefont {You}, \citenamefont {Liu},\ and\ \citenamefont {Wu}}]{Jin:2023wri}%
  \BibitemOpen
  \bibfield  {author} {\bibinfo {author} {\bibfnamefont {J.-H.}\ \bibnamefont {Jin}}, \bibinfo {author} {\bibfnamefont {Z.-C.}\ \bibnamefont {Chen}}, \bibinfo {author} {\bibfnamefont {Z.}~\bibnamefont {Yi}}, \bibinfo {author} {\bibfnamefont {Z.-Q.}\ \bibnamefont {You}}, \bibinfo {author} {\bibfnamefont {L.}~\bibnamefont {Liu}}, \ and\ \bibinfo {author} {\bibfnamefont {Y.}~\bibnamefont {Wu}},\ }\href {\doibase 10.1088/1475-7516/2023/09/016} {\bibfield  {journal} {\bibinfo  {journal} {JCAP}\ }\textbf {\bibinfo {volume} {09}},\ \bibinfo {pages} {016} (\bibinfo {year} {2023})},\ \Eprint {http://arxiv.org/abs/2307.08687} {arXiv:2307.08687 [astro-ph.CO]} \BibitemShut {NoStop}%
\bibitem [{\citenamefont {Zhang}\ \emph {et~al.}(2023)\citenamefont {Zhang}, \citenamefont {Cai}, \citenamefont {Su}, \citenamefont {Wang}, \citenamefont {Yu},\ and\ \citenamefont {Zhang}}]{Zhang:2023nrs}%
  \BibitemOpen
  \bibfield  {author} {\bibinfo {author} {\bibfnamefont {Z.}~\bibnamefont {Zhang}}, \bibinfo {author} {\bibfnamefont {C.}~\bibnamefont {Cai}}, \bibinfo {author} {\bibfnamefont {Y.-H.}\ \bibnamefont {Su}}, \bibinfo {author} {\bibfnamefont {S.}~\bibnamefont {Wang}}, \bibinfo {author} {\bibfnamefont {Z.-H.}\ \bibnamefont {Yu}}, \ and\ \bibinfo {author} {\bibfnamefont {H.-H.}\ \bibnamefont {Zhang}},\ }\href {\doibase 10.1103/PhysRevD.108.095037} {\bibfield  {journal} {\bibinfo  {journal} {Phys. Rev. D}\ }\textbf {\bibinfo {volume} {108}},\ \bibinfo {pages} {095037} (\bibinfo {year} {2023})},\ \Eprint {http://arxiv.org/abs/2307.11495} {arXiv:2307.11495 [hep-ph]} \BibitemShut {NoStop}%
\bibitem [{\citenamefont {Choudhury}(2024)}]{Choudhury:2023kam}%
  \BibitemOpen
  \bibfield  {author} {\bibinfo {author} {\bibfnamefont {S.}~\bibnamefont {Choudhury}},\ }\href {\doibase 10.1140/epjc/s10052-024-12625-9} {\bibfield  {journal} {\bibinfo  {journal} {Eur. Phys. J. C}\ }\textbf {\bibinfo {volume} {84}},\ \bibinfo {pages} {278} (\bibinfo {year} {2024})},\ \Eprint {http://arxiv.org/abs/2307.03249} {arXiv:2307.03249 [astro-ph.CO]} \BibitemShut {NoStop}%
\bibitem [{\citenamefont {Gorji}\ \emph {et~al.}(2023)\citenamefont {Gorji}, \citenamefont {Sasaki},\ and\ \citenamefont {Suyama}}]{Gorji:2023sil}%
  \BibitemOpen
  \bibfield  {author} {\bibinfo {author} {\bibfnamefont {M.~A.}\ \bibnamefont {Gorji}}, \bibinfo {author} {\bibfnamefont {M.}~\bibnamefont {Sasaki}}, \ and\ \bibinfo {author} {\bibfnamefont {T.}~\bibnamefont {Suyama}},\ }\href {\doibase 10.1016/j.physletb.2023.138214} {\bibfield  {journal} {\bibinfo  {journal} {Phys. Lett. B}\ }\textbf {\bibinfo {volume} {846}},\ \bibinfo {pages} {138214} (\bibinfo {year} {2023})},\ \Eprint {http://arxiv.org/abs/2307.13109} {arXiv:2307.13109 [astro-ph.CO]} \BibitemShut {NoStop}%
\bibitem [{\citenamefont {Das}\ \emph {et~al.}(2023)\citenamefont {Das}, \citenamefont {Jaman},\ and\ \citenamefont {Sami}}]{Das:2023nmm}%
  \BibitemOpen
  \bibfield  {author} {\bibinfo {author} {\bibfnamefont {B.}~\bibnamefont {Das}}, \bibinfo {author} {\bibfnamefont {N.}~\bibnamefont {Jaman}}, \ and\ \bibinfo {author} {\bibfnamefont {M.}~\bibnamefont {Sami}},\ }\href {\doibase 10.1103/PhysRevD.108.103510} {\bibfield  {journal} {\bibinfo  {journal} {Phys. Rev. D}\ }\textbf {\bibinfo {volume} {108}},\ \bibinfo {pages} {103510} (\bibinfo {year} {2023})},\ \Eprint {http://arxiv.org/abs/2307.12913} {arXiv:2307.12913 [gr-qc]} \BibitemShut {NoStop}%
\bibitem [{\citenamefont {Yi}\ \emph {et~al.}(2024)\citenamefont {Yi}, \citenamefont {You},\ and\ \citenamefont {Wu}}]{Yi:2023tdk}%
  \BibitemOpen
  \bibfield  {author} {\bibinfo {author} {\bibfnamefont {Z.}~\bibnamefont {Yi}}, \bibinfo {author} {\bibfnamefont {Z.-Q.}\ \bibnamefont {You}}, \ and\ \bibinfo {author} {\bibfnamefont {Y.}~\bibnamefont {Wu}},\ }\href {\doibase 10.1088/1475-7516/2024/01/066} {\bibfield  {journal} {\bibinfo  {journal} {JCAP}\ }\textbf {\bibinfo {volume} {01}},\ \bibinfo {pages} {066} (\bibinfo {year} {2024})},\ \Eprint {http://arxiv.org/abs/2308.05632} {arXiv:2308.05632 [astro-ph.CO]} \BibitemShut {NoStop}%
\bibitem [{\citenamefont {Ellis}\ \emph {et~al.}(2024)\citenamefont {Ellis}, \citenamefont {Fairbairn}, \citenamefont {Franciolini}, \citenamefont {H\"utsi}, \citenamefont {Iovino}, \citenamefont {Lewicki}, \citenamefont {Raidal}, \citenamefont {Urrutia}, \citenamefont {Vaskonen},\ and\ \citenamefont {Veerm\"ae}}]{Ellis:2023oxs}%
  \BibitemOpen
  \bibfield  {author} {\bibinfo {author} {\bibfnamefont {J.}~\bibnamefont {Ellis}}, \bibinfo {author} {\bibfnamefont {M.}~\bibnamefont {Fairbairn}}, \bibinfo {author} {\bibfnamefont {G.}~\bibnamefont {Franciolini}}, \bibinfo {author} {\bibfnamefont {G.}~\bibnamefont {H\"utsi}}, \bibinfo {author} {\bibfnamefont {A.}~\bibnamefont {Iovino}}, \bibinfo {author} {\bibfnamefont {M.}~\bibnamefont {Lewicki}}, \bibinfo {author} {\bibfnamefont {M.}~\bibnamefont {Raidal}}, \bibinfo {author} {\bibfnamefont {J.}~\bibnamefont {Urrutia}}, \bibinfo {author} {\bibfnamefont {V.}~\bibnamefont {Vaskonen}}, \ and\ \bibinfo {author} {\bibfnamefont {H.}~\bibnamefont {Veerm\"ae}},\ }\href {\doibase 10.1103/PhysRevD.109.023522} {\bibfield  {journal} {\bibinfo  {journal} {Phys. Rev. D}\ }\textbf {\bibinfo {volume} {109}},\ \bibinfo {pages} {023522} (\bibinfo {year} {2024})},\ \Eprint {http://arxiv.org/abs/2308.08546} {arXiv:2308.08546 [astro-ph.CO]} \BibitemShut {NoStop}%
\bibitem [{\citenamefont {He}\ \emph {et~al.}(2023)\citenamefont {He}, \citenamefont {Li}, \citenamefont {Wang},\ and\ \citenamefont {Wang}}]{He:2023ado}%
  \BibitemOpen
  \bibfield  {author} {\bibinfo {author} {\bibfnamefont {S.}~\bibnamefont {He}}, \bibinfo {author} {\bibfnamefont {L.}~\bibnamefont {Li}}, \bibinfo {author} {\bibfnamefont {S.}~\bibnamefont {Wang}}, \ and\ \bibinfo {author} {\bibfnamefont {S.-J.}\ \bibnamefont {Wang}},\ }\href@noop {} {\  (\bibinfo {year} {2023})},\ \Eprint {http://arxiv.org/abs/2308.07257} {arXiv:2308.07257 [hep-ph]} \BibitemShut {NoStop}%
\bibitem [{\citenamefont {Balaji}\ \emph {et~al.}(2023)\citenamefont {Balaji}, \citenamefont {Dom\`enech},\ and\ \citenamefont {Franciolini}}]{Balaji:2023ehk}%
  \BibitemOpen
  \bibfield  {author} {\bibinfo {author} {\bibfnamefont {S.}~\bibnamefont {Balaji}}, \bibinfo {author} {\bibfnamefont {G.}~\bibnamefont {Dom\`enech}}, \ and\ \bibinfo {author} {\bibfnamefont {G.}~\bibnamefont {Franciolini}},\ }\href {\doibase 10.1088/1475-7516/2023/10/041} {\bibfield  {journal} {\bibinfo  {journal} {JCAP}\ }\textbf {\bibinfo {volume} {10}},\ \bibinfo {pages} {041} (\bibinfo {year} {2023})},\ \Eprint {http://arxiv.org/abs/2307.08552} {arXiv:2307.08552 [gr-qc]} \BibitemShut {NoStop}%
\bibitem [{\citenamefont {Cannizzaro}\ \emph {et~al.}(2024)\citenamefont {Cannizzaro}, \citenamefont {Franciolini},\ and\ \citenamefont {Pani}}]{Cannizzaro:2023mgc}%
  \BibitemOpen
  \bibfield  {author} {\bibinfo {author} {\bibfnamefont {E.}~\bibnamefont {Cannizzaro}}, \bibinfo {author} {\bibfnamefont {G.}~\bibnamefont {Franciolini}}, \ and\ \bibinfo {author} {\bibfnamefont {P.}~\bibnamefont {Pani}},\ }\href {\doibase 10.1088/1475-7516/2024/04/056} {\bibfield  {journal} {\bibinfo  {journal} {JCAP}\ }\textbf {\bibinfo {volume} {04}},\ \bibinfo {pages} {056} (\bibinfo {year} {2024})},\ \Eprint {http://arxiv.org/abs/2307.11665} {arXiv:2307.11665 [gr-qc]} \BibitemShut {NoStop}%
\bibitem [{\citenamefont {Maji}\ and\ \citenamefont {Park}(2024)}]{Maji:2023fhv}%
  \BibitemOpen
  \bibfield  {author} {\bibinfo {author} {\bibfnamefont {R.}~\bibnamefont {Maji}}\ and\ \bibinfo {author} {\bibfnamefont {W.-I.}\ \bibnamefont {Park}},\ }\href {\doibase 10.1088/1475-7516/2024/01/015} {\bibfield  {journal} {\bibinfo  {journal} {JCAP}\ }\textbf {\bibinfo {volume} {01}},\ \bibinfo {pages} {015} (\bibinfo {year} {2024})},\ \Eprint {http://arxiv.org/abs/2308.11439} {arXiv:2308.11439 [hep-ph]} \BibitemShut {NoStop}%
\bibitem [{\citenamefont {Bhaumik}\ \emph {et~al.}(2023)\citenamefont {Bhaumik}, \citenamefont {Jain},\ and\ \citenamefont {Lewicki}}]{Bhaumik:2023wmw}%
  \BibitemOpen
  \bibfield  {author} {\bibinfo {author} {\bibfnamefont {N.}~\bibnamefont {Bhaumik}}, \bibinfo {author} {\bibfnamefont {R.~K.}\ \bibnamefont {Jain}}, \ and\ \bibinfo {author} {\bibfnamefont {M.}~\bibnamefont {Lewicki}},\ }\href {\doibase 10.1103/PhysRevD.108.123532} {\bibfield  {journal} {\bibinfo  {journal} {Phys. Rev. D}\ }\textbf {\bibinfo {volume} {108}},\ \bibinfo {pages} {123532} (\bibinfo {year} {2023})},\ \Eprint {http://arxiv.org/abs/2308.07912} {arXiv:2308.07912 [astro-ph.CO]} \BibitemShut {NoStop}%
\bibitem [{\citenamefont {Zhu}\ \emph {et~al.}(2023)\citenamefont {Zhu}, \citenamefont {Ye},\ and\ \citenamefont {Cai}}]{Zhu:2023lbf}%
  \BibitemOpen
  \bibfield  {author} {\bibinfo {author} {\bibfnamefont {M.}~\bibnamefont {Zhu}}, \bibinfo {author} {\bibfnamefont {G.}~\bibnamefont {Ye}}, \ and\ \bibinfo {author} {\bibfnamefont {Y.}~\bibnamefont {Cai}},\ }\href {\doibase 10.1140/epjc/s10052-023-11963-4} {\bibfield  {journal} {\bibinfo  {journal} {Eur. Phys. J. C}\ }\textbf {\bibinfo {volume} {83}},\ \bibinfo {pages} {816} (\bibinfo {year} {2023})},\ \Eprint {http://arxiv.org/abs/2307.16211} {arXiv:2307.16211 [astro-ph.CO]} \BibitemShut {NoStop}%
\bibitem [{\citenamefont {Basilakos}\ \emph {et~al.}(2024)\citenamefont {Basilakos}, \citenamefont {Nanopoulos}, \citenamefont {Papanikolaou}, \citenamefont {Saridakis},\ and\ \citenamefont {Tzerefos}}]{Basilakos:2023xof}%
  \BibitemOpen
  \bibfield  {author} {\bibinfo {author} {\bibfnamefont {S.}~\bibnamefont {Basilakos}}, \bibinfo {author} {\bibfnamefont {D.~V.}\ \bibnamefont {Nanopoulos}}, \bibinfo {author} {\bibfnamefont {T.}~\bibnamefont {Papanikolaou}}, \bibinfo {author} {\bibfnamefont {E.~N.}\ \bibnamefont {Saridakis}}, \ and\ \bibinfo {author} {\bibfnamefont {C.}~\bibnamefont {Tzerefos}},\ }\href {\doibase 10.1016/j.physletb.2024.138507} {\bibfield  {journal} {\bibinfo  {journal} {Phys. Lett. B}\ }\textbf {\bibinfo {volume} {850}},\ \bibinfo {pages} {138507} (\bibinfo {year} {2024})},\ \Eprint {http://arxiv.org/abs/2307.08601} {arXiv:2307.08601 [hep-th]} \BibitemShut {NoStop}%
\bibitem [{\citenamefont {Huang}\ \emph {et~al.}(2023)\citenamefont {Huang}, \citenamefont {Cai}, \citenamefont {Jiang}, \citenamefont {Zhang},\ and\ \citenamefont {Piao}}]{Huang:2023chx}%
  \BibitemOpen
  \bibfield  {author} {\bibinfo {author} {\bibfnamefont {H.-L.}\ \bibnamefont {Huang}}, \bibinfo {author} {\bibfnamefont {Y.}~\bibnamefont {Cai}}, \bibinfo {author} {\bibfnamefont {J.-Q.}\ \bibnamefont {Jiang}}, \bibinfo {author} {\bibfnamefont {J.}~\bibnamefont {Zhang}}, \ and\ \bibinfo {author} {\bibfnamefont {Y.-S.}\ \bibnamefont {Piao}},\ }\href@noop {} {\  (\bibinfo {year} {2023})},\ \Eprint {http://arxiv.org/abs/2306.17577} {arXiv:2306.17577 [gr-qc]} \BibitemShut {NoStop}%
\bibitem [{\citenamefont {Jiang}\ \emph {et~al.}(2023)\citenamefont {Jiang}, \citenamefont {Cai}, \citenamefont {Ye},\ and\ \citenamefont {Piao}}]{Jiang:2023gfe}%
  \BibitemOpen
  \bibfield  {author} {\bibinfo {author} {\bibfnamefont {J.-Q.}\ \bibnamefont {Jiang}}, \bibinfo {author} {\bibfnamefont {Y.}~\bibnamefont {Cai}}, \bibinfo {author} {\bibfnamefont {G.}~\bibnamefont {Ye}}, \ and\ \bibinfo {author} {\bibfnamefont {Y.-S.}\ \bibnamefont {Piao}},\ }\href@noop {} {\  (\bibinfo {year} {2023})},\ \Eprint {http://arxiv.org/abs/2307.15547} {arXiv:2307.15547 [astro-ph.CO]} \BibitemShut {NoStop}%
\bibitem [{\citenamefont {Agazie}\ \emph {et~al.}(2023{\natexlab{c}})\citenamefont {Agazie} \emph {et~al.}}]{InternationalPulsarTimingArray:2023mzf}%
  \BibitemOpen
  \bibfield  {author} {\bibinfo {author} {\bibfnamefont {G.}~\bibnamefont {Agazie}} \emph {et~al.} (\bibinfo {collaboration} {International Pulsar Timing Array}),\ }\href@noop {} {\  (\bibinfo {year} {2023}{\natexlab{c}})},\ \Eprint {http://arxiv.org/abs/2309.00693} {arXiv:2309.00693 [astro-ph.HE]} \BibitemShut {NoStop}%
\bibitem [{\citenamefont {Harigaya}\ \emph {et~al.}(2023)\citenamefont {Harigaya}, \citenamefont {Inomata},\ and\ \citenamefont {Terada}}]{Harigaya:2023pmw}%
  \BibitemOpen
  \bibfield  {author} {\bibinfo {author} {\bibfnamefont {K.}~\bibnamefont {Harigaya}}, \bibinfo {author} {\bibfnamefont {K.}~\bibnamefont {Inomata}}, \ and\ \bibinfo {author} {\bibfnamefont {T.}~\bibnamefont {Terada}},\ }\href {\doibase 10.1103/PhysRevD.108.123538} {\bibfield  {journal} {\bibinfo  {journal} {Phys. Rev. D}\ }\textbf {\bibinfo {volume} {108}},\ \bibinfo {pages} {123538} (\bibinfo {year} {2023})},\ \Eprint {http://arxiv.org/abs/2309.00228} {arXiv:2309.00228 [astro-ph.CO]} \BibitemShut {NoStop}%
\bibitem [{\citenamefont {Lozanov}\ \emph {et~al.}(2023)\citenamefont {Lozanov}, \citenamefont {Pi}, \citenamefont {Sasaki}, \citenamefont {Takhistov},\ and\ \citenamefont {Wang}}]{Lozanov:2023rcd}%
  \BibitemOpen
  \bibfield  {author} {\bibinfo {author} {\bibfnamefont {K.~D.}\ \bibnamefont {Lozanov}}, \bibinfo {author} {\bibfnamefont {S.}~\bibnamefont {Pi}}, \bibinfo {author} {\bibfnamefont {M.}~\bibnamefont {Sasaki}}, \bibinfo {author} {\bibfnamefont {V.}~\bibnamefont {Takhistov}}, \ and\ \bibinfo {author} {\bibfnamefont {A.}~\bibnamefont {Wang}},\ }\href@noop {} {\  (\bibinfo {year} {2023})},\ \Eprint {http://arxiv.org/abs/2310.03594} {arXiv:2310.03594 [astro-ph.CO]} \BibitemShut {NoStop}%
\bibitem [{\citenamefont {Choudhury}\ \emph {et~al.}(2023)\citenamefont {Choudhury}, \citenamefont {Dey}, \citenamefont {Karde}, \citenamefont {Panda},\ and\ \citenamefont {Sami}}]{Choudhury:2023fwk}%
  \BibitemOpen
  \bibfield  {author} {\bibinfo {author} {\bibfnamefont {S.}~\bibnamefont {Choudhury}}, \bibinfo {author} {\bibfnamefont {K.}~\bibnamefont {Dey}}, \bibinfo {author} {\bibfnamefont {A.}~\bibnamefont {Karde}}, \bibinfo {author} {\bibfnamefont {S.}~\bibnamefont {Panda}}, \ and\ \bibinfo {author} {\bibfnamefont {M.}~\bibnamefont {Sami}},\ }\href@noop {} {\  (\bibinfo {year} {2023})},\ \Eprint {http://arxiv.org/abs/2310.11034} {arXiv:2310.11034 [astro-ph.CO]} \BibitemShut {NoStop}%
\bibitem [{\citenamefont {Cang}\ \emph {et~al.}(2023)\citenamefont {Cang}, \citenamefont {Gao}, \citenamefont {Liu},\ and\ \citenamefont {Sun}}]{Cang:2023ysz}%
  \BibitemOpen
  \bibfield  {author} {\bibinfo {author} {\bibfnamefont {J.}~\bibnamefont {Cang}}, \bibinfo {author} {\bibfnamefont {Y.}~\bibnamefont {Gao}}, \bibinfo {author} {\bibfnamefont {Y.}~\bibnamefont {Liu}}, \ and\ \bibinfo {author} {\bibfnamefont {S.}~\bibnamefont {Sun}},\ }\href@noop {} {\  (\bibinfo {year} {2023})},\ \Eprint {http://arxiv.org/abs/2309.15069} {arXiv:2309.15069 [astro-ph.CO]} \BibitemShut {NoStop}%
\bibitem [{\citenamefont {Mu}\ \emph {et~al.}(2023)\citenamefont {Mu}, \citenamefont {Liu}, \citenamefont {Cheng},\ and\ \citenamefont {Guo}}]{Mu:2023wdt}%
  \BibitemOpen
  \bibfield  {author} {\bibinfo {author} {\bibfnamefont {B.}~\bibnamefont {Mu}}, \bibinfo {author} {\bibfnamefont {J.}~\bibnamefont {Liu}}, \bibinfo {author} {\bibfnamefont {G.}~\bibnamefont {Cheng}}, \ and\ \bibinfo {author} {\bibfnamefont {Z.-K.}\ \bibnamefont {Guo}},\ }\href@noop {} {\  (\bibinfo {year} {2023})},\ \Eprint {http://arxiv.org/abs/2310.20564} {arXiv:2310.20564 [astro-ph.CO]} \BibitemShut {NoStop}%
\bibitem [{\citenamefont {Chen}\ \emph {et~al.}(2024)\citenamefont {Chen}, \citenamefont {Li}, \citenamefont {Wu},\ and\ \citenamefont {Yu}}]{Chen:2023bms}%
  \BibitemOpen
  \bibfield  {author} {\bibinfo {author} {\bibfnamefont {Z.-C.}\ \bibnamefont {Chen}}, \bibinfo {author} {\bibfnamefont {S.-L.}\ \bibnamefont {Li}}, \bibinfo {author} {\bibfnamefont {P.}~\bibnamefont {Wu}}, \ and\ \bibinfo {author} {\bibfnamefont {H.}~\bibnamefont {Yu}},\ }\href {\doibase 10.1103/PhysRevD.109.043022} {\bibfield  {journal} {\bibinfo  {journal} {Phys. Rev. D}\ }\textbf {\bibinfo {volume} {109}},\ \bibinfo {pages} {043022} (\bibinfo {year} {2024})},\ \Eprint {http://arxiv.org/abs/2312.01824} {arXiv:2312.01824 [astro-ph.CO]} \BibitemShut {NoStop}%
\bibitem [{\citenamefont {Liu}\ \emph {et~al.}(2024{\natexlab{b}})\citenamefont {Liu}, \citenamefont {Wu},\ and\ \citenamefont {Chen}}]{Liu:2023hpw}%
  \BibitemOpen
  \bibfield  {author} {\bibinfo {author} {\bibfnamefont {L.}~\bibnamefont {Liu}}, \bibinfo {author} {\bibfnamefont {Y.}~\bibnamefont {Wu}}, \ and\ \bibinfo {author} {\bibfnamefont {Z.-C.}\ \bibnamefont {Chen}},\ }\href {\doibase 10.1088/1475-7516/2024/04/011} {\bibfield  {journal} {\bibinfo  {journal} {JCAP}\ }\textbf {\bibinfo {volume} {04}},\ \bibinfo {pages} {011} (\bibinfo {year} {2024}{\natexlab{b}})},\ \Eprint {http://arxiv.org/abs/2310.16500} {arXiv:2310.16500 [astro-ph.CO]} \BibitemShut {NoStop}%
\bibitem [{\citenamefont {Chao}\ \emph {et~al.}(2023)\citenamefont {Chao}, \citenamefont {Feng}, \citenamefont {Guo},\ and\ \citenamefont {Li}}]{Chao:2023lox}%
  \BibitemOpen
  \bibfield  {author} {\bibinfo {author} {\bibfnamefont {W.}~\bibnamefont {Chao}}, \bibinfo {author} {\bibfnamefont {J.-j.}\ \bibnamefont {Feng}}, \bibinfo {author} {\bibfnamefont {H.-k.}\ \bibnamefont {Guo}}, \ and\ \bibinfo {author} {\bibfnamefont {T.}~\bibnamefont {Li}},\ }\href@noop {} {\  (\bibinfo {year} {2023})},\ \Eprint {http://arxiv.org/abs/2312.04017} {arXiv:2312.04017 [hep-ph]} \BibitemShut {NoStop}%
\bibitem [{\citenamefont {Fei}(2024)}]{Fei:2023iel}%
  \BibitemOpen
  \bibfield  {author} {\bibinfo {author} {\bibfnamefont {Q.}~\bibnamefont {Fei}},\ }\href {\doibase 10.1088/1572-9494/ad1988} {\bibfield  {journal} {\bibinfo  {journal} {Commun. Theor. Phys.}\ }\textbf {\bibinfo {volume} {76}},\ \bibinfo {pages} {015404} (\bibinfo {year} {2024})},\ \Eprint {http://arxiv.org/abs/2310.17199} {arXiv:2310.17199 [gr-qc]} \BibitemShut {NoStop}%
\bibitem [{\citenamefont {Maiti}\ \emph {et~al.}(2024)\citenamefont {Maiti}, \citenamefont {Maity},\ and\ \citenamefont {Sriramkumar}}]{Maiti:2024nhv}%
  \BibitemOpen
  \bibfield  {author} {\bibinfo {author} {\bibfnamefont {S.}~\bibnamefont {Maiti}}, \bibinfo {author} {\bibfnamefont {D.}~\bibnamefont {Maity}}, \ and\ \bibinfo {author} {\bibfnamefont {L.}~\bibnamefont {Sriramkumar}},\ }\href@noop {} {\  (\bibinfo {year} {2024})},\ \Eprint {http://arxiv.org/abs/2401.01864} {arXiv:2401.01864 [gr-qc]} \BibitemShut {NoStop}%
\bibitem [{\citenamefont {Belotsky}\ \emph {et~al.}(2014)\citenamefont {Belotsky}, \citenamefont {Dmitriev}, \citenamefont {Esipova}, \citenamefont {Gani}, \citenamefont {Grobov}, \citenamefont {Khlopov}, \citenamefont {Kirillov}, \citenamefont {Rubin},\ and\ \citenamefont {Svadkovsky}}]{Belotsky:2014kca}%
  \BibitemOpen
  \bibfield  {author} {\bibinfo {author} {\bibfnamefont {K.~M.}\ \bibnamefont {Belotsky}}, \bibinfo {author} {\bibfnamefont {A.~D.}\ \bibnamefont {Dmitriev}}, \bibinfo {author} {\bibfnamefont {E.~A.}\ \bibnamefont {Esipova}}, \bibinfo {author} {\bibfnamefont {V.~A.}\ \bibnamefont {Gani}}, \bibinfo {author} {\bibfnamefont {A.~V.}\ \bibnamefont {Grobov}}, \bibinfo {author} {\bibfnamefont {M.~Y.}\ \bibnamefont {Khlopov}}, \bibinfo {author} {\bibfnamefont {A.~A.}\ \bibnamefont {Kirillov}}, \bibinfo {author} {\bibfnamefont {S.~G.}\ \bibnamefont {Rubin}}, \ and\ \bibinfo {author} {\bibfnamefont {I.~V.}\ \bibnamefont {Svadkovsky}},\ }\href {\doibase 10.1142/S0217732314400057} {\bibfield  {journal} {\bibinfo  {journal} {Mod. Phys. Lett. A}\ }\textbf {\bibinfo {volume} {29}},\ \bibinfo {pages} {1440005} (\bibinfo {year} {2014})},\ \Eprint {http://arxiv.org/abs/1410.0203} {arXiv:1410.0203 [astro-ph.CO]} \BibitemShut {NoStop}%
\bibitem [{\citenamefont {Wang}\ \emph {et~al.}(2018)\citenamefont {Wang}, \citenamefont {Wang}, \citenamefont {Huang},\ and\ \citenamefont {Li}}]{Wang:2016ana}%
  \BibitemOpen
  \bibfield  {author} {\bibinfo {author} {\bibfnamefont {S.}~\bibnamefont {Wang}}, \bibinfo {author} {\bibfnamefont {Y.-F.}\ \bibnamefont {Wang}}, \bibinfo {author} {\bibfnamefont {Q.-G.}\ \bibnamefont {Huang}}, \ and\ \bibinfo {author} {\bibfnamefont {T.~G.~F.}\ \bibnamefont {Li}},\ }\href {\doibase 10.1103/PhysRevLett.120.191102} {\bibfield  {journal} {\bibinfo  {journal} {Phys. Rev. Lett.}\ }\textbf {\bibinfo {volume} {120}},\ \bibinfo {pages} {191102} (\bibinfo {year} {2018})},\ \Eprint {http://arxiv.org/abs/1610.08725} {arXiv:1610.08725 [astro-ph.CO]} \BibitemShut {NoStop}%
\bibitem [{\citenamefont {Carr}\ \emph {et~al.}(2016)\citenamefont {Carr}, \citenamefont {Kuhnel},\ and\ \citenamefont {Sandstad}}]{Carr:2016drx}%
  \BibitemOpen
  \bibfield  {author} {\bibinfo {author} {\bibfnamefont {B.}~\bibnamefont {Carr}}, \bibinfo {author} {\bibfnamefont {F.}~\bibnamefont {Kuhnel}}, \ and\ \bibinfo {author} {\bibfnamefont {M.}~\bibnamefont {Sandstad}},\ }\href {\doibase 10.1103/PhysRevD.94.083504} {\bibfield  {journal} {\bibinfo  {journal} {Phys. Rev. D}\ }\textbf {\bibinfo {volume} {94}},\ \bibinfo {pages} {083504} (\bibinfo {year} {2016})},\ \Eprint {http://arxiv.org/abs/1607.06077} {arXiv:1607.06077 [astro-ph.CO]} \BibitemShut {NoStop}%
\bibitem [{\citenamefont {Garcia-Bellido}\ and\ \citenamefont {Ruiz~Morales}(2017)}]{Garcia-Bellido:2017mdw}%
  \BibitemOpen
  \bibfield  {author} {\bibinfo {author} {\bibfnamefont {J.}~\bibnamefont {Garcia-Bellido}}\ and\ \bibinfo {author} {\bibfnamefont {E.}~\bibnamefont {Ruiz~Morales}},\ }\href {\doibase 10.1016/j.dark.2017.09.007} {\bibfield  {journal} {\bibinfo  {journal} {Phys. Dark Univ.}\ }\textbf {\bibinfo {volume} {18}},\ \bibinfo {pages} {47} (\bibinfo {year} {2017})},\ \Eprint {http://arxiv.org/abs/1702.03901} {arXiv:1702.03901 [astro-ph.CO]} \BibitemShut {NoStop}%
\bibitem [{\citenamefont {Carr}\ \emph {et~al.}(2017)\citenamefont {Carr}, \citenamefont {Raidal}, \citenamefont {Tenkanen}, \citenamefont {Vaskonen},\ and\ \citenamefont {Veerm\"ae}}]{Carr:2017jsz}%
  \BibitemOpen
  \bibfield  {author} {\bibinfo {author} {\bibfnamefont {B.}~\bibnamefont {Carr}}, \bibinfo {author} {\bibfnamefont {M.}~\bibnamefont {Raidal}}, \bibinfo {author} {\bibfnamefont {T.}~\bibnamefont {Tenkanen}}, \bibinfo {author} {\bibfnamefont {V.}~\bibnamefont {Vaskonen}}, \ and\ \bibinfo {author} {\bibfnamefont {H.}~\bibnamefont {Veerm\"ae}},\ }\href {\doibase 10.1103/PhysRevD.96.023514} {\bibfield  {journal} {\bibinfo  {journal} {Phys. Rev. D}\ }\textbf {\bibinfo {volume} {96}},\ \bibinfo {pages} {023514} (\bibinfo {year} {2017})},\ \Eprint {http://arxiv.org/abs/1705.05567} {arXiv:1705.05567 [astro-ph.CO]} \BibitemShut {NoStop}%
\bibitem [{\citenamefont {Germani}\ and\ \citenamefont {Prokopec}(2017)}]{Germani:2017bcs}%
  \BibitemOpen
  \bibfield  {author} {\bibinfo {author} {\bibfnamefont {C.}~\bibnamefont {Germani}}\ and\ \bibinfo {author} {\bibfnamefont {T.}~\bibnamefont {Prokopec}},\ }\href {\doibase 10.1016/j.dark.2017.09.001} {\bibfield  {journal} {\bibinfo  {journal} {Phys. Dark Univ.}\ }\textbf {\bibinfo {volume} {18}},\ \bibinfo {pages} {6} (\bibinfo {year} {2017})},\ \Eprint {http://arxiv.org/abs/1706.04226} {arXiv:1706.04226 [astro-ph.CO]} \BibitemShut {NoStop}%
\bibitem [{\citenamefont {Liu}\ \emph {et~al.}(2019{\natexlab{a}})\citenamefont {Liu}, \citenamefont {Guo},\ and\ \citenamefont {Cai}}]{Liu:2018ess}%
  \BibitemOpen
  \bibfield  {author} {\bibinfo {author} {\bibfnamefont {L.}~\bibnamefont {Liu}}, \bibinfo {author} {\bibfnamefont {Z.-K.}\ \bibnamefont {Guo}}, \ and\ \bibinfo {author} {\bibfnamefont {R.-G.}\ \bibnamefont {Cai}},\ }\href {\doibase 10.1103/PhysRevD.99.063523} {\bibfield  {journal} {\bibinfo  {journal} {Phys. Rev. D}\ }\textbf {\bibinfo {volume} {99}},\ \bibinfo {pages} {063523} (\bibinfo {year} {2019}{\natexlab{a}})},\ \Eprint {http://arxiv.org/abs/1812.05376} {arXiv:1812.05376 [astro-ph.CO]} \BibitemShut {NoStop}%
\bibitem [{\citenamefont {Chen}\ and\ \citenamefont {Huang}(2018)}]{Chen:2018czv}%
  \BibitemOpen
  \bibfield  {author} {\bibinfo {author} {\bibfnamefont {Z.-C.}\ \bibnamefont {Chen}}\ and\ \bibinfo {author} {\bibfnamefont {Q.-G.}\ \bibnamefont {Huang}},\ }\href {\doibase 10.3847/1538-4357/aad6e2} {\bibfield  {journal} {\bibinfo  {journal} {Astrophys. J.}\ }\textbf {\bibinfo {volume} {864}},\ \bibinfo {pages} {61} (\bibinfo {year} {2018})},\ \Eprint {http://arxiv.org/abs/1801.10327} {arXiv:1801.10327 [astro-ph.CO]} \BibitemShut {NoStop}%
\bibitem [{\citenamefont {Liu}\ \emph {et~al.}(2019{\natexlab{b}})\citenamefont {Liu}, \citenamefont {Guo},\ and\ \citenamefont {Cai}}]{Liu:2019rnx}%
  \BibitemOpen
  \bibfield  {author} {\bibinfo {author} {\bibfnamefont {L.}~\bibnamefont {Liu}}, \bibinfo {author} {\bibfnamefont {Z.-K.}\ \bibnamefont {Guo}}, \ and\ \bibinfo {author} {\bibfnamefont {R.-G.}\ \bibnamefont {Cai}},\ }\href {\doibase 10.1140/epjc/s10052-019-7227-0} {\bibfield  {journal} {\bibinfo  {journal} {Eur. Phys. J. C}\ }\textbf {\bibinfo {volume} {79}},\ \bibinfo {pages} {717} (\bibinfo {year} {2019}{\natexlab{b}})},\ \Eprint {http://arxiv.org/abs/1901.07672} {arXiv:1901.07672 [astro-ph.CO]} \BibitemShut {NoStop}%
\bibitem [{\citenamefont {Fu}\ \emph {et~al.}(2019)\citenamefont {Fu}, \citenamefont {Wu},\ and\ \citenamefont {Yu}}]{Fu:2019ttf}%
  \BibitemOpen
  \bibfield  {author} {\bibinfo {author} {\bibfnamefont {C.}~\bibnamefont {Fu}}, \bibinfo {author} {\bibfnamefont {P.}~\bibnamefont {Wu}}, \ and\ \bibinfo {author} {\bibfnamefont {H.}~\bibnamefont {Yu}},\ }\href {\doibase 10.1103/PhysRevD.100.063532} {\bibfield  {journal} {\bibinfo  {journal} {Phys. Rev. D}\ }\textbf {\bibinfo {volume} {100}},\ \bibinfo {pages} {063532} (\bibinfo {year} {2019})},\ \Eprint {http://arxiv.org/abs/1907.05042} {arXiv:1907.05042 [astro-ph.CO]} \BibitemShut {NoStop}%
\bibitem [{\citenamefont {Wang}\ \emph {et~al.}(2019)\citenamefont {Wang}, \citenamefont {Terada},\ and\ \citenamefont {Kohri}}]{Wang:2019kaf}%
  \BibitemOpen
  \bibfield  {author} {\bibinfo {author} {\bibfnamefont {S.}~\bibnamefont {Wang}}, \bibinfo {author} {\bibfnamefont {T.}~\bibnamefont {Terada}}, \ and\ \bibinfo {author} {\bibfnamefont {K.}~\bibnamefont {Kohri}},\ }\href {\doibase 10.1103/PhysRevD.99.103531} {\bibfield  {journal} {\bibinfo  {journal} {Phys. Rev. D}\ }\textbf {\bibinfo {volume} {99}},\ \bibinfo {pages} {103531} (\bibinfo {year} {2019})},\ \bibinfo {note} {[Erratum: Phys.Rev.D 101, 069901 (2020)]},\ \Eprint {http://arxiv.org/abs/1903.05924} {arXiv:1903.05924 [astro-ph.CO]} \BibitemShut {NoStop}%
\bibitem [{\citenamefont {Liu}\ \emph {et~al.}(2020{\natexlab{a}})\citenamefont {Liu}, \citenamefont {Guo},\ and\ \citenamefont {Cai}}]{Liu:2019lul}%
  \BibitemOpen
  \bibfield  {author} {\bibinfo {author} {\bibfnamefont {J.}~\bibnamefont {Liu}}, \bibinfo {author} {\bibfnamefont {Z.-K.}\ \bibnamefont {Guo}}, \ and\ \bibinfo {author} {\bibfnamefont {R.-G.}\ \bibnamefont {Cai}},\ }\href {\doibase 10.1103/PhysRevD.101.023513} {\bibfield  {journal} {\bibinfo  {journal} {Phys. Rev. D}\ }\textbf {\bibinfo {volume} {101}},\ \bibinfo {pages} {023513} (\bibinfo {year} {2020}{\natexlab{a}})},\ \Eprint {http://arxiv.org/abs/1908.02662} {arXiv:1908.02662 [astro-ph.CO]} \BibitemShut {NoStop}%
\bibitem [{\citenamefont {Cai}\ \emph {et~al.}(2020{\natexlab{a}})\citenamefont {Cai}, \citenamefont {Guo}, \citenamefont {Liu}, \citenamefont {Liu},\ and\ \citenamefont {Yang}}]{Cai:2019bmk}%
  \BibitemOpen
  \bibfield  {author} {\bibinfo {author} {\bibfnamefont {R.-G.}\ \bibnamefont {Cai}}, \bibinfo {author} {\bibfnamefont {Z.-K.}\ \bibnamefont {Guo}}, \bibinfo {author} {\bibfnamefont {J.}~\bibnamefont {Liu}}, \bibinfo {author} {\bibfnamefont {L.}~\bibnamefont {Liu}}, \ and\ \bibinfo {author} {\bibfnamefont {X.-Y.}\ \bibnamefont {Yang}},\ }\href {\doibase 10.1088/1475-7516/2020/06/013} {\bibfield  {journal} {\bibinfo  {journal} {JCAP}\ }\textbf {\bibinfo {volume} {06}},\ \bibinfo {pages} {013} (\bibinfo {year} {2020}{\natexlab{a}})},\ \Eprint {http://arxiv.org/abs/1912.10437} {arXiv:1912.10437 [astro-ph.CO]} \BibitemShut {NoStop}%
\bibitem [{\citenamefont {Liu}\ \emph {et~al.}(2020{\natexlab{b}})\citenamefont {Liu}, \citenamefont {Guo}, \citenamefont {Cai},\ and\ \citenamefont {Kim}}]{Liu:2020cds}%
  \BibitemOpen
  \bibfield  {author} {\bibinfo {author} {\bibfnamefont {L.}~\bibnamefont {Liu}}, \bibinfo {author} {\bibfnamefont {Z.-K.}\ \bibnamefont {Guo}}, \bibinfo {author} {\bibfnamefont {R.-G.}\ \bibnamefont {Cai}}, \ and\ \bibinfo {author} {\bibfnamefont {S.~P.}\ \bibnamefont {Kim}},\ }\href {\doibase 10.1103/PhysRevD.102.043508} {\bibfield  {journal} {\bibinfo  {journal} {Phys. Rev. D}\ }\textbf {\bibinfo {volume} {102}},\ \bibinfo {pages} {043508} (\bibinfo {year} {2020}{\natexlab{b}})},\ \Eprint {http://arxiv.org/abs/2001.02984} {arXiv:2001.02984 [astro-ph.CO]} \BibitemShut {NoStop}%
\bibitem [{\citenamefont {Fu}\ \emph {et~al.}(2020)\citenamefont {Fu}, \citenamefont {Wu},\ and\ \citenamefont {Yu}}]{Fu:2020lob}%
  \BibitemOpen
  \bibfield  {author} {\bibinfo {author} {\bibfnamefont {C.}~\bibnamefont {Fu}}, \bibinfo {author} {\bibfnamefont {P.}~\bibnamefont {Wu}}, \ and\ \bibinfo {author} {\bibfnamefont {H.}~\bibnamefont {Yu}},\ }\href {\doibase 10.1103/PhysRevD.102.043527} {\bibfield  {journal} {\bibinfo  {journal} {Phys. Rev. D}\ }\textbf {\bibinfo {volume} {102}},\ \bibinfo {pages} {043527} (\bibinfo {year} {2020})},\ \Eprint {http://arxiv.org/abs/2006.03768} {arXiv:2006.03768 [astro-ph.CO]} \BibitemShut {NoStop}%
\bibitem [{\citenamefont {Wu}(2020)}]{Wu:2020drm}%
  \BibitemOpen
  \bibfield  {author} {\bibinfo {author} {\bibfnamefont {Y.}~\bibnamefont {Wu}},\ }\href {\doibase 10.1103/PhysRevD.101.083008} {\bibfield  {journal} {\bibinfo  {journal} {Phys. Rev. D}\ }\textbf {\bibinfo {volume} {101}},\ \bibinfo {pages} {083008} (\bibinfo {year} {2020})},\ \Eprint {http://arxiv.org/abs/2001.03833} {arXiv:2001.03833 [astro-ph.CO]} \BibitemShut {NoStop}%
\bibitem [{\citenamefont {De~Luca}\ \emph {et~al.}(2021{\natexlab{a}})\citenamefont {De~Luca}, \citenamefont {Desjacques}, \citenamefont {Franciolini}, \citenamefont {Pani},\ and\ \citenamefont {Riotto}}]{DeLuca:2020sae}%
  \BibitemOpen
  \bibfield  {author} {\bibinfo {author} {\bibfnamefont {V.}~\bibnamefont {De~Luca}}, \bibinfo {author} {\bibfnamefont {V.}~\bibnamefont {Desjacques}}, \bibinfo {author} {\bibfnamefont {G.}~\bibnamefont {Franciolini}}, \bibinfo {author} {\bibfnamefont {P.}~\bibnamefont {Pani}}, \ and\ \bibinfo {author} {\bibfnamefont {A.}~\bibnamefont {Riotto}},\ }\href {\doibase 10.1103/PhysRevLett.126.051101} {\bibfield  {journal} {\bibinfo  {journal} {Phys. Rev. Lett.}\ }\textbf {\bibinfo {volume} {126}},\ \bibinfo {pages} {051101} (\bibinfo {year} {2021}{\natexlab{a}})},\ \Eprint {http://arxiv.org/abs/2009.01728} {arXiv:2009.01728 [astro-ph.CO]} \BibitemShut {NoStop}%
\bibitem [{\citenamefont {Vaskonen}\ and\ \citenamefont {Veerm\"ae}(2021)}]{Vaskonen:2020lbd}%
  \BibitemOpen
  \bibfield  {author} {\bibinfo {author} {\bibfnamefont {V.}~\bibnamefont {Vaskonen}}\ and\ \bibinfo {author} {\bibfnamefont {H.}~\bibnamefont {Veerm\"ae}},\ }\href {\doibase 10.1103/PhysRevLett.126.051303} {\bibfield  {journal} {\bibinfo  {journal} {Phys. Rev. Lett.}\ }\textbf {\bibinfo {volume} {126}},\ \bibinfo {pages} {051303} (\bibinfo {year} {2021})},\ \Eprint {http://arxiv.org/abs/2009.07832} {arXiv:2009.07832 [astro-ph.CO]} \BibitemShut {NoStop}%
\bibitem [{\citenamefont {De~Luca}\ \emph {et~al.}(2021{\natexlab{b}})\citenamefont {De~Luca}, \citenamefont {Franciolini},\ and\ \citenamefont {Riotto}}]{DeLuca:2020agl}%
  \BibitemOpen
  \bibfield  {author} {\bibinfo {author} {\bibfnamefont {V.}~\bibnamefont {De~Luca}}, \bibinfo {author} {\bibfnamefont {G.}~\bibnamefont {Franciolini}}, \ and\ \bibinfo {author} {\bibfnamefont {A.}~\bibnamefont {Riotto}},\ }\href {\doibase 10.1103/PhysRevLett.126.041303} {\bibfield  {journal} {\bibinfo  {journal} {Phys. Rev. Lett.}\ }\textbf {\bibinfo {volume} {126}},\ \bibinfo {pages} {041303} (\bibinfo {year} {2021}{\natexlab{b}})},\ \Eprint {http://arxiv.org/abs/2009.08268} {arXiv:2009.08268 [astro-ph.CO]} \BibitemShut {NoStop}%
\bibitem [{\citenamefont {Dom\`enech}\ \emph {et~al.}(2021)\citenamefont {Dom\`enech}, \citenamefont {Lin},\ and\ \citenamefont {Sasaki}}]{Domenech:2020ssp}%
  \BibitemOpen
  \bibfield  {author} {\bibinfo {author} {\bibfnamefont {G.}~\bibnamefont {Dom\`enech}}, \bibinfo {author} {\bibfnamefont {C.}~\bibnamefont {Lin}}, \ and\ \bibinfo {author} {\bibfnamefont {M.}~\bibnamefont {Sasaki}},\ }\href {\doibase 10.1088/1475-7516/2021/11/E01} {\bibfield  {journal} {\bibinfo  {journal} {JCAP}\ }\textbf {\bibinfo {volume} {04}},\ \bibinfo {pages} {062} (\bibinfo {year} {2021})},\ \bibinfo {note} {[Erratum: JCAP 11, E01 (2021)]},\ \Eprint {http://arxiv.org/abs/2012.08151} {arXiv:2012.08151 [gr-qc]} \BibitemShut {NoStop}%
\bibitem [{\citenamefont {Dom\`enech}\ and\ \citenamefont {Pi}(2022)}]{Domenech:2020ers}%
  \BibitemOpen
  \bibfield  {author} {\bibinfo {author} {\bibfnamefont {G.}~\bibnamefont {Dom\`enech}}\ and\ \bibinfo {author} {\bibfnamefont {S.}~\bibnamefont {Pi}},\ }\href {\doibase 10.1007/s11433-021-1839-6} {\bibfield  {journal} {\bibinfo  {journal} {Sci. China Phys. Mech. Astron.}\ }\textbf {\bibinfo {volume} {65}},\ \bibinfo {pages} {230411} (\bibinfo {year} {2022})},\ \Eprint {http://arxiv.org/abs/2010.03976} {arXiv:2010.03976 [astro-ph.CO]} \BibitemShut {NoStop}%
\bibitem [{\citenamefont {H\"utsi}\ \emph {et~al.}(2021)\citenamefont {H\"utsi}, \citenamefont {Raidal}, \citenamefont {Vaskonen},\ and\ \citenamefont {Veerm\"ae}}]{Hutsi:2020sol}%
  \BibitemOpen
  \bibfield  {author} {\bibinfo {author} {\bibfnamefont {G.}~\bibnamefont {H\"utsi}}, \bibinfo {author} {\bibfnamefont {M.}~\bibnamefont {Raidal}}, \bibinfo {author} {\bibfnamefont {V.}~\bibnamefont {Vaskonen}}, \ and\ \bibinfo {author} {\bibfnamefont {H.}~\bibnamefont {Veerm\"ae}},\ }\href {\doibase 10.1088/1475-7516/2021/03/068} {\bibfield  {journal} {\bibinfo  {journal} {JCAP}\ }\textbf {\bibinfo {volume} {03}},\ \bibinfo {pages} {068} (\bibinfo {year} {2021})},\ \Eprint {http://arxiv.org/abs/2012.02786} {arXiv:2012.02786 [astro-ph.CO]} \BibitemShut {NoStop}%
\bibitem [{\citenamefont {Kawai}\ and\ \citenamefont {Kim}(2021)}]{Kawai:2021edk}%
  \BibitemOpen
  \bibfield  {author} {\bibinfo {author} {\bibfnamefont {S.}~\bibnamefont {Kawai}}\ and\ \bibinfo {author} {\bibfnamefont {J.}~\bibnamefont {Kim}},\ }\href {\doibase 10.1103/PhysRevD.104.083545} {\bibfield  {journal} {\bibinfo  {journal} {Phys. Rev. D}\ }\textbf {\bibinfo {volume} {104}},\ \bibinfo {pages} {083545} (\bibinfo {year} {2021})},\ \Eprint {http://arxiv.org/abs/2108.01340} {arXiv:2108.01340 [astro-ph.CO]} \BibitemShut {NoStop}%
\bibitem [{\citenamefont {Braglia}\ \emph {et~al.}(2021)\citenamefont {Braglia}, \citenamefont {Garcia-Bellido},\ and\ \citenamefont {Kuroyanagi}}]{Braglia:2021wwa}%
  \BibitemOpen
  \bibfield  {author} {\bibinfo {author} {\bibfnamefont {M.}~\bibnamefont {Braglia}}, \bibinfo {author} {\bibfnamefont {J.}~\bibnamefont {Garcia-Bellido}}, \ and\ \bibinfo {author} {\bibfnamefont {S.}~\bibnamefont {Kuroyanagi}},\ }\href {\doibase 10.1088/1475-7516/2021/12/012} {\bibfield  {journal} {\bibinfo  {journal} {JCAP}\ }\textbf {\bibinfo {volume} {12}},\ \bibinfo {pages} {012} (\bibinfo {year} {2021})},\ \Eprint {http://arxiv.org/abs/2110.07488} {arXiv:2110.07488 [astro-ph.CO]} \BibitemShut {NoStop}%
\bibitem [{\citenamefont {Cai}\ \emph {et~al.}(2021)\citenamefont {Cai}, \citenamefont {Chen},\ and\ \citenamefont {Fu}}]{Cai:2021wzd}%
  \BibitemOpen
  \bibfield  {author} {\bibinfo {author} {\bibfnamefont {R.-G.}\ \bibnamefont {Cai}}, \bibinfo {author} {\bibfnamefont {C.}~\bibnamefont {Chen}}, \ and\ \bibinfo {author} {\bibfnamefont {C.}~\bibnamefont {Fu}},\ }\href {\doibase 10.1103/PhysRevD.104.083537} {\bibfield  {journal} {\bibinfo  {journal} {Phys. Rev. D}\ }\textbf {\bibinfo {volume} {104}},\ \bibinfo {pages} {083537} (\bibinfo {year} {2021})},\ \Eprint {http://arxiv.org/abs/2108.03422} {arXiv:2108.03422 [astro-ph.CO]} \BibitemShut {NoStop}%
\bibitem [{\citenamefont {Liu}\ \emph {et~al.}(2023{\natexlab{c}})\citenamefont {Liu}, \citenamefont {Yang}, \citenamefont {Guo},\ and\ \citenamefont {Cai}}]{Liu:2021jnw}%
  \BibitemOpen
  \bibfield  {author} {\bibinfo {author} {\bibfnamefont {L.}~\bibnamefont {Liu}}, \bibinfo {author} {\bibfnamefont {X.-Y.}\ \bibnamefont {Yang}}, \bibinfo {author} {\bibfnamefont {Z.-K.}\ \bibnamefont {Guo}}, \ and\ \bibinfo {author} {\bibfnamefont {R.-G.}\ \bibnamefont {Cai}},\ }\href {\doibase 10.1088/1475-7516/2023/01/006} {\bibfield  {journal} {\bibinfo  {journal} {JCAP}\ }\textbf {\bibinfo {volume} {01}},\ \bibinfo {pages} {006} (\bibinfo {year} {2023}{\natexlab{c}})},\ \Eprint {http://arxiv.org/abs/2112.05473} {arXiv:2112.05473 [astro-ph.CO]} \BibitemShut {NoStop}%
\bibitem [{\citenamefont {Braglia}\ \emph {et~al.}(2023)\citenamefont {Braglia}, \citenamefont {Garcia-Bellido},\ and\ \citenamefont {Kuroyanagi}}]{Braglia:2022icu}%
  \BibitemOpen
  \bibfield  {author} {\bibinfo {author} {\bibfnamefont {M.}~\bibnamefont {Braglia}}, \bibinfo {author} {\bibfnamefont {J.}~\bibnamefont {Garcia-Bellido}}, \ and\ \bibinfo {author} {\bibfnamefont {S.}~\bibnamefont {Kuroyanagi}},\ }\href {\doibase 10.1093/mnras/stad082} {\bibfield  {journal} {\bibinfo  {journal} {Mon. Not. Roy. Astron. Soc.}\ }\textbf {\bibinfo {volume} {519}},\ \bibinfo {pages} {6008} (\bibinfo {year} {2023})},\ \Eprint {http://arxiv.org/abs/2201.13414} {arXiv:2201.13414 [astro-ph.CO]} \BibitemShut {NoStop}%
\bibitem [{\citenamefont {Chen}\ \emph {et~al.}(2023{\natexlab{c}})\citenamefont {Chen}, \citenamefont {Kim},\ and\ \citenamefont {Liu}}]{Chen:2022qvg}%
  \BibitemOpen
  \bibfield  {author} {\bibinfo {author} {\bibfnamefont {Z.-C.}\ \bibnamefont {Chen}}, \bibinfo {author} {\bibfnamefont {S.~P.}\ \bibnamefont {Kim}}, \ and\ \bibinfo {author} {\bibfnamefont {L.}~\bibnamefont {Liu}},\ }\href {\doibase 10.1088/1572-9494/acce98} {\bibfield  {journal} {\bibinfo  {journal} {Commun. Theor. Phys.}\ }\textbf {\bibinfo {volume} {75}},\ \bibinfo {pages} {065401} (\bibinfo {year} {2023}{\natexlab{c}})},\ \Eprint {http://arxiv.org/abs/2210.15564} {arXiv:2210.15564 [gr-qc]} \BibitemShut {NoStop}%
\bibitem [{\citenamefont {Inomata}\ \emph {et~al.}(2023)\citenamefont {Inomata}, \citenamefont {Braglia}, \citenamefont {Chen},\ and\ \citenamefont {Renaux-Petel}}]{Inomata:2022yte}%
  \BibitemOpen
  \bibfield  {author} {\bibinfo {author} {\bibfnamefont {K.}~\bibnamefont {Inomata}}, \bibinfo {author} {\bibfnamefont {M.}~\bibnamefont {Braglia}}, \bibinfo {author} {\bibfnamefont {X.}~\bibnamefont {Chen}}, \ and\ \bibinfo {author} {\bibfnamefont {S.}~\bibnamefont {Renaux-Petel}},\ }\href {\doibase 10.1088/1475-7516/2023/04/011} {\bibfield  {journal} {\bibinfo  {journal} {JCAP}\ }\textbf {\bibinfo {volume} {04}},\ \bibinfo {pages} {011} (\bibinfo {year} {2023})},\ \bibinfo {note} {[Erratum: JCAP 09, E01 (2023)]},\ \Eprint {http://arxiv.org/abs/2211.02586} {arXiv:2211.02586 [astro-ph.CO]} \BibitemShut {NoStop}%
\bibitem [{\citenamefont {Guo}\ \emph {et~al.}(2023)\citenamefont {Guo}, \citenamefont {Khlopov}, \citenamefont {Liu}, \citenamefont {Wu}, \citenamefont {Wu},\ and\ \citenamefont {Zhu}}]{Guo:2023hyp}%
  \BibitemOpen
  \bibfield  {author} {\bibinfo {author} {\bibfnamefont {S.-Y.}\ \bibnamefont {Guo}}, \bibinfo {author} {\bibfnamefont {M.}~\bibnamefont {Khlopov}}, \bibinfo {author} {\bibfnamefont {X.}~\bibnamefont {Liu}}, \bibinfo {author} {\bibfnamefont {L.}~\bibnamefont {Wu}}, \bibinfo {author} {\bibfnamefont {Y.}~\bibnamefont {Wu}}, \ and\ \bibinfo {author} {\bibfnamefont {B.}~\bibnamefont {Zhu}},\ }\href@noop {} {\  (\bibinfo {year} {2023})},\ \Eprint {http://arxiv.org/abs/2306.17022} {arXiv:2306.17022 [hep-ph]} \BibitemShut {NoStop}%
\bibitem [{\citenamefont {Cai}\ \emph {et~al.}(2023{\natexlab{b}})\citenamefont {Cai}, \citenamefont {Zhu},\ and\ \citenamefont {Piao}}]{Cai:2023uhc}%
  \BibitemOpen
  \bibfield  {author} {\bibinfo {author} {\bibfnamefont {Y.}~\bibnamefont {Cai}}, \bibinfo {author} {\bibfnamefont {M.}~\bibnamefont {Zhu}}, \ and\ \bibinfo {author} {\bibfnamefont {Y.-S.}\ \bibnamefont {Piao}},\ }\href@noop {} {\  (\bibinfo {year} {2023}{\natexlab{b}})},\ \Eprint {http://arxiv.org/abs/2305.10933} {arXiv:2305.10933 [gr-qc]} \BibitemShut {NoStop}%
\bibitem [{\citenamefont {Meng}\ \emph {et~al.}(2023)\citenamefont {Meng}, \citenamefont {Yuan},\ and\ \citenamefont {Huang}}]{Meng:2022low}%
  \BibitemOpen
  \bibfield  {author} {\bibinfo {author} {\bibfnamefont {D.-S.}\ \bibnamefont {Meng}}, \bibinfo {author} {\bibfnamefont {C.}~\bibnamefont {Yuan}}, \ and\ \bibinfo {author} {\bibfnamefont {Q.-G.}\ \bibnamefont {Huang}},\ }\href {\doibase 10.1007/s11433-022-2095-5} {\bibfield  {journal} {\bibinfo  {journal} {Sci. China Phys. Mech. Astron.}\ }\textbf {\bibinfo {volume} {66}},\ \bibinfo {pages} {280411} (\bibinfo {year} {2023})},\ \Eprint {http://arxiv.org/abs/2212.03577} {arXiv:2212.03577 [astro-ph.CO]} \BibitemShut {NoStop}%
\bibitem [{\citenamefont {Gu}\ \emph {et~al.}(2023)\citenamefont {Gu}, \citenamefont {Shu},\ and\ \citenamefont {Yang}}]{Gu:2023mmd}%
  \BibitemOpen
  \bibfield  {author} {\bibinfo {author} {\bibfnamefont {B.-M.}\ \bibnamefont {Gu}}, \bibinfo {author} {\bibfnamefont {F.-W.}\ \bibnamefont {Shu}}, \ and\ \bibinfo {author} {\bibfnamefont {K.}~\bibnamefont {Yang}},\ }\href@noop {} {\  (\bibinfo {year} {2023})},\ \Eprint {http://arxiv.org/abs/2307.00510} {arXiv:2307.00510 [astro-ph.CO]} \BibitemShut {NoStop}%
\bibitem [{\citenamefont {Sasaki}\ \emph {et~al.}(2018)\citenamefont {Sasaki}, \citenamefont {Suyama}, \citenamefont {Tanaka},\ and\ \citenamefont {Yokoyama}}]{Sasaki:2018dmp}%
  \BibitemOpen
  \bibfield  {author} {\bibinfo {author} {\bibfnamefont {M.}~\bibnamefont {Sasaki}}, \bibinfo {author} {\bibfnamefont {T.}~\bibnamefont {Suyama}}, \bibinfo {author} {\bibfnamefont {T.}~\bibnamefont {Tanaka}}, \ and\ \bibinfo {author} {\bibfnamefont {S.}~\bibnamefont {Yokoyama}},\ }\href {\doibase 10.1088/1361-6382/aaa7b4} {\bibfield  {journal} {\bibinfo  {journal} {Class. Quant. Grav.}\ }\textbf {\bibinfo {volume} {35}},\ \bibinfo {pages} {063001} (\bibinfo {year} {2018})},\ \Eprint {http://arxiv.org/abs/1801.05235} {arXiv:1801.05235 [astro-ph.CO]} \BibitemShut {NoStop}%
\bibitem [{\citenamefont {Carr}\ \emph {et~al.}(2021)\citenamefont {Carr}, \citenamefont {Kohri}, \citenamefont {Sendouda},\ and\ \citenamefont {Yokoyama}}]{Carr:2020gox}%
  \BibitemOpen
  \bibfield  {author} {\bibinfo {author} {\bibfnamefont {B.}~\bibnamefont {Carr}}, \bibinfo {author} {\bibfnamefont {K.}~\bibnamefont {Kohri}}, \bibinfo {author} {\bibfnamefont {Y.}~\bibnamefont {Sendouda}}, \ and\ \bibinfo {author} {\bibfnamefont {J.}~\bibnamefont {Yokoyama}},\ }\href {\doibase 10.1088/1361-6633/ac1e31} {\bibfield  {journal} {\bibinfo  {journal} {Rept. Prog. Phys.}\ }\textbf {\bibinfo {volume} {84}},\ \bibinfo {pages} {116902} (\bibinfo {year} {2021})},\ \Eprint {http://arxiv.org/abs/2002.12778} {arXiv:2002.12778 [astro-ph.CO]} \BibitemShut {NoStop}%
\bibitem [{\citenamefont {Carr}\ and\ \citenamefont {Kuhnel}(2020)}]{Carr:2020xqk}%
  \BibitemOpen
  \bibfield  {author} {\bibinfo {author} {\bibfnamefont {B.}~\bibnamefont {Carr}}\ and\ \bibinfo {author} {\bibfnamefont {F.}~\bibnamefont {Kuhnel}},\ }\href {\doibase 10.1146/annurev-nucl-050520-125911} {\bibfield  {journal} {\bibinfo  {journal} {Ann. Rev. Nucl. Part. Sci.}\ }\textbf {\bibinfo {volume} {70}},\ \bibinfo {pages} {355} (\bibinfo {year} {2020})},\ \Eprint {http://arxiv.org/abs/2006.02838} {arXiv:2006.02838 [astro-ph.CO]} \BibitemShut {NoStop}%
\bibitem [{\citenamefont {Ananda}\ \emph {et~al.}(2007)\citenamefont {Ananda}, \citenamefont {Clarkson},\ and\ \citenamefont {Wands}}]{Ananda:2006af}%
  \BibitemOpen
  \bibfield  {author} {\bibinfo {author} {\bibfnamefont {K.~N.}\ \bibnamefont {Ananda}}, \bibinfo {author} {\bibfnamefont {C.}~\bibnamefont {Clarkson}}, \ and\ \bibinfo {author} {\bibfnamefont {D.}~\bibnamefont {Wands}},\ }\href {\doibase 10.1103/PhysRevD.75.123518} {\bibfield  {journal} {\bibinfo  {journal} {Phys. Rev. D}\ }\textbf {\bibinfo {volume} {75}},\ \bibinfo {pages} {123518} (\bibinfo {year} {2007})},\ \Eprint {http://arxiv.org/abs/gr-qc/0612013} {arXiv:gr-qc/0612013} \BibitemShut {NoStop}%
\bibitem [{\citenamefont {Baumann}\ \emph {et~al.}(2007)\citenamefont {Baumann}, \citenamefont {Steinhardt}, \citenamefont {Takahashi},\ and\ \citenamefont {Ichiki}}]{Baumann:2007zm}%
  \BibitemOpen
  \bibfield  {author} {\bibinfo {author} {\bibfnamefont {D.}~\bibnamefont {Baumann}}, \bibinfo {author} {\bibfnamefont {P.~J.}\ \bibnamefont {Steinhardt}}, \bibinfo {author} {\bibfnamefont {K.}~\bibnamefont {Takahashi}}, \ and\ \bibinfo {author} {\bibfnamefont {K.}~\bibnamefont {Ichiki}},\ }\href {\doibase 10.1103/PhysRevD.76.084019} {\bibfield  {journal} {\bibinfo  {journal} {Phys. Rev. D}\ }\textbf {\bibinfo {volume} {76}},\ \bibinfo {pages} {084019} (\bibinfo {year} {2007})},\ \Eprint {http://arxiv.org/abs/hep-th/0703290} {arXiv:hep-th/0703290} \BibitemShut {NoStop}%
\bibitem [{\citenamefont {Garcia-Bellido}\ \emph {et~al.}(2016)\citenamefont {Garcia-Bellido}, \citenamefont {Peloso},\ and\ \citenamefont {Unal}}]{Garcia-Bellido:2016dkw}%
  \BibitemOpen
  \bibfield  {author} {\bibinfo {author} {\bibfnamefont {J.}~\bibnamefont {Garcia-Bellido}}, \bibinfo {author} {\bibfnamefont {M.}~\bibnamefont {Peloso}}, \ and\ \bibinfo {author} {\bibfnamefont {C.}~\bibnamefont {Unal}},\ }\href {\doibase 10.1088/1475-7516/2016/12/031} {\bibfield  {journal} {\bibinfo  {journal} {JCAP}\ }\textbf {\bibinfo {volume} {12}},\ \bibinfo {pages} {031} (\bibinfo {year} {2016})},\ \Eprint {http://arxiv.org/abs/1610.03763} {arXiv:1610.03763 [astro-ph.CO]} \BibitemShut {NoStop}%
\bibitem [{\citenamefont {Inomata}\ \emph {et~al.}(2017)\citenamefont {Inomata}, \citenamefont {Kawasaki}, \citenamefont {Mukaida}, \citenamefont {Tada},\ and\ \citenamefont {Yanagida}}]{Inomata:2016rbd}%
  \BibitemOpen
  \bibfield  {author} {\bibinfo {author} {\bibfnamefont {K.}~\bibnamefont {Inomata}}, \bibinfo {author} {\bibfnamefont {M.}~\bibnamefont {Kawasaki}}, \bibinfo {author} {\bibfnamefont {K.}~\bibnamefont {Mukaida}}, \bibinfo {author} {\bibfnamefont {Y.}~\bibnamefont {Tada}}, \ and\ \bibinfo {author} {\bibfnamefont {T.~T.}\ \bibnamefont {Yanagida}},\ }\href {\doibase 10.1103/PhysRevD.95.123510} {\bibfield  {journal} {\bibinfo  {journal} {Phys. Rev. D}\ }\textbf {\bibinfo {volume} {95}},\ \bibinfo {pages} {123510} (\bibinfo {year} {2017})},\ \Eprint {http://arxiv.org/abs/1611.06130} {arXiv:1611.06130 [astro-ph.CO]} \BibitemShut {NoStop}%
\bibitem [{\citenamefont {Garcia-Bellido}\ \emph {et~al.}(2017)\citenamefont {Garcia-Bellido}, \citenamefont {Peloso},\ and\ \citenamefont {Unal}}]{Garcia-Bellido:2017aan}%
  \BibitemOpen
  \bibfield  {author} {\bibinfo {author} {\bibfnamefont {J.}~\bibnamefont {Garcia-Bellido}}, \bibinfo {author} {\bibfnamefont {M.}~\bibnamefont {Peloso}}, \ and\ \bibinfo {author} {\bibfnamefont {C.}~\bibnamefont {Unal}},\ }\href {\doibase 10.1088/1475-7516/2017/09/013} {\bibfield  {journal} {\bibinfo  {journal} {JCAP}\ }\textbf {\bibinfo {volume} {09}},\ \bibinfo {pages} {013} (\bibinfo {year} {2017})},\ \Eprint {http://arxiv.org/abs/1707.02441} {arXiv:1707.02441 [astro-ph.CO]} \BibitemShut {NoStop}%
\bibitem [{\citenamefont {Kohri}\ and\ \citenamefont {Terada}(2018)}]{Kohri:2018awv}%
  \BibitemOpen
  \bibfield  {author} {\bibinfo {author} {\bibfnamefont {K.}~\bibnamefont {Kohri}}\ and\ \bibinfo {author} {\bibfnamefont {T.}~\bibnamefont {Terada}},\ }\href {\doibase 10.1103/PhysRevD.97.123532} {\bibfield  {journal} {\bibinfo  {journal} {Phys. Rev. D}\ }\textbf {\bibinfo {volume} {97}},\ \bibinfo {pages} {123532} (\bibinfo {year} {2018})},\ \Eprint {http://arxiv.org/abs/1804.08577} {arXiv:1804.08577 [gr-qc]} \BibitemShut {NoStop}%
\bibitem [{\citenamefont {Cai}\ \emph {et~al.}(2019)\citenamefont {Cai}, \citenamefont {Pi},\ and\ \citenamefont {Sasaki}}]{Cai:2018dig}%
  \BibitemOpen
  \bibfield  {author} {\bibinfo {author} {\bibfnamefont {R.-g.}\ \bibnamefont {Cai}}, \bibinfo {author} {\bibfnamefont {S.}~\bibnamefont {Pi}}, \ and\ \bibinfo {author} {\bibfnamefont {M.}~\bibnamefont {Sasaki}},\ }\href {\doibase 10.1103/PhysRevLett.122.201101} {\bibfield  {journal} {\bibinfo  {journal} {Phys. Rev. Lett.}\ }\textbf {\bibinfo {volume} {122}},\ \bibinfo {pages} {201101} (\bibinfo {year} {2019})},\ \Eprint {http://arxiv.org/abs/1810.11000} {arXiv:1810.11000 [astro-ph.CO]} \BibitemShut {NoStop}%
\bibitem [{\citenamefont {Lu}\ \emph {et~al.}(2019)\citenamefont {Lu}, \citenamefont {Gong}, \citenamefont {Yi},\ and\ \citenamefont {Zhang}}]{Lu:2019sti}%
  \BibitemOpen
  \bibfield  {author} {\bibinfo {author} {\bibfnamefont {Y.}~\bibnamefont {Lu}}, \bibinfo {author} {\bibfnamefont {Y.}~\bibnamefont {Gong}}, \bibinfo {author} {\bibfnamefont {Z.}~\bibnamefont {Yi}}, \ and\ \bibinfo {author} {\bibfnamefont {F.}~\bibnamefont {Zhang}},\ }\href {\doibase 10.1088/1475-7516/2019/12/031} {\bibfield  {journal} {\bibinfo  {journal} {JCAP}\ }\textbf {\bibinfo {volume} {12}},\ \bibinfo {pages} {031} (\bibinfo {year} {2019})},\ \Eprint {http://arxiv.org/abs/1907.11896} {arXiv:1907.11896 [gr-qc]} \BibitemShut {NoStop}%
\bibitem [{\citenamefont {Yuan}\ \emph {et~al.}(2020{\natexlab{a}})\citenamefont {Yuan}, \citenamefont {Chen},\ and\ \citenamefont {Huang}}]{Yuan:2019wwo}%
  \BibitemOpen
  \bibfield  {author} {\bibinfo {author} {\bibfnamefont {C.}~\bibnamefont {Yuan}}, \bibinfo {author} {\bibfnamefont {Z.-C.}\ \bibnamefont {Chen}}, \ and\ \bibinfo {author} {\bibfnamefont {Q.-G.}\ \bibnamefont {Huang}},\ }\href {\doibase 10.1103/PhysRevD.101.043019} {\bibfield  {journal} {\bibinfo  {journal} {Phys. Rev. D}\ }\textbf {\bibinfo {volume} {101}},\ \bibinfo {pages} {043019} (\bibinfo {year} {2020}{\natexlab{a}})},\ \Eprint {http://arxiv.org/abs/1910.09099} {arXiv:1910.09099 [astro-ph.CO]} \BibitemShut {NoStop}%
\bibitem [{\citenamefont {Chen}\ \emph {et~al.}(2020)\citenamefont {Chen}, \citenamefont {Yuan},\ and\ \citenamefont {Huang}}]{Chen:2019xse}%
  \BibitemOpen
  \bibfield  {author} {\bibinfo {author} {\bibfnamefont {Z.-C.}\ \bibnamefont {Chen}}, \bibinfo {author} {\bibfnamefont {C.}~\bibnamefont {Yuan}}, \ and\ \bibinfo {author} {\bibfnamefont {Q.-G.}\ \bibnamefont {Huang}},\ }\href {\doibase 10.1103/PhysRevLett.124.251101} {\bibfield  {journal} {\bibinfo  {journal} {Phys. Rev. Lett.}\ }\textbf {\bibinfo {volume} {124}},\ \bibinfo {pages} {251101} (\bibinfo {year} {2020})},\ \Eprint {http://arxiv.org/abs/1910.12239} {arXiv:1910.12239 [astro-ph.CO]} \BibitemShut {NoStop}%
\bibitem [{\citenamefont {Xu}\ \emph {et~al.}(2020)\citenamefont {Xu}, \citenamefont {Liu}, \citenamefont {Gao},\ and\ \citenamefont {Guo}}]{Xu:2019bdp}%
  \BibitemOpen
  \bibfield  {author} {\bibinfo {author} {\bibfnamefont {W.-T.}\ \bibnamefont {Xu}}, \bibinfo {author} {\bibfnamefont {J.}~\bibnamefont {Liu}}, \bibinfo {author} {\bibfnamefont {T.-J.}\ \bibnamefont {Gao}}, \ and\ \bibinfo {author} {\bibfnamefont {Z.-K.}\ \bibnamefont {Guo}},\ }\href {\doibase 10.1103/PhysRevD.101.023505} {\bibfield  {journal} {\bibinfo  {journal} {Phys. Rev. D}\ }\textbf {\bibinfo {volume} {101}},\ \bibinfo {pages} {023505} (\bibinfo {year} {2020})},\ \Eprint {http://arxiv.org/abs/1907.05213} {arXiv:1907.05213 [astro-ph.CO]} \BibitemShut {NoStop}%
\bibitem [{\citenamefont {Cai}\ \emph {et~al.}(2020{\natexlab{b}})\citenamefont {Cai}, \citenamefont {Pi},\ and\ \citenamefont {Sasaki}}]{Cai:2019cdl}%
  \BibitemOpen
  \bibfield  {author} {\bibinfo {author} {\bibfnamefont {R.-G.}\ \bibnamefont {Cai}}, \bibinfo {author} {\bibfnamefont {S.}~\bibnamefont {Pi}}, \ and\ \bibinfo {author} {\bibfnamefont {M.}~\bibnamefont {Sasaki}},\ }\href {\doibase 10.1103/PhysRevD.102.083528} {\bibfield  {journal} {\bibinfo  {journal} {Phys. Rev. D}\ }\textbf {\bibinfo {volume} {102}},\ \bibinfo {pages} {083528} (\bibinfo {year} {2020}{\natexlab{b}})},\ \Eprint {http://arxiv.org/abs/1909.13728} {arXiv:1909.13728 [astro-ph.CO]} \BibitemShut {NoStop}%
\bibitem [{\citenamefont {Yuan}\ \emph {et~al.}(2020{\natexlab{b}})\citenamefont {Yuan}, \citenamefont {Chen},\ and\ \citenamefont {Huang}}]{Yuan:2019fwv}%
  \BibitemOpen
  \bibfield  {author} {\bibinfo {author} {\bibfnamefont {C.}~\bibnamefont {Yuan}}, \bibinfo {author} {\bibfnamefont {Z.-C.}\ \bibnamefont {Chen}}, \ and\ \bibinfo {author} {\bibfnamefont {Q.-G.}\ \bibnamefont {Huang}},\ }\href {\doibase 10.1103/PhysRevD.101.063018} {\bibfield  {journal} {\bibinfo  {journal} {Phys. Rev. D}\ }\textbf {\bibinfo {volume} {101}},\ \bibinfo {pages} {063018} (\bibinfo {year} {2020}{\natexlab{b}})},\ \Eprint {http://arxiv.org/abs/1912.00885} {arXiv:1912.00885 [astro-ph.CO]} \BibitemShut {NoStop}%
\bibitem [{\citenamefont {Yi}\ \emph {et~al.}(2021{\natexlab{a}})\citenamefont {Yi}, \citenamefont {Gong}, \citenamefont {Wang},\ and\ \citenamefont {Zhu}}]{Yi:2020kmq}%
  \BibitemOpen
  \bibfield  {author} {\bibinfo {author} {\bibfnamefont {Z.}~\bibnamefont {Yi}}, \bibinfo {author} {\bibfnamefont {Y.}~\bibnamefont {Gong}}, \bibinfo {author} {\bibfnamefont {B.}~\bibnamefont {Wang}}, \ and\ \bibinfo {author} {\bibfnamefont {Z.-h.}\ \bibnamefont {Zhu}},\ }\href {\doibase 10.1103/PhysRevD.103.063535} {\bibfield  {journal} {\bibinfo  {journal} {Phys. Rev. D}\ }\textbf {\bibinfo {volume} {103}},\ \bibinfo {pages} {063535} (\bibinfo {year} {2021}{\natexlab{a}})},\ \Eprint {http://arxiv.org/abs/2007.09957} {arXiv:2007.09957 [gr-qc]} \BibitemShut {NoStop}%
\bibitem [{\citenamefont {Yi}\ \emph {et~al.}(2021{\natexlab{b}})\citenamefont {Yi}, \citenamefont {Gao}, \citenamefont {Gong},\ and\ \citenamefont {Zhu}}]{Yi:2020cut}%
  \BibitemOpen
  \bibfield  {author} {\bibinfo {author} {\bibfnamefont {Z.}~\bibnamefont {Yi}}, \bibinfo {author} {\bibfnamefont {Q.}~\bibnamefont {Gao}}, \bibinfo {author} {\bibfnamefont {Y.}~\bibnamefont {Gong}}, \ and\ \bibinfo {author} {\bibfnamefont {Z.-h.}\ \bibnamefont {Zhu}},\ }\href {\doibase 10.1103/PhysRevD.103.063534} {\bibfield  {journal} {\bibinfo  {journal} {Phys. Rev. D}\ }\textbf {\bibinfo {volume} {103}},\ \bibinfo {pages} {063534} (\bibinfo {year} {2021}{\natexlab{b}})},\ \Eprint {http://arxiv.org/abs/2011.10606} {arXiv:2011.10606 [astro-ph.CO]} \BibitemShut {NoStop}%
\bibitem [{\citenamefont {Liu}\ \emph {et~al.}(2020{\natexlab{c}})\citenamefont {Liu}, \citenamefont {Guo},\ and\ \citenamefont {Cai}}]{Liu:2020oqe}%
  \BibitemOpen
  \bibfield  {author} {\bibinfo {author} {\bibfnamefont {J.}~\bibnamefont {Liu}}, \bibinfo {author} {\bibfnamefont {Z.-K.}\ \bibnamefont {Guo}}, \ and\ \bibinfo {author} {\bibfnamefont {R.-G.}\ \bibnamefont {Cai}},\ }\href {\doibase 10.1103/PhysRevD.101.083535} {\bibfield  {journal} {\bibinfo  {journal} {Phys. Rev. D}\ }\textbf {\bibinfo {volume} {101}},\ \bibinfo {pages} {083535} (\bibinfo {year} {2020}{\natexlab{c}})},\ \Eprint {http://arxiv.org/abs/2003.02075} {arXiv:2003.02075 [astro-ph.CO]} \BibitemShut {NoStop}%
\bibitem [{\citenamefont {Gao}\ \emph {et~al.}(2021)\citenamefont {Gao}, \citenamefont {Gong},\ and\ \citenamefont {Yi}}]{Gao:2020tsa}%
  \BibitemOpen
  \bibfield  {author} {\bibinfo {author} {\bibfnamefont {Q.}~\bibnamefont {Gao}}, \bibinfo {author} {\bibfnamefont {Y.}~\bibnamefont {Gong}}, \ and\ \bibinfo {author} {\bibfnamefont {Z.}~\bibnamefont {Yi}},\ }\href {\doibase 10.1016/j.nuclphysb.2021.115480} {\bibfield  {journal} {\bibinfo  {journal} {Nucl. Phys. B}\ }\textbf {\bibinfo {volume} {969}},\ \bibinfo {pages} {115480} (\bibinfo {year} {2021})},\ \Eprint {http://arxiv.org/abs/2012.03856} {arXiv:2012.03856 [gr-qc]} \BibitemShut {NoStop}%
\bibitem [{\citenamefont {Yi}\ and\ \citenamefont {Zhu}(2022)}]{Yi:2021lxc}%
  \BibitemOpen
  \bibfield  {author} {\bibinfo {author} {\bibfnamefont {Z.}~\bibnamefont {Yi}}\ and\ \bibinfo {author} {\bibfnamefont {Z.-H.}\ \bibnamefont {Zhu}},\ }\href {\doibase 10.1088/1475-7516/2022/05/046} {\bibfield  {journal} {\bibinfo  {journal} {JCAP}\ }\textbf {\bibinfo {volume} {05}},\ \bibinfo {pages} {046} (\bibinfo {year} {2022})},\ \Eprint {http://arxiv.org/abs/2105.01943} {arXiv:2105.01943 [gr-qc]} \BibitemShut {NoStop}%
\bibitem [{\citenamefont {Yi}(2023)}]{Yi:2022anu}%
  \BibitemOpen
  \bibfield  {author} {\bibinfo {author} {\bibfnamefont {Z.}~\bibnamefont {Yi}},\ }\href {\doibase 10.1088/1475-7516/2023/03/048} {\bibfield  {journal} {\bibinfo  {journal} {JCAP}\ }\textbf {\bibinfo {volume} {03}},\ \bibinfo {pages} {048} (\bibinfo {year} {2023})},\ \Eprint {http://arxiv.org/abs/2206.01039} {arXiv:2206.01039 [gr-qc]} \BibitemShut {NoStop}%
\bibitem [{\citenamefont {Yi}\ and\ \citenamefont {Fei}(2023)}]{Yi:2022ymw}%
  \BibitemOpen
  \bibfield  {author} {\bibinfo {author} {\bibfnamefont {Z.}~\bibnamefont {Yi}}\ and\ \bibinfo {author} {\bibfnamefont {Q.}~\bibnamefont {Fei}},\ }\href {\doibase 10.1140/epjc/s10052-023-11233-3} {\bibfield  {journal} {\bibinfo  {journal} {Eur. Phys. J. C}\ }\textbf {\bibinfo {volume} {83}},\ \bibinfo {pages} {82} (\bibinfo {year} {2023})},\ \Eprint {http://arxiv.org/abs/2210.03641} {arXiv:2210.03641 [astro-ph.CO]} \BibitemShut {NoStop}%
\bibitem [{\citenamefont {Yuan}\ \emph {et~al.}(2019)\citenamefont {Yuan}, \citenamefont {Chen},\ and\ \citenamefont {Huang}}]{Yuan:2019udt}%
  \BibitemOpen
  \bibfield  {author} {\bibinfo {author} {\bibfnamefont {C.}~\bibnamefont {Yuan}}, \bibinfo {author} {\bibfnamefont {Z.-C.}\ \bibnamefont {Chen}}, \ and\ \bibinfo {author} {\bibfnamefont {Q.-G.}\ \bibnamefont {Huang}},\ }\href {\doibase 10.1103/PhysRevD.100.081301} {\bibfield  {journal} {\bibinfo  {journal} {Phys. Rev. D}\ }\textbf {\bibinfo {volume} {100}},\ \bibinfo {pages} {081301} (\bibinfo {year} {2019})},\ \Eprint {http://arxiv.org/abs/1906.11549} {arXiv:1906.11549 [astro-ph.CO]} \BibitemShut {NoStop}%
\bibitem [{\citenamefont {Inui}\ \emph {et~al.}(2023)\citenamefont {Inui}, \citenamefont {Jaraba}, \citenamefont {Kuroyanagi},\ and\ \citenamefont {Yokoyama}}]{Inui:2023qsd}%
  \BibitemOpen
  \bibfield  {author} {\bibinfo {author} {\bibfnamefont {R.}~\bibnamefont {Inui}}, \bibinfo {author} {\bibfnamefont {S.}~\bibnamefont {Jaraba}}, \bibinfo {author} {\bibfnamefont {S.}~\bibnamefont {Kuroyanagi}}, \ and\ \bibinfo {author} {\bibfnamefont {S.}~\bibnamefont {Yokoyama}},\ }\href@noop {} {\  (\bibinfo {year} {2023})},\ \Eprint {http://arxiv.org/abs/2311.05423} {arXiv:2311.05423 [astro-ph.CO]} \BibitemShut {NoStop}%
\bibitem [{\citenamefont {Dom\`enech}(2021)}]{Domenech:2021ztg}%
  \BibitemOpen
  \bibfield  {author} {\bibinfo {author} {\bibfnamefont {G.}~\bibnamefont {Dom\`enech}},\ }\href {\doibase 10.3390/universe7110398} {\bibfield  {journal} {\bibinfo  {journal} {Universe}\ }\textbf {\bibinfo {volume} {7}},\ \bibinfo {pages} {398} (\bibinfo {year} {2021})},\ \Eprint {http://arxiv.org/abs/2109.01398} {arXiv:2109.01398 [gr-qc]} \BibitemShut {NoStop}%
\bibitem [{\citenamefont {Abbott}\ \emph {et~al.}(2017{\natexlab{a}})\citenamefont {Abbott} \emph {et~al.}}]{LIGOScientific:2017vwq}%
  \BibitemOpen
  \bibfield  {author} {\bibinfo {author} {\bibfnamefont {B.~P.}\ \bibnamefont {Abbott}} \emph {et~al.} (\bibinfo {collaboration} {LIGO Scientific, Virgo}),\ }\href {\doibase 10.1103/PhysRevLett.119.161101} {\bibfield  {journal} {\bibinfo  {journal} {Phys. Rev. Lett.}\ }\textbf {\bibinfo {volume} {119}},\ \bibinfo {pages} {161101} (\bibinfo {year} {2017}{\natexlab{a}})},\ \Eprint {http://arxiv.org/abs/1710.05832} {arXiv:1710.05832 [gr-qc]} \BibitemShut {NoStop}%
\bibitem [{\citenamefont {Abbott}\ \emph {et~al.}(2017{\natexlab{b}})\citenamefont {Abbott} \emph {et~al.}}]{LIGOScientific:2017zic}%
  \BibitemOpen
  \bibfield  {author} {\bibinfo {author} {\bibfnamefont {B.~P.}\ \bibnamefont {Abbott}} \emph {et~al.} (\bibinfo {collaboration} {LIGO Scientific, Virgo, Fermi-GBM, INTEGRAL}),\ }\href {\doibase 10.3847/2041-8213/aa920c} {\bibfield  {journal} {\bibinfo  {journal} {Astrophys. J. Lett.}\ }\textbf {\bibinfo {volume} {848}},\ \bibinfo {pages} {L13} (\bibinfo {year} {2017}{\natexlab{b}})},\ \Eprint {http://arxiv.org/abs/1710.05834} {arXiv:1710.05834 [astro-ph.HE]} \BibitemShut {NoStop}%
\bibitem [{\citenamefont {de~Rham}\ \emph {et~al.}(2011)\citenamefont {de~Rham}, \citenamefont {Gabadadze},\ and\ \citenamefont {Tolley}}]{deRham:2010kj}%
  \BibitemOpen
  \bibfield  {author} {\bibinfo {author} {\bibfnamefont {C.}~\bibnamefont {de~Rham}}, \bibinfo {author} {\bibfnamefont {G.}~\bibnamefont {Gabadadze}}, \ and\ \bibinfo {author} {\bibfnamefont {A.~J.}\ \bibnamefont {Tolley}},\ }\href {\doibase 10.1103/PhysRevLett.106.231101} {\bibfield  {journal} {\bibinfo  {journal} {Phys. Rev. Lett.}\ }\textbf {\bibinfo {volume} {106}},\ \bibinfo {pages} {231101} (\bibinfo {year} {2011})},\ \Eprint {http://arxiv.org/abs/1011.1232} {arXiv:1011.1232 [hep-th]} \BibitemShut {NoStop}%
\bibitem [{\citenamefont {Hassan}\ and\ \citenamefont {Rosen}(2012{\natexlab{a}})}]{Hassan:2011hr}%
  \BibitemOpen
  \bibfield  {author} {\bibinfo {author} {\bibfnamefont {S.~F.}\ \bibnamefont {Hassan}}\ and\ \bibinfo {author} {\bibfnamefont {R.~A.}\ \bibnamefont {Rosen}},\ }\href {\doibase 10.1103/PhysRevLett.108.041101} {\bibfield  {journal} {\bibinfo  {journal} {Phys. Rev. Lett.}\ }\textbf {\bibinfo {volume} {108}},\ \bibinfo {pages} {041101} (\bibinfo {year} {2012}{\natexlab{a}})},\ \Eprint {http://arxiv.org/abs/1106.3344} {arXiv:1106.3344 [hep-th]} \BibitemShut {NoStop}%
\bibitem [{\citenamefont {Hassan}\ and\ \citenamefont {Rosen}(2012{\natexlab{b}})}]{Hassan:2011zd}%
  \BibitemOpen
  \bibfield  {author} {\bibinfo {author} {\bibfnamefont {S.~F.}\ \bibnamefont {Hassan}}\ and\ \bibinfo {author} {\bibfnamefont {R.~A.}\ \bibnamefont {Rosen}},\ }\href {\doibase 10.1007/JHEP02(2012)126} {\bibfield  {journal} {\bibinfo  {journal} {JHEP}\ }\textbf {\bibinfo {volume} {02}},\ \bibinfo {pages} {126} (\bibinfo {year} {2012}{\natexlab{b}})},\ \Eprint {http://arxiv.org/abs/1109.3515} {arXiv:1109.3515 [hep-th]} \BibitemShut {NoStop}%
\bibitem [{\citenamefont {Schmidt-May}\ and\ \citenamefont {von Strauss}(2016)}]{Schmidt-May:2015vnx}%
  \BibitemOpen
  \bibfield  {author} {\bibinfo {author} {\bibfnamefont {A.}~\bibnamefont {Schmidt-May}}\ and\ \bibinfo {author} {\bibfnamefont {M.}~\bibnamefont {von Strauss}},\ }\href {\doibase 10.1088/1751-8113/49/18/183001} {\bibfield  {journal} {\bibinfo  {journal} {J. Phys. A}\ }\textbf {\bibinfo {volume} {49}},\ \bibinfo {pages} {183001} (\bibinfo {year} {2016})},\ \Eprint {http://arxiv.org/abs/1512.00021} {arXiv:1512.00021 [hep-th]} \BibitemShut {NoStop}%
\bibitem [{\citenamefont {Dvali}\ \emph {et~al.}(2000)\citenamefont {Dvali}, \citenamefont {Gabadadze},\ and\ \citenamefont {Porrati}}]{Dvali:2000hr}%
  \BibitemOpen
  \bibfield  {author} {\bibinfo {author} {\bibfnamefont {G.~R.}\ \bibnamefont {Dvali}}, \bibinfo {author} {\bibfnamefont {G.}~\bibnamefont {Gabadadze}}, \ and\ \bibinfo {author} {\bibfnamefont {M.}~\bibnamefont {Porrati}},\ }\href {\doibase 10.1016/S0370-2693(00)00669-9} {\bibfield  {journal} {\bibinfo  {journal} {Phys. Lett. B}\ }\textbf {\bibinfo {volume} {485}},\ \bibinfo {pages} {208} (\bibinfo {year} {2000})},\ \Eprint {http://arxiv.org/abs/hep-th/0005016} {arXiv:hep-th/0005016} \BibitemShut {NoStop}%
\bibitem [{\citenamefont {Weinberg}(2004)}]{Weinberg:2003ur}%
  \BibitemOpen
  \bibfield  {author} {\bibinfo {author} {\bibfnamefont {S.}~\bibnamefont {Weinberg}},\ }\href {\doibase 10.1103/PhysRevD.69.023503} {\bibfield  {journal} {\bibinfo  {journal} {Phys. Rev. D}\ }\textbf {\bibinfo {volume} {69}},\ \bibinfo {pages} {023503} (\bibinfo {year} {2004})},\ \Eprint {http://arxiv.org/abs/astro-ph/0306304} {arXiv:astro-ph/0306304} \BibitemShut {NoStop}%
\bibitem [{\citenamefont {Watanabe}\ and\ \citenamefont {Komatsu}(2006)}]{Watanabe:2006qe}%
  \BibitemOpen
  \bibfield  {author} {\bibinfo {author} {\bibfnamefont {Y.}~\bibnamefont {Watanabe}}\ and\ \bibinfo {author} {\bibfnamefont {E.}~\bibnamefont {Komatsu}},\ }\href {\doibase 10.1103/PhysRevD.73.123515} {\bibfield  {journal} {\bibinfo  {journal} {Phys. Rev. D}\ }\textbf {\bibinfo {volume} {73}},\ \bibinfo {pages} {123515} (\bibinfo {year} {2006})},\ \Eprint {http://arxiv.org/abs/astro-ph/0604176} {arXiv:astro-ph/0604176} \BibitemShut {NoStop}%
\bibitem [{\citenamefont {Li}\ and\ \citenamefont {Guo}(2023)}]{Li:2023uhu}%
  \BibitemOpen
  \bibfield  {author} {\bibinfo {author} {\bibfnamefont {J.}~\bibnamefont {Li}}\ and\ \bibinfo {author} {\bibfnamefont {G.-H.}\ \bibnamefont {Guo}},\ }\href@noop {} {\  (\bibinfo {year} {2023})},\ \Eprint {http://arxiv.org/abs/2312.04589} {arXiv:2312.04589 [gr-qc]} \BibitemShut {NoStop}%
\bibitem [{\citenamefont {Espinosa}\ \emph {et~al.}(2018)\citenamefont {Espinosa}, \citenamefont {Racco},\ and\ \citenamefont {Riotto}}]{Espinosa:2018eve}%
  \BibitemOpen
  \bibfield  {author} {\bibinfo {author} {\bibfnamefont {J.~R.}\ \bibnamefont {Espinosa}}, \bibinfo {author} {\bibfnamefont {D.}~\bibnamefont {Racco}}, \ and\ \bibinfo {author} {\bibfnamefont {A.}~\bibnamefont {Riotto}},\ }\href {\doibase 10.1088/1475-7516/2018/09/012} {\bibfield  {journal} {\bibinfo  {journal} {JCAP}\ }\textbf {\bibinfo {volume} {09}},\ \bibinfo {pages} {012} (\bibinfo {year} {2018})},\ \Eprint {http://arxiv.org/abs/1804.07732} {arXiv:1804.07732 [hep-ph]} \BibitemShut {NoStop}%
\bibitem [{\citenamefont {Thrane}\ and\ \citenamefont {Romano}(2013)}]{Thrane:2013oya}%
  \BibitemOpen
  \bibfield  {author} {\bibinfo {author} {\bibfnamefont {E.}~\bibnamefont {Thrane}}\ and\ \bibinfo {author} {\bibfnamefont {J.~D.}\ \bibnamefont {Romano}},\ }\href {\doibase 10.1103/PhysRevD.88.124032} {\bibfield  {journal} {\bibinfo  {journal} {Phys. Rev. D}\ }\textbf {\bibinfo {volume} {88}},\ \bibinfo {pages} {124032} (\bibinfo {year} {2013})},\ \Eprint {http://arxiv.org/abs/1310.5300} {arXiv:1310.5300 [astro-ph.IM]} \BibitemShut {NoStop}%
\bibitem [{\citenamefont {Hellings}\ and\ \citenamefont {Downs}(1983)}]{Hellings:1983fr}%
  \BibitemOpen
  \bibfield  {author} {\bibinfo {author} {\bibfnamefont {R.~w.}\ \bibnamefont {Hellings}}\ and\ \bibinfo {author} {\bibfnamefont {G.~s.}\ \bibnamefont {Downs}},\ }\href {\doibase 10.1086/183954} {\bibfield  {journal} {\bibinfo  {journal} {Astrophys. J. Lett.}\ }\textbf {\bibinfo {volume} {265}},\ \bibinfo {pages} {L39} (\bibinfo {year} {1983})}\BibitemShut {NoStop}%
\bibitem [{\citenamefont {Bernardo}\ and\ \citenamefont {Ng}(2023{\natexlab{a}})}]{Bernardo:2023zna}%
  \BibitemOpen
  \bibfield  {author} {\bibinfo {author} {\bibfnamefont {R.~C.}\ \bibnamefont {Bernardo}}\ and\ \bibinfo {author} {\bibfnamefont {K.-W.}\ \bibnamefont {Ng}},\ }\href@noop {} {\  (\bibinfo {year} {2023}{\natexlab{a}})},\ \Eprint {http://arxiv.org/abs/2310.07537} {arXiv:2310.07537 [gr-qc]} \BibitemShut {NoStop}%
\bibitem [{\citenamefont {Bernardo}\ and\ \citenamefont {Ng}(2023{\natexlab{b}})}]{Bernardo:2023mxc}%
  \BibitemOpen
  \bibfield  {author} {\bibinfo {author} {\bibfnamefont {R.~C.}\ \bibnamefont {Bernardo}}\ and\ \bibinfo {author} {\bibfnamefont {K.-W.}\ \bibnamefont {Ng}},\ }\href {\doibase 10.1103/PhysRevD.107.L101502} {\bibfield  {journal} {\bibinfo  {journal} {Phys. Rev. D}\ }\textbf {\bibinfo {volume} {107}},\ \bibinfo {pages} {L101502} (\bibinfo {year} {2023}{\natexlab{b}})},\ \Eprint {http://arxiv.org/abs/2302.11796} {arXiv:2302.11796 [gr-qc]} \BibitemShut {NoStop}%
\bibitem [{\citenamefont {Antoniadis}\ \emph {et~al.}(2023{\natexlab{d}})\citenamefont {Antoniadis} \emph {et~al.}}]{Antoniadis:2023rey}%
  \BibitemOpen
  \bibfield  {author} {\bibinfo {author} {\bibfnamefont {J.}~\bibnamefont {Antoniadis}} \emph {et~al.} (\bibinfo {collaboration} {EPTA, InPTA:}),\ }\href {\doibase 10.1051/0004-6361/202346844} {\bibfield  {journal} {\bibinfo  {journal} {Astron. Astrophys.}\ }\textbf {\bibinfo {volume} {678}},\ \bibinfo {pages} {A50} (\bibinfo {year} {2023}{\natexlab{d}})},\ \Eprint {http://arxiv.org/abs/2306.16214} {arXiv:2306.16214 [astro-ph.HE]} \BibitemShut {NoStop}%
\bibitem [{\citenamefont {Aghanim}\ \emph {et~al.}(2020)\citenamefont {Aghanim} \emph {et~al.}}]{Planck:2018vyg}%
  \BibitemOpen
  \bibfield  {author} {\bibinfo {author} {\bibfnamefont {N.}~\bibnamefont {Aghanim}} \emph {et~al.} (\bibinfo {collaboration} {Planck}),\ }\href {\doibase 10.1051/0004-6361/201833910} {\bibfield  {journal} {\bibinfo  {journal} {Astron. Astrophys.}\ }\textbf {\bibinfo {volume} {641}},\ \bibinfo {pages} {A6} (\bibinfo {year} {2020})},\ \bibinfo {note} {[Erratum: Astron.Astrophys. 652, C4 (2021)]},\ \Eprint {http://arxiv.org/abs/1807.06209} {arXiv:1807.06209 [astro-ph.CO]} \BibitemShut {NoStop}%
\bibitem [{\citenamefont {Moore}\ and\ \citenamefont {Vecchio}(2021)}]{Moore:2021ibq}%
  \BibitemOpen
  \bibfield  {author} {\bibinfo {author} {\bibfnamefont {C.~J.}\ \bibnamefont {Moore}}\ and\ \bibinfo {author} {\bibfnamefont {A.}~\bibnamefont {Vecchio}},\ }\href {\doibase 10.1038/s41550-021-01489-8} {\bibfield  {journal} {\bibinfo  {journal} {Nature Astron.}\ }\textbf {\bibinfo {volume} {5}},\ \bibinfo {pages} {1268} (\bibinfo {year} {2021})},\ \Eprint {http://arxiv.org/abs/2104.15130} {arXiv:2104.15130 [astro-ph.CO]} \BibitemShut {NoStop}%
\bibitem [{\citenamefont {Lamb}\ \emph {et~al.}(2023)\citenamefont {Lamb}, \citenamefont {Taylor},\ and\ \citenamefont {van Haasteren}}]{Lamb:2023jls}%
  \BibitemOpen
  \bibfield  {author} {\bibinfo {author} {\bibfnamefont {W.~G.}\ \bibnamefont {Lamb}}, \bibinfo {author} {\bibfnamefont {S.~R.}\ \bibnamefont {Taylor}}, \ and\ \bibinfo {author} {\bibfnamefont {R.}~\bibnamefont {van Haasteren}},\ }\href {\doibase 10.1103/PhysRevD.108.103019} {\bibfield  {journal} {\bibinfo  {journal} {Phys. Rev. D}\ }\textbf {\bibinfo {volume} {108}},\ \bibinfo {pages} {103019} (\bibinfo {year} {2023})},\ \Eprint {http://arxiv.org/abs/2303.15442} {arXiv:2303.15442 [astro-ph.HE]} \BibitemShut {NoStop}%
\bibitem [{\citenamefont {Speagle}(2020)}]{Speagle:2019ivv}%
  \BibitemOpen
  \bibfield  {author} {\bibinfo {author} {\bibfnamefont {J.~S.}\ \bibnamefont {Speagle}},\ }\href {\doibase 10.1093/mnras/staa278} {\bibfield  {journal} {\bibinfo  {journal} {Mon. Not. Roy. Astron. Soc.}\ }\textbf {\bibinfo {volume} {493}},\ \bibinfo {pages} {3132} (\bibinfo {year} {2020})},\ \Eprint {http://arxiv.org/abs/1904.02180} {arXiv:1904.02180 [astro-ph.IM]} \BibitemShut {NoStop}%
\bibitem [{\citenamefont {Ashton}\ \emph {et~al.}(2019)\citenamefont {Ashton} \emph {et~al.}}]{Ashton:2018jfp}%
  \BibitemOpen
  \bibfield  {author} {\bibinfo {author} {\bibfnamefont {G.}~\bibnamefont {Ashton}} \emph {et~al.},\ }\href {\doibase 10.3847/1538-4365/ab06fc} {\bibfield  {journal} {\bibinfo  {journal} {Astrophys. J. Suppl.}\ }\textbf {\bibinfo {volume} {241}},\ \bibinfo {pages} {27} (\bibinfo {year} {2019})},\ \Eprint {http://arxiv.org/abs/1811.02042} {arXiv:1811.02042 [astro-ph.IM]} \BibitemShut {NoStop}%
\bibitem [{\citenamefont {Romero-Shaw}\ \emph {et~al.}(2020)\citenamefont {Romero-Shaw} \emph {et~al.}}]{Romero-Shaw:2020owr}%
  \BibitemOpen
  \bibfield  {author} {\bibinfo {author} {\bibfnamefont {I.~M.}\ \bibnamefont {Romero-Shaw}} \emph {et~al.},\ }\href {\doibase 10.1093/mnras/staa2850} {\bibfield  {journal} {\bibinfo  {journal} {Mon. Not. Roy. Astron. Soc.}\ }\textbf {\bibinfo {volume} {499}},\ \bibinfo {pages} {3295} (\bibinfo {year} {2020})},\ \Eprint {http://arxiv.org/abs/2006.00714} {arXiv:2006.00714 [astro-ph.IM]} \BibitemShut {NoStop}%
\bibitem [{\citenamefont {de~Rham}\ and\ \citenamefont {Melville}(2018)}]{deRham:2018red}%
  \BibitemOpen
  \bibfield  {author} {\bibinfo {author} {\bibfnamefont {C.}~\bibnamefont {de~Rham}}\ and\ \bibinfo {author} {\bibfnamefont {S.}~\bibnamefont {Melville}},\ }\href {\doibase 10.1103/PhysRevLett.121.221101} {\bibfield  {journal} {\bibinfo  {journal} {Phys. Rev. Lett.}\ }\textbf {\bibinfo {volume} {121}},\ \bibinfo {pages} {221101} (\bibinfo {year} {2018})},\ \Eprint {http://arxiv.org/abs/1806.09417} {arXiv:1806.09417 [hep-th]} \BibitemShut {NoStop}%
\bibitem [{\citenamefont {Baker}\ \emph {et~al.}(2022)\citenamefont {Baker} \emph {et~al.}}]{LISACosmologyWorkingGroup:2022wjo}%
  \BibitemOpen
  \bibfield  {author} {\bibinfo {author} {\bibfnamefont {T.}~\bibnamefont {Baker}} \emph {et~al.} (\bibinfo {collaboration} {LISA Cosmology Working Group}),\ }\href {\doibase 10.1088/1475-7516/2022/08/031} {\bibfield  {journal} {\bibinfo  {journal} {JCAP}\ }\textbf {\bibinfo {volume} {08}},\ \bibinfo {pages} {031} (\bibinfo {year} {2022})},\ \Eprint {http://arxiv.org/abs/2203.00566} {arXiv:2203.00566 [gr-qc]} \BibitemShut {NoStop}%
\bibitem [{\citenamefont {Lazio}(2013)}]{Lazio:2013mea}%
  \BibitemOpen
  \bibfield  {author} {\bibinfo {author} {\bibfnamefont {T.~J.~W.}\ \bibnamefont {Lazio}},\ }\href {\doibase 10.1088/0264-9381/30/22/224011} {\bibfield  {journal} {\bibinfo  {journal} {Class. Quant. Grav.}\ }\textbf {\bibinfo {volume} {30}},\ \bibinfo {pages} {224011} (\bibinfo {year} {2013})}\BibitemShut {NoStop}%
\end{thebibliography}%

\end{document}